\documentclass[%
reprint,
nofootinbib,
amsmath,amssymb,
aps,
showkeys,
onecolumn,
pra,
]{revtex4-2}
\usepackage[utf8]{inputenc} 
\usepackage[T1]{fontenc}    
\usepackage{hyperref}       
\usepackage{url}            
\usepackage{booktabs}       
\usepackage{amsfonts}       
\usepackage{nicefrac}       
\usepackage{microtype}      
\usepackage{xcolor}         
\usepackage{amsthm}
\usepackage[normalem]{ulem} 
\usepackage{array}
\usepackage{amsmath}
\usepackage{amssymb}
\usepackage{graphicx}
\usepackage{mathtools}
\usepackage{footnote}
\usepackage{comment}
\usepackage{cprotect}
\usepackage{accents}
\usepackage{placeins}
\usepackage{float}

\usepackage{algorithm}

\usepackage{bbold}
\usepackage{algpseudocode}
\setcitestyle{numbers,square}

\DeclareMathOperator{\R}{\mathbb{R}}

\DeclareMathOperator{\hrho}{\hat{\rho}}

\newcommand{\x}{\underbar{x}}
\newcommand{\z}{\underbar{z}}
\newcommand{\B}{\underbar{B}}

\newcommand{\dd}{\mathop{}\!\mathrm{d}}
\newcommand{\ii}{\mathrm{i}}

\usepackage{dcolumn}
\usepackage{bm}

\usepackage{indentfirst}
\begin{document}

\preprint{APS/123-QED}

\title{Eigenvalue Crossings and Discontinuous BBP Transitions in Spiked Euclidean Random Matrices} 

\author{Pasquale Casaburi}
\email{pasquale.casaburi@kcl.ac.uk}

\author{Pierpaolo Vivo}

\affiliation{Department of Mathematics, King’s College London, The Strand, London WC2R 2LS, UK}

\begin{abstract}
We study the recovery of a rank-one signal embedded in a symmetric Euclidean random matrix, whose dense and geometrically correlated entries generally lack rotational invariance. We show that signal recovery through the leading eigenvector is governed by a competition between the principal directions of the noise and the signal. In particular, the signal may already be encoded in a subleading eigendirection, while remaining undetected by the leading principal component. As the signal strength increases, a crossing between competing eigenvalues can change their ordering, causing the leading eigenvector to switch abruptly from a noise-dominated to a signal-bearing direction. This mechanism gives rise to a discontinuous BBP-type transition, in which the overlap with the signal jumps from zero to a finite value. Using the replica method, we reduce the problem to a Fredholm eigenvalue equation and derive general criteria for the existence of a positive recovery threshold, expressed in terms of the unperturbed spectral edge, the spike mean and the spatial averages of the principal eigenfunctions. We obtain analytically the limiting largest eigenvalue, eigenvector-component distribution and signal overlap, and illustrate the results with explicit solutions for quadratic distance and rank-two kernels, finding excellent agreement with numerical diagonalisation.
\end{abstract}


\maketitle

\section{Introduction}\label{sec:introduction}

Extracting a structured signal from noisy observations is a central problem in statistical inference. A common approach is to construct a matrix from the observations and use its leading eigenvectors to estimate latent directions. Principal component analysis (PCA) provides a familiar example: the eigenvectors of a sample covariance matrix identify mutually orthogonal directions of decreasing empirical variance, allowing a small number of components to summarize a large dataset~\cite{JolliffeCadima2016}. Related spectral approaches are used to distinguish collective correlations from sampling noise in financial time series and to detect communities in networks~\cite{Laloux1999,Plerou1999,NadakuditiNewman2012,Krzakala2013}. In these applications, the usefulness of an eigenvector depends on whether it captures the structure of interest or rather a feature of the random background.
A standard formulation of rank-one signal recovery is the additive spiked model
\begin{equation}\label{eq:intro_rank_one_spike}
    X^{(\theta)}
    = X^{(0)} + \frac{\theta}{N}\vec y\,\vec y^{\,\top},
\end{equation}
where $X^{(0)}$ is a random matrix representing the noise, $\vec y\in\mathbb R^N$ is the unknown signal vector, and $\theta>0$ controls its strength. Within this framework, recovering the signal through the leading principal component of standard PCA amounts to estimating the direction of $\vec y$ using the top eigenvector of $X^{(\theta)}$, with the overlap between the two vectors measuring the quality of recovery. The principal eigenpair of $X^{(\theta)}$, consisting of the largest eigenvalue and its associated eigenvector, is therefore the key object for understanding how the performance of this estimator depends on the signal strength~\cite{Perry2018}.

Such spiked models provide an analytically tractable framework for studying signal recovery, with early studies focusing on dense random noise matrices and establishing the Baik, Ben Arous and P\'ech\'e (BBP) transition at a critical signal strength $\theta_c$~\cite{BBP2005,Peche2006,FeralPeche2007}. For $\theta<\theta_c$, the largest eigenvalue remains at the upper edge of the noise spectrum and the top eigenvector has vanishing overlap with the signal, whereas for $\theta>\theta_c$ the eigenvalue separates from the spectrum and its eigenvector acquires a nonzero overlap, allowing partial recovery~\cite{Paul2007,BenaychGeorgesNadakuditi2011}. The transition thus identifies a recovery threshold for standard PCA, whose performance can differ from what is achievable through other inference procedures~\cite{Perry2018}.

The behavior of extreme eigenvalues under finite-rank additive perturbations has been studied for Gaussian Hermitian and more general Wigner backgrounds~\cite{Peche2006,FeralPeche2007,Capitaine2009}. A general framework treats additive and multiplicative perturbations of any fixed finite rank for symmetric or Hermitian noise backgrounds without requiring Gaussian or independent entries~\cite{BenaychGeorgesNadakuditi2011}. Assuming that the background and perturbation are independent and at least one is rotationally invariant, this framework characterizes limiting outlier locations and eigenvector alignments through integral transforms of the limiting spectrum, extending BBP criteria beyond the classical Wigner, Wishart and Jacobi ensembles. Exact results for rank-one perturbations in several classical ensembles are reviewed in Ref.~\cite{Forrester2023}.

In the classical ensembles discussed above, the BBP
transition is continuous, with the overlap between the top
eigenvector and the signal tending to zero as $\theta$
approaches $\theta_c$ from above~\cite{Nadler2008}. More recently, a
qualitatively different regime of spectral recovery has
been characterized through discontinuous BBP
transitions~\cite{Bocchi2026}. In this regime, the overlap
jumps from zero to a positive value as $\theta$ crosses
$\theta_c$ in the large-matrix limit, so that recovery through PCA begins abruptly rather than gradually.
Such transitions have
been established in deformed Gaussian and reweighted Wishart
ensembles when the continuous spectral density vanishes
faster than linearly at the edge. They are accompanied by
pronounced finite-size effects and sample-to-sample
fluctuations. Analogous behavior has also been identified at
the lower spectral edge of the loss Hessian in simple neural
networks at random initialization, where increasing
overparameterization can turn a continuous transition into
a discontinuous one~\cite{Annesi2025}.

Discontinuous transitions have also been found in spiked
models with sparse noise, both for weighted adjacency
matrices of undirected random graphs with finite average
connectivity~\cite{Adomaityte2025} and for Wishart matrices
formed from sparse rectangular matrices~\cite{Forer2026}. In both settings, the
eigenvalue associated with the signal can separate from the
spectral bulk and remain below a structural outlier already
present in the noise. The squared overlap between the top
eigenvector and the signal jumps from zero to a positive
value when the signal eigenvalue overtakes this outlier.
These models relax the assumption of dense noise, which can
be restrictive in applications such as video surveillance,
face recognition, financial data analysis and ecological
network inference, where sparse corruptions or sparse
dependence structures can coexist with a dense signal
representing global effects that one seeks to recover~\cite{Candes2011,Vary2021,Mantegna1999,FanLiaoLiu2016,Deng2012,Kurtz2015}.

Euclidean random matrices provide a natural framework for
studying signal recovery in a less explored setting, where
the noise is dense, the entries are dependent, and the matrix
ensemble generally lacks rotational invariance. Given $N$
random points $\x_1,\ldots,\x_N\in\mathbb R^d$ with joint
probability distribution $P(\x_1,\ldots,\x_N)$, a Euclidean
random matrix is defined through a deterministic kernel as
\begin{equation}\label{eq:general_X}
    X_{ij}=f(\x_i,\x_j)\ .
\end{equation}
At fixed $d$, the $O(N^2)$ matrix entries are generated by only $O(N)$ random degrees of freedom, so that even independent point positions generally produce dependencies and geometric constraints across different rows and columns. Together with the general absence of rotational invariance, these constraints can make the analysis of extreme eigenvalues and their associated eigenvectors more challenging~\cite{casaburi2026}. For the recovery problem, this means that the results reviewed above do not apply directly to a general Euclidean background, and the geometric correlations must be accounted for explicitly.

These matrices are also of direct physical interest, as they arise in models of vibrational spectra and instantaneous normal modes in disordered materials~\cite{Parisi2006,CicutaKrausserMilkusZaccone2018}, Anderson localization~\cite{Ciliberti2005}, and cooperative photon emission in random atomic clouds~\cite{Viggiano2023,viggiano_2025}. Distance matrices provide a further example, with spectral and localization properties shaped by the geometry of the sampled points~\cite{Bogomolny2003}. Euclidean random matrices also arise as adjacency matrices of random geometric graphs, with applications to spatial networks~\cite{DettmannGeorgiouKnight2017,DettmannKnight2018,Inwood2026}.

Much of the Euclidean spectral literature focuses on the density of states and limiting spectral measures. These properties have been investigated through field-theoretic and high-density expansions~\cite{Grigera_2011}, while rigorous results establish limiting spectral measures in several spatial regimes~\cite{MezardParisiZee1999,Grigera_2011,Bordenave2008,GoetschySkipetrov2013}. Largest-eigenvalue fluctuations have also been characterized for Lipschitz kernels with uniform inputs on the unit interval, with applications to graphon-based random graphs~\cite{ChatterjeeHuang2025}. Random kernel matrices with both the input dimension and the number of observations tending to infinity exhibit kernel-dependent scalings and limiting spectral laws~\cite{ElKaroui2010,ChengSinger2013,bordenave_2,jiang_palle}.

The technical challenges posed by noise outside the usual independent-entry and rotationally invariant settings, together with the broad range of applications of Euclidean random matrices, motivate studying signal recovery through standard PCA in the spiked model in Eq.~\eqref{eq:intro_rank_one_spike} with a symmetric Euclidean background and an independent rank-one signal. Throughout this work, we consider PCA in the broader sense of a leading-eigenvector spectral estimator for a symmetric matrix, which need not be positive semidefinite. The questions are whether alignment of the leading eigenvector requires a strictly positive threshold $\theta_c$ or occurs for every $\theta>0$, which properties of the background noise and the spike distribution determine that threshold when it exists, and whether alignment then emerges gradually or abruptly. These questions amount to determining the existence and nature of a BBP transition in Euclidean random matrices, including the possibility of a discontinuous onset of recovery. Addressing them requires studying the principal eigenpair, consisting of the largest eigenvalue and its associated top eigenvector. An important distinction is that the presence of a signal-bearing eigendirection does not necessarily imply its recovery through the leading eigenvector. A sufficiently strong background mode may dominate the spectral ordering even when the signal is already represented by another eigenvector~\cite{Colwell2014}. Understanding signal recovery therefore requires not only identifying the eigenvalues and eigenvectors of the perturbed problem, but also determining how their ordering changes with
the signal strength.

To address these questions, we employ the replica method from the physics of disordered systems~\cite{pedestrian}. It allows us to average over the randomness in the matrix and derive equations for the average largest eigenvalue, the distribution of the top eigenvector's components, and the average overlap between the top eigenvector and the signal in the large-$N$ limit.
Early applications yielded the spectrum of Gaussian symmetric matrices and the separation of an eigenvalue induced by a nonzero mean~\cite{EdwardsJones1976}. Subsequent replica calculations addressed the density of states of sparse symmetric matrices~\cite{RodgersBray1988}. Related supersymmetric formulations provided an alternative treatment~\cite{RodgersDeDominicis1990}. Replica approaches were also developed for sparse covariance matrices~\cite{NagaoTanaka2007}. In the context of sparse random graphs, representing replica-symmetric order parameters as superpositions of Gaussians yields distributional equations that can be solved through population dynamics~\cite{Kuhn2008}. Replica field theories have also been applied directly to Euclidean spectra and to superpositions of independent Euclidean matrices~\cite{MezardParisiZee1999,BattistaMonasson2020}. Cavity methods provide a complementary approach for sparse matrices~\cite{Rogers2008}, with their relation to replica calculations and numerical implementation reviewed in~\cite{SuscaReview2021}.

For extremal observables, a useful statistical-mechanical formulation starts from optimization of a quadratic form on a sphere~\cite{Doussal_replica}. Replica and cavity approaches have been developed to study the largest eigenvalue and its eigenvector in sparse symmetric ensembles~\cite{kabashima_sparse,KabashimaTakahashi2012,Susca2019}. Replica calculations have also characterized the top eigenpair of diluted Wishart matrices~\cite{Budnick2025}. Recent extensions address spiked recovery problems with sparse symmetric and sparse Wishart backgrounds~\cite{Adomaityte2025,Forer2026}.

In our previous work on symmetric Euclidean matrices~\cite{casaburi2026}, the replica saddle-point equations were reduced to an eigenvalue problem for a Fredholm integral operator determined by the kernel and point distribution. Its principal eigenvalue equals the limiting average largest eigenvalue of the matrix, while its principal eigenfunction determines the probability density of the top eigenvector's components. This formulation, applicable when the principal eigenvalue is strictly positive, supplies the starting point for the present analysis.

In the present work, we add an independent rank-one perturbation to the Euclidean background and show how the replica framework of Ref.~\cite{casaburi2026} can be adapted to determine the principal eigenpair and the overlap between the top eigenvector and the signal. The resulting self-consistency equations depend on the unperturbed Fredholm operator and the first two moments of the spike distribution.

We find that a BBP transition is not always present: recovery can require a positive threshold $\theta_c$ or occur for every $\theta>0$, depending on the unperturbed upper spectral edge, the spike mean and the spatial averages of the principal eigenfunctions. These criteria and the corresponding threshold formulas provide a general analytical solution to the transition problem, without the need to specify the kernel or the point and spike distributions. 
Whenever a transition occurs, it is
discontinuous, with the principal eigenvalue switching from an
unperturbed branch to a competing branch carrying nonzero
overlap with the signal. Below the threshold, the signal is
already present in a subleading component, so it is missed
when PCA keeps only the first principal component. As the signal
grows stronger relative to the noise, its eigenvalue overtakes
the leading unperturbed eigenvalue. PCA then selects the
component carrying the signal, and the overlap jumps from zero
to a positive value. This crossing takes place between isolated
positive eigenvalues of a compact operator, giving a mechanism
distinct from the one involving a continuous spectral density
in deformed Gaussian and reweighted Wishart
ensembles~\cite{Bocchi2026}.

We illustrate and test the general results using the quadratic distance kernel and a rank-two kernel, with numerical simulations showing excellent agreement with the theoretical predictions.

The paper is organized as follows. Section~\ref{sec:replica_method} introduces the replica representations for the average largest eigenvalue, the density of the top eigenvector's components, and the average overlap with the signal. Section~\ref{sec:recap_results} recalls the results for the unperturbed model, introduces the spiked model, and presents the general transition classification. Sections~\ref{sec:results_quadratic_distance} and~\ref{sec:results_rank2} treat the quadratic distance and rank-two kernels, respectively, comparing the theoretical predictions with numerical simulations. Appendix~A contains the spectral-edge arguments, Appendix~B provides the replica derivation of the three observables for the general spiked model, and Appendix~C derives the general transition criteria and thresholds. Appendices~D and~E provide the detailed calculations for the two kernels, recovering the general results of Appendix~C.

\section{Observables of interest via Replica method}\label{sec:replica_method}

In this section we introduce the replica representations for the three observables of interest: the average largest eigenvalue, the density of the top eigenvector's components, and its average overlap with the signal. Following Ref.~\cite{casaburi2026}, we first formulate the problem for a generic real symmetric random matrix $X$, with angular brackets denoting the average over the matrix ensemble.

\subsection{Largest eigenvalue}
By the Courant--Fischer variational principle, the largest eigenvalue of a symmetric matrix $X$ can be written as
\begin{equation}\label{eq:general_variational_eigenvalue}
    \lambda_{\max}(X)
    =
    \frac{1}{N}
    \max_{\lVert\vec v\rVert^2=N}
    (\vec v,X\vec v)
    \ .
\end{equation}
We therefore introduce the auxiliary partition function
\begin{equation}\label{eq:general_partition_function}
    Z_N(\beta)
    \coloneqq
    \int \dd\vec v\,
    \exp\left[
        \frac{\beta}{2}(\vec v,X\vec v)
    \right]
    \delta\left(\lVert\vec v\rVert^2-N\right)
    \ ,
\end{equation}
whose dominant contributions in the zero-temperature limit $\beta\to\infty$ come from the maximizers of the quadratic form, namely the top eigenvectors of $X$ with the prescribed normalization. Applying Laplace's method and averaging over the ensemble consequently gives the following relation in the large-$N$ limit
\begin{equation}
    \lim_{N\to\infty}
    \left\langle\lambda_{\max}(X)\right\rangle
    =
    \lim_{N\to\infty}
    \lim_{\beta\to\infty}
    \frac{2}{\beta N}
    \left\langle\log Z_N(\beta)\right\rangle
    \ .
\end{equation}
To compute the average of the logarithm, we employ the replica trick and obtain
\begin{equation}\label{eq:general_eigenvalue_replica}
    \lim_{N\to\infty}
    \left\langle\lambda_{\max}(X)\right\rangle
    =
    \lim_{N\to\infty} \frac{1}{N}
    \lim_{\beta\to\infty} \frac{2}{\beta}
    \lim_{n\to0}
    \frac{1}{n}
    \log\left\langle Z_N(\beta)^n\right\rangle
    \ .
\end{equation}
For positive integer $n$, the replicated partition function involves $n$ copies of the integration variables sharing the same matrix realisation. Its disorder average is evaluated at integer $n$ and then analytically continued to a neighborhood of $n=0$. As in Ref.~\cite{casaburi2026}, we assume that the large-$N$ limit can be taken before the remaining limits, allowing a saddle-point evaluation.

\subsection{Eigenvector observables}

Properties of the top eigenvector can be accessed within a common framework by adding a source term to the auxiliary partition function~\cite{Adomaityte2025,Forer2026,Susca2019}. Let $\mathcal O:\mathbb R\times\mathbb R^N\times\mathbb R^N\to\mathbb R$
be an observable depending on a scalar argument $u$, a vector $\vec v$, and the signal $\vec y$ introduced in Eq.~\eqref{eq:intro_rank_one_spike}. Denoting the top eigenvector of $X$ by $\vec v^{\, (1)}$, we are interested in the large-$N$ limit of the ensemble average, $\langle\mathcal O(u,\vec v^{\, (1)},\vec y)\rangle$.
To compute this quantity, we introduce the auxiliary partition function
\begin{equation}\label{eq:general_source_partition}
\begin{aligned}
    Z_N(\beta;t\mathcal O)
    \coloneqq
    \int \dd\vec v\,
    \exp\left[
        \frac{\beta}{2}(\vec v,X\vec v)
        +\beta tN\mathcal O(u,\vec v,\vec y)
    \right]
    \delta\left(\lVert\vec v\rVert^2-N\right)
    \ ,
\end{aligned}
\end{equation}
where setting $t=0$ recovers the partition function $Z_N(\beta)$ defined in Eq.~\eqref{eq:general_partition_function}. Taking the zero-temperature limit and applying the replica identity then gives
\begin{align}
    &\lim_{N\to\infty}
    \left\langle
        \mathcal O(u,\vec v^{\, (1)},\vec y)
    \right\rangle
    \nonumber\\
    &\qquad=
    \lim_{N\to\infty}
    \lim_{\beta\to\infty}
    \left.
    \frac{1}{\beta N}
    \frac{\partial}{\partial t}
    \left\langle
        \log Z_N(\beta;t\mathcal O)
    \right\rangle
    \right|_{t=0}
    \nonumber\\
    &\qquad=
    \lim_{N\to\infty}
    \lim_{\beta\to\infty}
    \lim_{n\to0}
    \left.
    \frac{1}{n\beta N}
    \frac{\partial}{\partial t}
    \log\left\langle
        \left[Z_N(\beta;t\mathcal O)\right]^n
    \right\rangle
    \right|_{t=0}
    \ .
    \label{eq:general_eigenvector_replica}
\end{align}
The two eigenvector observables of interest follow from different choices of the source function\footnote{In Appendix~\ref{appendix_replica_der}, we carry out the calculation using a smooth regulariser $\delta_\varepsilon$ for the Dirac delta appearing in the source functions, with an additional limit $\varepsilon\to0^+$ to be taken in Eq.~\eqref{eq:general_eigenvector_replica} after differentiation with respect to $t$.
}
\begin{align}
    \mathcal O_{\mathrm{vec}}(u,\vec v)
    &\coloneqq
    \frac{1}{N}\sum_{i=1}^{N}
    \delta(u-v_i)
    \ ,
    \nonumber\\
    \mathcal O_{\mathrm{ov}}(u,\vec v,\vec y)
    &\coloneqq
    \frac{1}{N}\sum_{i=1}^{N}
    \delta(u-y_i v_i)
    \ .
    \label{eq:general_source_observables}
\end{align}
Inserting $\mathcal O_{\mathrm{vec}}$ into Eq.~\eqref{eq:general_eigenvector_replica} yields the limiting density of the top eigenvector's components
\begin{equation}\label{eq:general_component_density}
    T(u)
    =
    \lim_{N\to\infty}
    \left\langle
        \frac{1}{N}\sum_{i=1}^{N}
        \delta\left(u-v_i^{\, (1)}\right)
    \right\rangle
    \ ,
\end{equation}
whereas choosing $\mathcal O_{\mathrm{ov}}$ gives the overlap density
\begin{equation}\label{eq:general_overlap_density}
    T_{\mathrm{ov}}(u)
    =
    \lim_{N\to\infty}
    \left\langle
        \frac{1}{N}\sum_{i=1}^{N}
        \delta\left(u-y_i v_i^{\, (1)}\right)
    \right\rangle
    \ .
\end{equation}
Its first moment determines the limiting average overlap between the top eigenvector and the signal
\begin{equation}\label{eq:general_average_overlap}
    \lim_{N\to\infty}
    \left\langle
        \frac{(\vec y,\vec v^{\, (1)})}{N}
    \right\rangle
    =
    \int \dd u\,u\,T_{\mathrm{ov}}(u)
    \ .
\end{equation}
Computing the three observables therefore requires evaluating the average replicated partition function $\langle[Z_N(\beta;t\mathcal O)]^n\rangle$. We carry out this calculation for the spiked Euclidean model in Appendix~B.

\section{Summary of results}\label{sec:recap_results}
In this section, we recall the results of Ref.~\cite{casaburi2026} for the unperturbed Euclidean model and summarize those for the spiked model, including the equations determining the three observables and the general classification of recovery transitions. The detailed derivation of the transition criteria and thresholds is deferred to Appendix~C.

\subsection{Unperturbed model}
In Ref.~\cite{casaburi2026}, we considered symmetric Euclidean random matrices of the form
\begin{equation}\label{eq:X_N}
    X_{ij}^{(0)} =
    \frac{f(\x_i,\x_j)}{N}\ , \qquad \forall i,j\in\{1,\dots,N\}\ ,
\end{equation}
where the random vectors $\{\x_i\}_{i=1}^{N}$ are independently sampled from a common probability distribution $P$ on $\mathbb{R}^{d}$, and the kernel $f:\mathbb{R}^{d}\times\mathbb{R}^{d}\to\mathbb{R}$ is symmetric. Using the replica method, we showed that, in the large-$N$ limit, the average largest eigenvalue and the associated top-eigenvector profile are determined by the spectral problem of the Fredholm integral operator
\begin{equation}\label{eq:general_Tf_action}
    (T_f\psi)(\x)
    :=
    \int \dd\x'\,P(\x')\,
    f(\x,\x')\psi(\x')\ .
\end{equation}
More precisely, one considers the spectral problem
\begin{equation}
    \lambda\,\psi(\x)
    =
    (T_f\psi)(\x)
    \  ,
\end{equation}
and denoting by $\bigl(\lambda_{\max}(T_f),\psi_{\max}\bigr)$ the principal eigenpair of $T_f$, the limiting average largest eigenvalue in Eq.~\eqref{eq:general_eigenvalue_replica} is given by
\begin{equation}\label{eq:lambda_max_f_piccolo}
    \lim_{N\to\infty}\left\langle\lambda_{\max}(X^{(0)})\right\rangle
    =
    \lambda_{\max}(T_f)
    \  ,
\end{equation}
while the probability density of the top eigenvector's components in Eq.~\eqref{eq:general_eigenvector_replica} reads
\begin{equation}\label{eq:Tu_f_piccolo}
    T(u)
    =
    \int \dd\x\,P(\x)
    \delta\left(
        u-
        \frac{\psi_{\max}(\x)}
        {\lambda_{\max}(T_f)}
    \right)
    \  .
\end{equation}
These results hold provided that $\lambda_{\max}(T_f)>0$ and that the corresponding principal eigenfunction is normalized according to
\begin{equation}\label{eq:psi_normalisation}
    1
    =
    \int \dd\x\,P(\x)
    \left(
        \frac{\psi_{\max}(\x)}
        {\lambda_{\max}(T_f)}
    \right)^2
    \  .
\end{equation}
For the Fredholm spectral problem to be well defined, the kernel must satisfy suitable regularity assumptions. We require
\begin{equation}\label{eq:kernel_L2_condition}
    \int \dd\x\,\dd\x'\,
    P(\x)P(\x')
    \left|f(\x,\x')\right|^2
    <\infty
    \  ,
\end{equation}
so that $T_f$ is a Hilbert–Schmidt operator and, consequently, a bounded and compact operator on $L^2(P)$ \cite{ReedSimon1980}. The symmetry of the kernel further implies that $T_f$ is self-adjoint with respect to the scalar product
\begin{equation}\label{eq:L2P_scalar product}
    (\phi,\psi)_P
    \coloneqq
    \int \dd\x\,P(\x)\phi(\x)\psi(\x)
    \  .
\end{equation}
It then follows from the spectral theorem for compact self-adjoint operators that all non-zero spectral values of $T_f$ are real eigenvalues with finite multiplicity, and that zero is their only possible accumulation point \cite{ReedSimon1980}. In particular, if the upper spectral edge
\begin{equation}\label{eq:unperturbed_spectral_edge}
    \Lambda_0
    \coloneqq
    \sup\operatorname{Spec}(T_f)
    \ ,
\end{equation}
is strictly positive, the spectral theorem guarantees that it is an eigenvalue with
at least one non-trivial principal eigenfunction.\footnote{
We distinguish the upper spectral edge $\Lambda_0$ from the
largest eigenvalue $\lambda_{\max}(T_f)$ because, if $\Lambda_0=0$
and $L^2(P)$ is infinite-dimensional, i.e., $P$ is not supported
on a finite number of points, the upper spectral edge need not
be an eigenvalue and a non-trivial principal eigenfunction may
not exist. The spectral theorem ensures that this complication
can occur only at zero, while in finite dimensions it is excluded
altogether, since every spectral value is an eigenvalue.
Even when a principal eigenfunction exists at $\Lambda_0=0$,
however, the representation given by the replica method for $T(u)$ in
Eq.~\eqref{eq:Tu_f_piccolo} remains inapplicable, because its
denominator vanishes.}
In this case
\begin{equation}\label{eq:unperturbed_principal_eigenvalue}
    \Lambda_0
    =
    \lambda_{\max}(T_f)
    >
    0
    \ ,
\end{equation}
and the replica expressions in
Eqs.~\eqref{eq:lambda_max_f_piccolo} and~\eqref{eq:Tu_f_piccolo}
apply, with the principal eigenfunction normalized according
to Eq.~\eqref{eq:psi_normalisation}.

\subsection{Spiked model}\label{sec:spiked model}
We now consider Euclidean random matrices with an additive rank-one perturbation of strength $\theta>0$
\begin{equation}\label{eq:spiked_matrix}
    X_{ij}^{(\theta)} =
    \frac{1}{N}[f(\x_i,\x_j)+\theta y_i y_j]\ , \qquad \forall i,j\in\{1,\dots,N\}\ .
\end{equation}
The vectors $\{\x_i\}_{i=1}^N$ are assumed to be i.i.d.\ with distribution $P$, while the variables $\{y_i\}_{i=1}^N$ are i.i.d.\ with distribution $G$ and independent of the $\x_i$'s. We further assume that $G$ has finite first and second moments
and is non-degenerate, so that its variance is finite and
strictly positive: $0<\sigma_y^2<\infty$. To cast Eq.~\eqref{eq:spiked_matrix} into the form of the symmetric Euclidean random matrix defined in Eq.~\eqref{eq:X_N}, we introduce the augmented vectors $\z_i=(\x_i,y_i)\in\mathbb R^{d+1}$, whose first $d$ coordinates coincide with those of $\x_i$ and whose last coordinate is $y_i$. The vectors $\{\z_i\}_{i=1}^N$ are then i.i.d. with product density $P(\x)G(y)$, and Eq.~\eqref{eq:spiked_matrix} can be written as
\begin{equation}
X^{(\theta)}_{ij}
=
\frac{1}{N}F_\theta(\z_i,\z_j)
\ ,
\end{equation}
where the augmented kernel is
\begin{equation}\label{eq:spiked_kernel}
F_\theta\bigl((\x,y),(\x',y')\bigr)
=
f(\x,\x')+\theta yy'
\ .
\end{equation}
Since $f$ is symmetric, $F_\theta$ is symmetric as well. The deformed model therefore belongs to the same class of symmetric Euclidean random matrices studied in \cite{casaburi2026}, with the underlying space $\mathbb R^d$ replaced by $\mathbb R^{d+1}$ and the sampling measure $P$ replaced by $P\otimes G$. The results of Ref.~\cite{casaburi2026} therefore apply to the augmented kernel $F_\theta$ and, as in the unperturbed case, the limiting top eigenpair is determined by the spectral problem
\begin{equation}\label{eq:Fredholm_problem_spiked}
\lambda\,\Psi(\x,y)
=
(T_\theta\Psi)(\x,y)
\ ,
\end{equation}
with the operator
\begin{equation}\label{eq:spiked_Fredholm_operator}
( T_\theta\Psi)(\x,y)
\coloneqq
\int \dd \x' \dd y' P(\x')\,G(y')\,
F_\theta\bigl((\x,y),(\x',y')\bigr)
\Psi(\x',y')
\ .
\end{equation}
If the unperturbed kernel $f$ satisfies
condition~\eqref{eq:kernel_L2_condition} and $G$ has a finite
second moment, then the augmented kernel $F_\theta$ satisfies
the analogous condition
\begin{equation}\label{eq:spiked_kernel_L2_condition}
\begin{aligned}
    &\int \dd\x\,\dd\x'\,\dd y\,\dd y'\,
    P(\x)P(\x')G(y)G(y')
    \left|
        F_\theta\bigl((\x,y),(\x',y')\bigr)
    \right|^2
    \\
    &\qquad\leq
    2\int \dd\x\,\dd\x'\,
    P(\x)P(\x')|f(\x,\x')|^2
    +2\theta^2\langle y^2\rangle_G^2
    <\infty
    \ .
\end{aligned}
\end{equation}
Consequently, $T_\theta$ is a Hilbert–Schmidt operator, and is therefore bounded and compact on $L^2(P\otimes G)$. Moreover, since $F_\theta$ is symmetric, $T_\theta$ is self-adjoint with respect to the scalar product
\begin{equation}
    (\Phi,\Psi)_{P\otimes G}
    \coloneqq
    \int \dd\x\,\dd y\,
    P(\x)G(y)\,
    \Phi(\x,y)\Psi(\x,y)
    \ .
\end{equation}
Therefore, provided that $\lambda_{\max}(T_{\theta})>0$, the analogues of Eqs.~\eqref{eq:lambda_max_f_piccolo} and \eqref{eq:Tu_f_piccolo} continue to hold. Denoting the principal eigenpair of $T_\theta$ by
$\bigl(\lambda_{\max}(T_\theta),\Psi_{\max}\bigr)$, we obtain
\begin{equation}\label{eq:spiked_eigenvalue_limit_general}
    \lim_{N\to\infty}\left\langle
        \lambda_{\max}\bigl(X^{(\theta)}\bigr)
    \right\rangle
    =
    \lambda_{\max}(T_\theta)
    \ ,
\end{equation}
and
\begin{equation}\label{eq:spiked_top_eigenvector_distribution}
    T(u)
    =
    \int \dd\x\,\dd y\,
    P(\x)G(y)\,
    \delta\left(
        u-
        \frac{\Psi_{\max}(\x,y)}
        {\lambda_{\max}(T_\theta)}
    \right)
    \ ,
\end{equation}
with the principal eigenfunction normalized according to
\begin{equation}\label{eq:spiked_psi_normalisation}
    1
    =
    \int \dd\x\,\dd y\,
    P(\x)G(y)
    \left(
        \frac{\Psi_{\max}(\x,y)}
        {\lambda_{\max}(T_\theta)}
    \right)^2
    \ .
\end{equation}
To solve the spectral problem, we use Eqs.~\eqref{eq:spiked_kernel} and \eqref{eq:spiked_Fredholm_operator} to decompose $T_\theta$ as
\begin{equation}
    T_\theta
    =
    T_0+S_\theta
    \ ,
\end{equation}
where the two operators act as
\begin{align}
(T_0\Psi)(\x,y)
&=
\int \dd\x'\,P(\x')\,f(\x,\x')
\int \dd y'\,G(y')\,\Psi(\x',y')
=
(T_f\bar{\Psi})(\x)
\ ,
\label{eq:T0_action}
\\
(S_\theta\Psi)(\x,y)
&=
\theta y
\int \dd\x'\,\dd y'\,
P(\x')G(y')\,
y'\Psi(\x',y')
=
\theta y\,q[\Psi]
\ .
\label{eq:S_theta_action}
\end{align}
The $G$-average $\bar{\Psi}(\x)$ and the linear scalar functional $q[\Psi]$ are defined respectively by
\begin{align}
    \bar{\Psi}(\x)
    &\coloneqq
    \left\langle\Psi(\x,y)\right\rangle_G
    =
    \int \dd y\,G(y)\,\Psi(\x,y)
    \ ,
    \label{eq:barpsi}
    \\
    q[\Psi]
    &\coloneqq
    \left\langle y\Psi(\x,y)\right\rangle_{P\otimes G}
    =
    \int \dd\x\,\dd y\,
    P(\x)G(y)\,
    y\Psi(\x,y)
    \ .
    \label{eq:q_psi}
\end{align}
Using this decomposition, the Fredholm equation~\eqref{eq:Fredholm_problem_spiked} and its average over $y$, taken with respect to its distribution $G$, read
\begin{align}
    \lambda\Psi(\x,y)
    &=
    (T_f\bar{\Psi})(\x)
    +
    \theta y\,q[\Psi]
    \ ,
    \\
    \lambda\bar{\Psi}(\x)
    &=
    (T_f\bar{\Psi})(\x)
    +
    \theta\langle y\rangle_G\,q[\Psi]
    \ .
    \label{eq:fredholm_psi_bar}
\end{align}
Subtracting the second equation from the first yields
\begin{equation}
    \lambda
    \left[
        \Psi(\x,y)-\bar{\Psi}(\x)
    \right]
    =
    \theta q[\Psi]
    \left(
        y-\langle y\rangle_G
    \right)
    \ .
\end{equation}
Therefore, every eigenfunction corresponding to a non-zero eigenvalue can be written as
\begin{equation}\label{eq:psi_complete_parametrisation}
    \Psi(\x,y)
    =
    \bar{\Psi}(\x)
    +
    \frac{\theta}{\lambda}
    q[\Psi]
    \left(
        y-\langle y\rangle_G
    \right)
    \ ,
\end{equation}
and is thus linear in $y$. Using the definition of $q[\Psi]$ in Eq.~\eqref{eq:q_psi}, this representation yields the self-consistency equation
\begin{equation}\label{eq:q_self_consistent}
    q[\Psi]
    =
    \langle y\rangle_G
    \left\langle\bar{\Psi}(\x)\right\rangle_P
    +
    \frac{\theta}{\lambda}
    \sigma_y^2\,q[\Psi]
    \ .
\end{equation}
Collecting the results above,  every eigenpair with a nonzero eigenvalue is determined by the self-consistent system given by Eqs.~\eqref{eq:fredholm_psi_bar} and \eqref{eq:q_self_consistent}
\begin{equation}\label{eq:final_sc_system}
    \begin{cases}
        \displaystyle
        \lambda\bar{\Psi}(\x)
        =
        (T_f\bar{\Psi})(\x)
        +
        \theta\langle y\rangle_G\,q[\Psi]
        \ ,
        \\[2mm]
        \displaystyle
        q[\Psi]\left(1-\frac{\theta}{\lambda}
        \sigma_y^2\right)
        =
        \langle y\rangle_G
        \left\langle\bar{\Psi}(\x)\right\rangle_P
        \ .
    \end{cases}
\end{equation}
We are interested in the largest solution $\lambda=\lambda_{\max}(T_{\theta})$, which we prove to be strictly positive for all $\theta>0$. In fact, denoting the upper spectral edges of the perturbed and unperturbed operators by
\begin{equation}\label{eq:main_definition_big_lambda}
    \Lambda_\theta\coloneqq\sup\operatorname{Spec}(T_\theta)
    \qquad\text{and}\qquad
    \Lambda_0\coloneqq\sup\operatorname{Spec}(T_f)
    \ ,
\end{equation}
in Appendix~\ref{appendix:spectral_edges} we prove that for every $\theta>0$
\begin{equation}\label{eq:spectral_edges_ordering}
    \Lambda_\theta\geq\Lambda_0    
    \ ,
\end{equation}
and
\begin{equation}\label{eq:perturbed_edge_positive}
    \Lambda_\theta\geq\theta\sigma_y^2>0
    \ .
\end{equation}
The strict positivity of $\Lambda_\theta$ guarantees that the principal eigenvalue is well defined and strictly positive, as required by the replica method\footnote{Equation~\eqref{eq:spectral_edges_ordering} also shows that adding
the perturbation cannot lower the upper spectral edge.
As discussed in Sec.~\ref{sec:recap_results}, when $\Lambda_0=0$
the unperturbed upper spectral edge might not be an eigenvalue,
and a non-trivial principal eigenfunction may not exist for the unperturbed operator $T_f$.
Nevertheless, Eq.~\eqref{eq:perturbed_edge_positive} guarantees
that $\Lambda_\theta>0$ for every $\theta>0$, so the perturbed
operator admits a strictly positive principal eigenvalue and
a corresponding normalized, non-trivial eigenfunction even
in this case. The replica results therefore apply to the perturbed problem
for every $\theta>0$.}. Indeed, since $T_\theta$ is compact and self-adjoint, its strictly positive upper spectral edge is necessarily an eigenvalue, and therefore
\begin{equation}\label{eq:lambda_max_positive}
    \lambda_{\max}(T_\theta)
    =
    \Lambda_\theta
    >
    0
    \ .
\end{equation}
Consequently, for every $\theta>0$, the self-consistent system
in Eq.~\eqref{eq:final_sc_system} admits a non-trivial
solution with $\lambda=\lambda_{\max}(T_\theta)>0$.
Any corresponding principal eigenfunction $\Psi_{\max}$
has the form given in Eq.~\eqref{eq:psi_complete_parametrisation},
with its normalization fixed by
Eq.~\eqref{eq:spiked_psi_normalisation}.
The positivity condition required for the replica results above
is therefore satisfied for every $\theta>0$.
The quantity $q[\Psi_{\max}]$, defined in Eq.~\eqref{eq:q_psi},
also determines the limiting average overlap between the top
eigenvector and the spike. Specifically, the replica calculation
in Appendix~\ref{appendix_replica_der} gives
\begin{equation}\label{eq:spiked_overlap_main}
    \lim_{N\to\infty}
    \left\langle
        \frac{(\vec y,\vec v^{\, (1)})}{N}
    \right\rangle
    =
    \frac{q[\Psi_{\max}]}{\lambda_{\max}(T_\theta)}
    \ .
\end{equation}
Since $\lambda_{\max}(T_\theta)>0$,
Eq.~\eqref{eq:spiked_overlap_main} identifies the branches
$q[\Psi_{\max}]=0$ and $q[\Psi_{\max}]\neq0$ with zero and nonzero
limiting average overlap between the top eigenvector and the
spike, respectively.

The eigenvalue equation and normalization leave the overall
sign of $\Psi_{\max}$ arbitrary. Since $q[\Psi_{\max}]$ is linear, changing
this sign also reverses $q[\Psi_{\max}]$, corresponding to the
freedom to choose the overall sign of the top eigenvector $\vec v^{\, (1)}$.
We fix the sign of $\Psi_{\max}$ so that $q[\Psi_{\max}]>0$
whenever this quantity is nonzero, corresponding to a positive
limiting average overlap whenever it does not vanish.

We therefore refer to a sharp transition at a strictly positive
threshold $\theta_c$ when the limiting average overlap vanishes
for $0<\theta<\theta_c$ and becomes nonzero for $\theta>\theta_c$.
The precise transition criteria and the corresponding thresholds
are derived in Appendix~\ref{appendix:transition_thresholds},
where we show that the existence of such a transition depends
on the spike mean and on the spectral properties of the
unperturbed operator $T_f$, as summarized in
Table~\ref{tab:overlap_transition_summary}.

The table shows that no transition occurs when $\Lambda_0\leq0$,
regardless of the spike mean. When $\Lambda_0>0$, a centered
spike always produces a transition, whereas for a non-centered
spike the answer depends on $\Pi_{E_1}\mathbf1$, the orthogonal
projection of the constant function $\mathbf1(\x)=1$ onto the
principal eigenspace $E_1$ of $T_f$.

To make this distinction precise, recall that when $\Lambda_0>0$,
it coincides with the principal eigenvalue $\lambda_{\max}(T_f)$.
Its eigenspace is therefore non-trivial, and we denote it by
\begin{equation}\label{eq:recap_principal_eigenspace}
    E_1
    \coloneqq
    \ker\!\left(\Lambda_0 I-T_f\right)
    \ .
\end{equation}
The orthogonal projector $\Pi_{E_1}$ from $L^2(P)$ onto $E_1$
satisfies
\begin{equation}
    \Pi_{E_1}\mathbf1=0
    \quad\Longleftrightarrow\quad
    (\mathbf1,\psi)_P=\langle\psi\rangle_P=0
    \quad
    \forall\psi\in E_1
    \ .
\end{equation}
Thus, for $\Lambda_0>0$, a non-centered spike produces a
transition precisely when every unperturbed principal
eigenfunction has zero $P$-average. Conversely, if at least
one eigenfunction in $E_1$ has nonzero $P$-average, so that
$\Pi_{E_1}\mathbf1\neq0$, a transition for a non-centered
spike is ruled out.

The threshold in the non-centered transition regime is expressed
in terms of a function $F$ defined through the spectral
decomposition of $T_f$. Since $T_f$ is compact and self-adjoint,
the spectral theorem guarantees the existence of a complete
orthonormal basis $\{e_k\}_k$ of $L^2(P)$ consisting of
eigenfunctions of $T_f$. We write
\begin{equation}\label{eq:recap_spectral_quantities}
    T_f e_k=\mu_k e_k
    \ ,\qquad
    (e_k,e_\ell)_P=\delta_{k\ell}
    \ ,\qquad
    c_k\coloneqq(\mathbf1,e_k)_P=\langle e_k\rangle_P
    \ .
\end{equation}
For $\lambda>\max\{\Lambda_0,0\}$, we define
\begin{equation}\label{eq:recap_F_definition}
    F(\lambda)\coloneqq
    \langle y\rangle_G^2
    \sum_k\frac{c_k^2}{\lambda-\mu_k}
    +\frac{\sigma_y^2}{\lambda}
    \ .
\end{equation}
For a non-centered spike, whenever the secular equation
\begin{equation}\label{eq:recap_secular_equation}
    F(\lambda)=\frac{1}{\theta}
    \ .
\end{equation}
admits a solution with $\lambda>\max\{\Lambda_0,0\}$,
we denote this solution by $\lambda_c(\theta)$.
As shown in Appendix~\ref{appendix:transition_thresholds},
it is unique and coincides with the principal eigenvalue
$\lambda_{\max}(T_\theta)$.
In the transition regime $\Lambda_0>0$ and
$\Pi_{E_1}\mathbf1=0$, such a solution exists only for
$\theta>\theta_c$.
In this case, the limit entering
the threshold in Table~\ref{tab:overlap_transition_summary}
is finite and strictly positive
\begin{equation}\label{eq:recap_F_edge}
    F(\Lambda_0^+)
    \coloneqq
    \lim_{\lambda\to\Lambda_0^+}F(\lambda)
    =
    \langle y\rangle_G^2
    \sum_{\mu_k<\Lambda_0}
    \frac{c_k^2}{\Lambda_0-\mu_k}
    +
    \frac{\sigma_y^2}{\Lambda_0}
    \ ,
\end{equation}
giving the threshold $\theta_c=[F(\Lambda_0^+)]^{-1}$.

For each transition regime identified in
Table~\ref{tab:overlap_transition_summary},
Table~\ref{tab:transition_observables} gives the principal
eigenvalue of $T_\theta$ and the limiting average overlap
on either side of $\theta_c$. Below the threshold, the
principal eigenvalue remains equal to $\Lambda_0$ and the
overlap vanishes. Above the threshold, the principal
eigenvalue exceeds $\Lambda_0$ and the overlap is nonzero.
For a centered spike, the overlap attains its maximal value
$\sigma_y=\sqrt{\langle y^2\rangle_G}$ for every
$\theta>\theta_c$. In the large-$N$ limit, the top eigenvector
is therefore fully aligned with the spike as soon as the
threshold is crossed.
For a non-centered spike, the principal eigenvalue
$\lambda_c(\theta)$ increases above the threshold and grows
asymptotically linearly. The alignment may not be perfect
just above the threshold, but the overlap is nondecreasing
and converges to its maximum $\sqrt{\langle y^2\rangle_G}$.
The top eigenvector therefore aligns fully with the spike
at least in the limit $\theta\to\infty$.
The right limit of the overlap at $\theta_c$ is $\sigma_y>0$
for a centered spike and is also strictly positive for a
non-centered spike, as shown in
Appendix~\ref{appendix:transition_thresholds}.
The transition is therefore discontinuous in both cases.
At the threshold itself, the principal eigenvalue
is degenerate, so the operator problem does not select a
unique normalized principal eigenfunction or overlap\footnote{
More generally, a degenerate principal eigenvalue of $T_\theta$
admits several orthogonal eigenfunctions, and any normalized
linear combination is also a principal eigenfunction.
Different choices may yield different densities $T(u)$ of
the top eigenvector components and different overlaps, so
neither quantity is in general uniquely determined by the
operator alone. Explicit examples of degeneracy at the
threshold are discussed in the Appendices.}.
In every regime without a transition, the overlap is
nonzero for all $\theta>0$.

The mechanism underlying the discontinuity is particularly
transparent for a centered spike when we track which direction PCA
selects as the spike strength increases.
In this case, as proven in Appendix~\ref{appendix:transition_thresholds}, the eigenfunctions of $T_f$
remain eigenfunctions of $T_\theta$ with the same eigenvalues.
They depend only on $\x$ and are therefore orthogonal to the
centered spike direction $y$, giving zero overlap. The spike
adds an eigenfunction proportional to $y$, with eigenvalue
$\theta\sigma_y^2$ and maximal overlap $\sigma_y$.
PCA selects the direction with the largest eigenvalue, and
the spike eigenvalue $\theta\sigma_y^2$ increases with both
the spike strength $\theta$ and the spike standard deviation
$\sigma_y$. 
Although this signal-bearing direction exists for every $\theta>0$, it is not necessarily the leading one. For $\theta\sigma_y^2<\Lambda_0$, the largest eigenvalue belongs to the noise background, while the signal direction lies in a subleading eigenspace. At $\theta_c\sigma_y^2=\Lambda_0$, the two eigenvalues cross and exchange their ordering. The leading eigenvector consequently switches from a direction orthogonal to the signal to the signal direction itself. The discontinuous onset of recovery thus reflects a change in the eigenvalue ordering and in the direction
associated with the leading eigenfunction, rather than the
appearance of a previously absent signal mode. Such a transition requires $\Lambda_0>0$, since for
$\Lambda_0\leq0$ the eigenvalue associated with the signal,
$\theta\sigma_y^2$, is the largest for every $\theta>0$,
implying that the signal can always be recovered.

For a non-centered spike, whenever a transition occurs, it
is driven by an eigenvalue crossing, as in the centered
case. The main difference is that an eigenpair of $T_f$ is now retained by
$T_\theta$ only when its eigenfunction $\psi$ satisfies
$(\mathbf1,\psi)_P=\langle\psi\rangle_P=0$. The retained
eigenfunctions depend only on $\x$ and have zero overlap
with the spike. 
For every $\theta>0$, the spike introduces a competing
eigenvalue given by the largest solution of Eq.~\eqref{eq:recap_secular_equation}.
This eigenvalue increases with $\theta$, and the corresponding
eigenfunction depends on $y$ and has nonzero overlap with
the spike.

For $\Lambda_0>0$, if $\Pi_{E_1}\mathbf1\neq0$, the spike
couples to a leading noise direction, and the solution of
Eq.~\eqref{eq:recap_secular_equation} lies above $\Lambda_0$
for every $\theta>0$. The signal-bearing branch is therefore
always dominant and the overlap is always nonzero, so no
transition occurs.
If $\Pi_{E_1}\mathbf1=0$, the entire leading eigenspace of
$T_f$ is retained at $\Lambda_0$. The signal-bearing
eigenvalue given by Eq.~\eqref{eq:recap_secular_equation}
is then subleading below $\theta_c$, crosses $\Lambda_0$
at $\theta_c$, and becomes leading above it. Across this
crossing, the overlap jumps from zero to a positive value,
giving a discontinuous transition as in the centered case.

\begin{figure}[H]
    \centering
    \includegraphics[width=\linewidth]{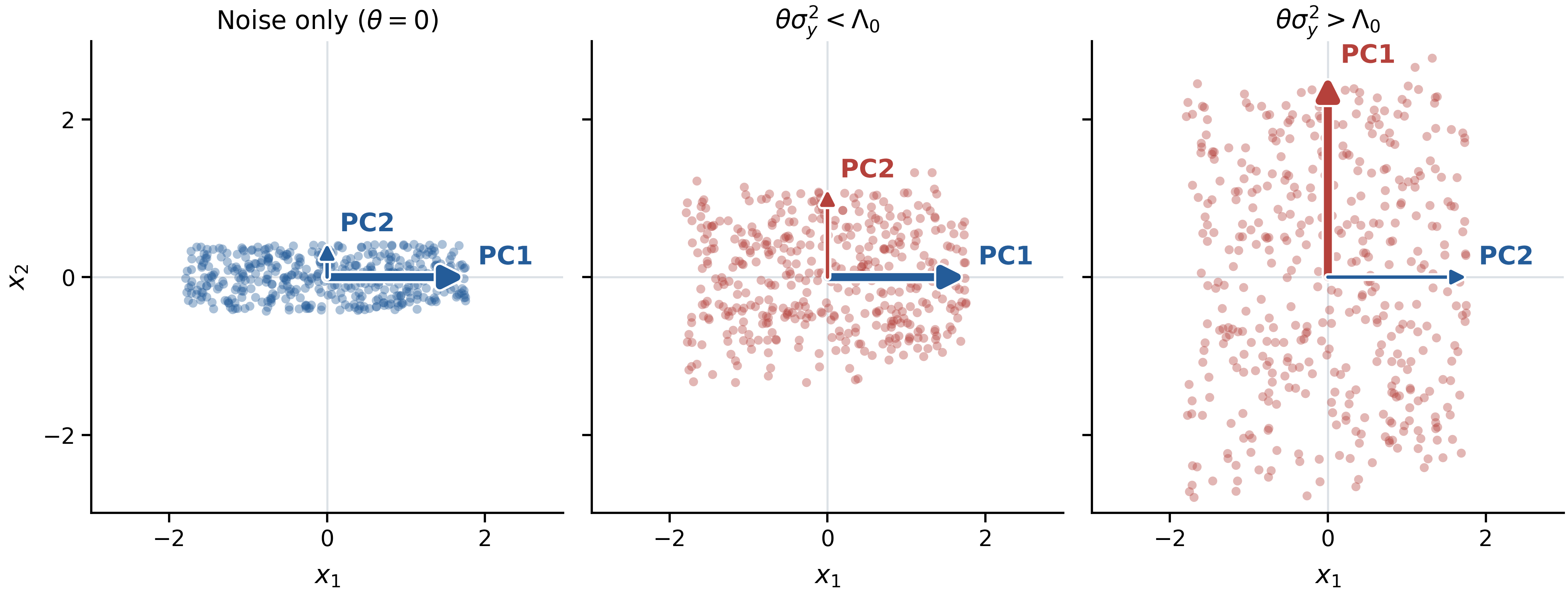}
    \caption{
    Schematic illustration of the transition mechanism.
    The blue points in the left panel ($\theta=0$) represent
    noise, with the first principal component (PC1) along
    $x_1$. The red clouds in the middle and right panels
    include a centered spike $\vec y$ along $x_2$. Below the threshold
    (middle), the spike direction is the second principal
    component (PC2), and the leading direction stays along
    $x_1$. Above the threshold (right), the eigenvalue along
    $x_2$ has overtaken the leading noise eigenvalue and is
    now the largest, so the principal component follows the
    spike. Thick and thin arrows mark the leading and
    subleading directions, respectively. The spike direction
    is shown in red.
    }
    \label{fig:pca_spike_transition}
\end{figure}

Figure~\ref{fig:pca_spike_transition} gives a schematic
illustration of the transition mechanism. We add a centered
spike to a synthetic noise cloud (blue points), producing
the red clouds at two signal-to-noise ratios. At the lower
ratio, the spike direction is subleading and is missed if
only the first principal component is retained. At the higher
ratio, the eigenvalue along the spike direction has overtaken
the leading noise eigenvalue, so the spike direction becomes
the leading PCA direction.

To obtain explicit expressions for the principal eigenpair and
test these general results, we need to specify the kernel $f$,
which determines the form of $\bar\Psi$ in the self-consistent
system in Eq.~\eqref{eq:final_sc_system}.  In the following sections, we consider the quadratic distance
kernel and a rank-two kernel, deriving the principal eigenpair
in each case. The corresponding transition conditions and
thresholds confirm the general results derived in
Appendix~\ref{appendix:transition_thresholds} and summarized
in Tables~\ref{tab:overlap_transition_summary}
and~\ref{tab:transition_observables}.
\begin{table}[H]
    \centering
    \small
    \renewcommand{\arraystretch}{1.5}
    \setlength{\tabcolsep}{6pt}
    \resizebox{0.60\linewidth}{!}{
    \begin{tabular}{@{}lccc@{}}
        \toprule
        & \multicolumn{3}{c}{Overlap transition criteria} \\
        \cmidrule(l){2-4}
        Spike
        &
        $\Lambda_0\leq0$
        &
        \shortstack{
            $\Lambda_0>0$\\
            $\Pi_{E_1}\mathbf1=0$
        }
        &
        \shortstack{
            $\Lambda_0>0$\\
            $\Pi_{E_1}\mathbf1\neq0$
        }
        \\
        \midrule
        \shortstack[l]{Centered\\ $\langle y\rangle_G=0$}
        &
        No transition
        &
        \shortstack{Transition\\ $\theta_c=\Lambda_0/\sigma_y^2$}
        &
        \shortstack{Transition\\ $\theta_c=\Lambda_0/\sigma_y^2$}
        \\[2mm]
        \shortstack[l]{Non-centered\\ $\langle y\rangle_G\neq0$}
        &
        No transition
        &
        \shortstack{Transition\\ $\theta_c=[F(\Lambda_0^+)]^{-1}$}
        &
        No transition
        \\
        \bottomrule
    \end{tabular}
    }
    \caption{
        Conditions and thresholds for a transition from zero
        to nonzero limiting average overlap between the top
        eigenvector and the spike. Here $\Lambda_0$ is the upper spectral edge of the
        unperturbed operator, defined in
        Eq.~\eqref{eq:main_definition_big_lambda}, and
        $F(\Lambda_0^+)$ is given in Eq.~\eqref{eq:recap_F_edge}.
    }
    \label{tab:overlap_transition_summary}
\end{table}

\begin{table}[H]
    \centering
    \small
    \renewcommand{\arraystretch}{1.5}
    \setlength{\tabcolsep}{6pt}

    \resizebox{\linewidth}{!}{%
        \begin{tabular}{@{}lcc|cc|c@{}}
            \toprule
            &
            \multicolumn{2}{c|}{
                Principal eigenvalue $\lambda_{\max}(T_\theta)$ ($\Lambda_0>0$)
            }
            &
            \multicolumn{2}{c|}{Limiting average overlap ($\Lambda_0>0$)}
            &
            \\
            \cmidrule(lr){2-3}\cmidrule(lr){4-5}
            Spike
            & $0<\theta<\theta_c$ & $\theta>\theta_c$
            & $0<\theta<\theta_c$ & $\theta>\theta_c$
            &
            \\
            \midrule
            $\langle y\rangle_G=0$
            & $\Lambda_0$ & $\theta\sigma_y^2$
            & $0$ & $\sigma_y$
            & $\theta_c=\Lambda_0/\sigma_y^2$
            \\[2mm]
            \shortstack[l]{
                $\langle y\rangle_G\neq0$,\\
                $\Pi_{E_1}\mathbf1=0$
            }
            & $\Lambda_0$ & $\lambda_c(\theta)$
            & $0$ & $\bigl[\theta\sqrt{-F'(\lambda_c(\theta))}\bigr]^{-1}$
            & $\theta_c=[F(\Lambda_0^+)]^{-1}$
            \\
            \bottomrule
        \end{tabular}%
    }

    \caption{
        Principal eigenvalue and limiting average overlap below
        and above the threshold in the transition regimes of
        Table~\ref{tab:overlap_transition_summary}, for which
        $\Lambda_0>0$.
        For a non-centered spike above the threshold,
        $\lambda_c(\theta)>\Lambda_0$ is the unique solution of
        Eq.~\eqref{eq:recap_secular_equation}, with $F$ defined
        in Eq.~\eqref{eq:recap_F_definition}.
        The overlap expressions are derived in
        Appendix~\ref{appendix:transition_thresholds}.
    }
    \label{tab:transition_observables}
\end{table}

\FloatBarrier

\section{Results for the quadratic distance kernel}\label{sec:results_quadratic_distance}
In this section we consider the quadratic distance kernel
\begin{equation}\label{eq:results_quadratic_kernel}
    f(\x,\x')=\lVert\x-\x'\rVert^2
    \ .
\end{equation}
We assume that the distribution $P$ of $\x$ has a finite fourth moment
and is not supported on a single point. The moment condition
ensures that Eq.~\eqref{eq:kernel_L2_condition} holds, so $T_f$
is a Hilbert–Schmidt operator and, by symmetry of the kernel,
is self-adjoint. Under these assumptions
\begin{equation}
    \Lambda_0=\lambda_{\max}(T_f)>0
    \ .
\end{equation}
Moreover, this principal eigenvalue is simple, and its
eigenfunction has nonzero $P$-average. The quadratic distance
kernel therefore belongs to the case $\Lambda_0>0$ and
$\Pi_{E_1}\mathbf1\neq0$ in
Table~\ref{tab:overlap_transition_summary}. The proofs of these
spectral properties are given in
Appendix~\ref{appendix:quadratic_distance_kernel}.
For the corresponding spiked kernel 
\begin{equation}\label{eq:results_quadratic_spiked_kernel}
    F_\theta\bigl((\x,y),(\x',y')\bigr)
    =
    \lVert\x-\x'\rVert^2+\theta yy'
    \ ,
\end{equation}
we retain the assumption that the distribution $G$ of the spike
components has a finite second moment and is non-degenerate,
so that $0<\sigma_y^2<\infty$. Together with the moment condition
on $P$, this ensures Eq.~\eqref{eq:spiked_kernel_L2_condition},
making the spiked spectral problem well defined. 
 
As shown in Appendix~\ref{appendix:quadratic_distance_kernel},
every eigenfunction associated with a nonzero eigenvalue
$\lambda$ of $T_\theta$ has the form
\begin{equation}\label{eq:summary_quadratic_general_form}
    \Psi(\x,y)
    =
    A\lVert\x\rVert^2+\B\cdot\x+C+\theta Dy
    \ ,\qquad
    D=\frac{q[\Psi]}{\lambda}
    \ ,
\end{equation}
where $A,C,D$ are scalar coefficients and $\B\in\mathbb R^d$,
with the expression for $D$ obtained by matching the coefficient
of $y$ in Eq.~\eqref{eq:psi_complete_parametrisation}.
In terms of the $d+3$ coefficients $(A,\B,C,D)$ of this
quadratic representation, the self-consistent system in
Eq.~\eqref{eq:final_sc_system} becomes
\begin{equation}\label{eq:summary_quadratic_system}
\left\{
\begin{aligned}
    \lambda A
    &=
    A\langle\lVert\x\rVert^2\rangle_P
    +\B\cdot\langle\x\rangle_P
    +C+\theta D\langle y\rangle_G
    \ ,\\
    \lambda B_\ell
    &=
    -2\left[
        A\langle x_\ell\lVert\x\rVert^2\rangle_P
        +\B\cdot\langle x_\ell\x\rangle_P
        +(C+\theta D\langle y\rangle_G)
            \langle x_\ell\rangle_P
    \right]
    \ ,
    \qquad \ell=1,\ldots,d
    \ ,\\
    \lambda C
    &=
    A\langle\lVert\x\rVert^4\rangle_P
    +\B\cdot\langle\x\lVert\x\rVert^2\rangle_P
    +(C+\theta D\langle y\rangle_G)
        \langle\lVert\x\rVert^2\rangle_P
    \ ,\\
    \lambda D
    &=
    \langle y\rangle_G
    \left[
        A\langle\lVert\x\rVert^2\rangle_P
        +\B\cdot\langle\x\rangle_P+C
    \right]
    +\theta D\langle y^2\rangle_G
    \ .
\end{aligned}
\right.
\end{equation}
This is an eigenvalue problem for a $(d+3)\times(d+3)$ matrix
whose entries depend on moments of the components of $\x$
up to fourth order and on the first two moments of $y$.
The principal eigenvalue is the largest value of $\lambda$
for which this system admits a non-trivial solution.
The principal eigenfunction must satisfy the normalization
condition in Eq.~\eqref{eq:spiked_psi_normalisation}, which,
in terms of its coefficients, reads
\begin{equation}\label{eq:summary_quadratic_normalization}
    2AC-\frac12\lVert\B\rVert^2+\theta D^2
    =
    \lambda_{\max}(T_\theta)
    \ .
\end{equation}
For this normalized principal solution, the coefficient $D$ in Eq.~\eqref{eq:summary_quadratic_general_form} is precisely the
limiting average overlap in Eq.~\eqref{eq:spiked_overlap_main}
\begin{equation}\label{eq:summary_quadratic_overlap}
    \lim_{N\to\infty}
    \left\langle
        \frac{(\vec y,\vec v^{\, (1)})}{N}
    \right\rangle
    =
    \frac{q[\Psi_{\max}]}{\lambda_{\max}(T_\theta)}
    =
    D
    \ .
\end{equation}

We now consider two classes of distributions for which
Eq.~\eqref{eq:summary_quadratic_system} simplifies:
isotropic distributions, $P(\x)=P(\lVert\x\rVert)$,
and distributions with independent and identically distributed
components drawn from an even marginal distribution,
$P(\x)=\prod_{\ell=1}^d P(x_\ell)$ with
$P(x_\ell)=P(-x_\ell)$.
For either class, these symmetries and the strict positivity
of $\lambda_{\max}(T_\theta)$ imply that $\B=0$
for the principal solution. Its eigenfunction therefore reduces to
\begin{equation}\label{eq:summary_quadratic_form}
    \Psi_{\max}(\x,y)
    =
    A\lVert\x\rVert^2+C+\theta Dy
    \ ,
\end{equation}
and the remaining coefficients satisfy the $3\times3$
eigenvalue problem
\begin{equation}\label{eq:summary_quadratic_reduced_system}
    \lambda
    \begin{pmatrix}
        A\\ C\\ D
    \end{pmatrix}
    =
    \begin{pmatrix}
        \langle\lVert\x\rVert^2\rangle_P
        & 1
        & \theta\langle y\rangle_G
        \\
        \langle\lVert\x\rVert^4\rangle_P
        & \langle\lVert\x\rVert^2\rangle_P
        & \theta\langle y\rangle_G
            \langle\lVert\x\rVert^2\rangle_P
        \\
        \langle y\rangle_G\langle\lVert\x\rVert^2\rangle_P
        & \langle y\rangle_G
        & \theta\langle y^2\rangle_G
    \end{pmatrix}
    \begin{pmatrix}
        A\\ C\\ D
    \end{pmatrix}
    \ .
\end{equation}
Diagonalizing this matrix gives $\lambda_{\max}(T_\theta)$
as its largest eigenvalue, while the normalization condition
\begin{equation}\label{eq:summary_quadratic_reduced_normalization}
    2AC+\theta D^2
    =
    \lambda_{\max}(T_\theta)
    \ ,
\end{equation}
fixes the scale of the corresponding principal eigenfunction
and hence its overlap through $D$.

At $\theta=0$, the same reduced system gives the unperturbed
principal eigenvalue
\begin{equation}\label{eq:summary_quadratic_unperturbed}
    \lambda_{\max}(T_f)
    =
    \langle\lVert\x\rVert^2\rangle_P
    +\sqrt{\langle\lVert\x\rVert^4\rangle_P}
    \ ,
\end{equation}
in agreement with the results of Ref.~\cite{casaburi2026}.
Below, we summarize the solutions for centered and non-centered
spikes separately. All derivations are given in
Appendix~\ref{appendix:quadratic_distance_kernel}.

\subsection{Centered spike}\label{sec:summary_quadratic_centered}
For $\langle y\rangle_G=0$, the general results of
Sec.~\ref{sec:recap_results}, summarized in
Table~\ref{tab:overlap_transition_summary}, predict a sharp
transition from zero to nonzero overlap at a positive threshold.
For the quadratic distance kernel, the explicit solution
derived in Appendix~\ref{appendix:quadratic_distance_kernel}
agrees with these general results.
The principal eigenvalue of $T_\theta$, which also gives
the large-$N$ limit of the average principal eigenvalue
of $X^{(\theta)}$ by
Eq.~\eqref{eq:spiked_eigenvalue_limit_general}, is
\begin{equation}\label{eq:summary_quadratic_centered_eigenvalue}
    \lambda_{\max}(T_\theta)
    =
    \begin{cases}
        \langle\lVert\x\rVert^2\rangle_P
        +\sqrt{\langle\lVert\x\rVert^4\rangle_P}\ ,
        & 0<\theta<\theta_c\ ,
        \\[1ex]
        \theta\sigma_y^2\ ,
        & \theta>\theta_c\ ,
    \end{cases}
\end{equation}
with transition threshold
\begin{equation}\label{eq:summary_quadratic_centered_threshold}
    \theta_c
    =
    \frac{
        \langle\lVert\x\rVert^2\rangle_P
        +\sqrt{\langle\lVert\x\rVert^4\rangle_P}
    }{\sigma_y^2}
    =
    \frac{\lambda_{\max}(T_f)}{\sigma_y^2}
    \ .
\end{equation}
Below the threshold, the principal eigenvalue therefore
coincides with the unperturbed value $\lambda_{\max}(T_f)$ in
Eq.~\eqref{eq:summary_quadratic_unperturbed}, whereas above
it the spike eigenvalue $\theta\sigma_y^2$ dominates. The principal eigenfunction normalized according to Eq.~\eqref{eq:summary_quadratic_reduced_normalization} is
\begin{equation}\label{eq:summary_quadratic_centered_eigenfunction}
    \Psi_{\max}(\x,y)
    =
    \begin{cases}
        \displaystyle
        \sqrt{
            \frac{
                \langle\lVert\x\rVert^2\rangle_P
                +\sqrt{\langle\lVert\x\rVert^4\rangle_P}
            }{
                2\sqrt{\langle\lVert\x\rVert^4\rangle_P}
            }
        }
        \left(
            \lVert\x\rVert^2
            +\sqrt{\langle\lVert\x\rVert^4\rangle_P}
        \right)\ ,
        & 0<\theta<\theta_c\ ,
        \\[3ex]
        \theta\sigma_y y\ ,
        & \theta>\theta_c\ ,
    \end{cases}
\end{equation}
and the limiting average overlap is
\begin{equation}\label{eq:summary_quadratic_centered_overlap}
   \lim_{N\to\infty}
    \left\langle
        \frac{(\vec y,\vec v^{\, (1)})}{N}
    \right\rangle = D=
    \begin{cases}
        0\ ,&0<\theta<\theta_c\ ,
        \\[1ex]
        \sigma_y\ ,&\theta>\theta_c\ .
    \end{cases}
\end{equation}
Thus, the principal eigenfunction is unchanged by the spike
below the threshold and becomes proportional to $y$ above it,
with the overlap displaying the sharp transition predicted
by the general theory. At $\theta=\theta_c$, the two branches coincide at
$\theta_c\sigma_y^2=\lambda_{\max}(T_f)$, giving a doubly
degenerate principal eigenvalue.

To test the analytical results obtained via the replica method,
we consider three isotropic distributions: the uniform
distribution on the sphere, the uniform distribution in the
ball, and the generalized radial exponential.
For these distributions, isotropy implies that the second
and fourth radial moments can be written in terms of two
coefficients $\sigma_x^2$ and $M$ as
\begin{equation}\label{eq:quadratic_isotropic_radial_moments}
    \langle\lVert\x\rVert^2\rangle_P=d\sigma_x^2
    \ ,\qquad
    \langle\lVert\x\rVert^4\rangle_P=d(d+2)M
    \ .
\end{equation}
The probability densities of the three considered distributions, 
together with the corresponding values of $\sigma_x^2$ and $M$,
are reported in Table~\ref{tab:isotropic}.

\begin{figure}[t]
    \centering
    \includegraphics[width=\textwidth]{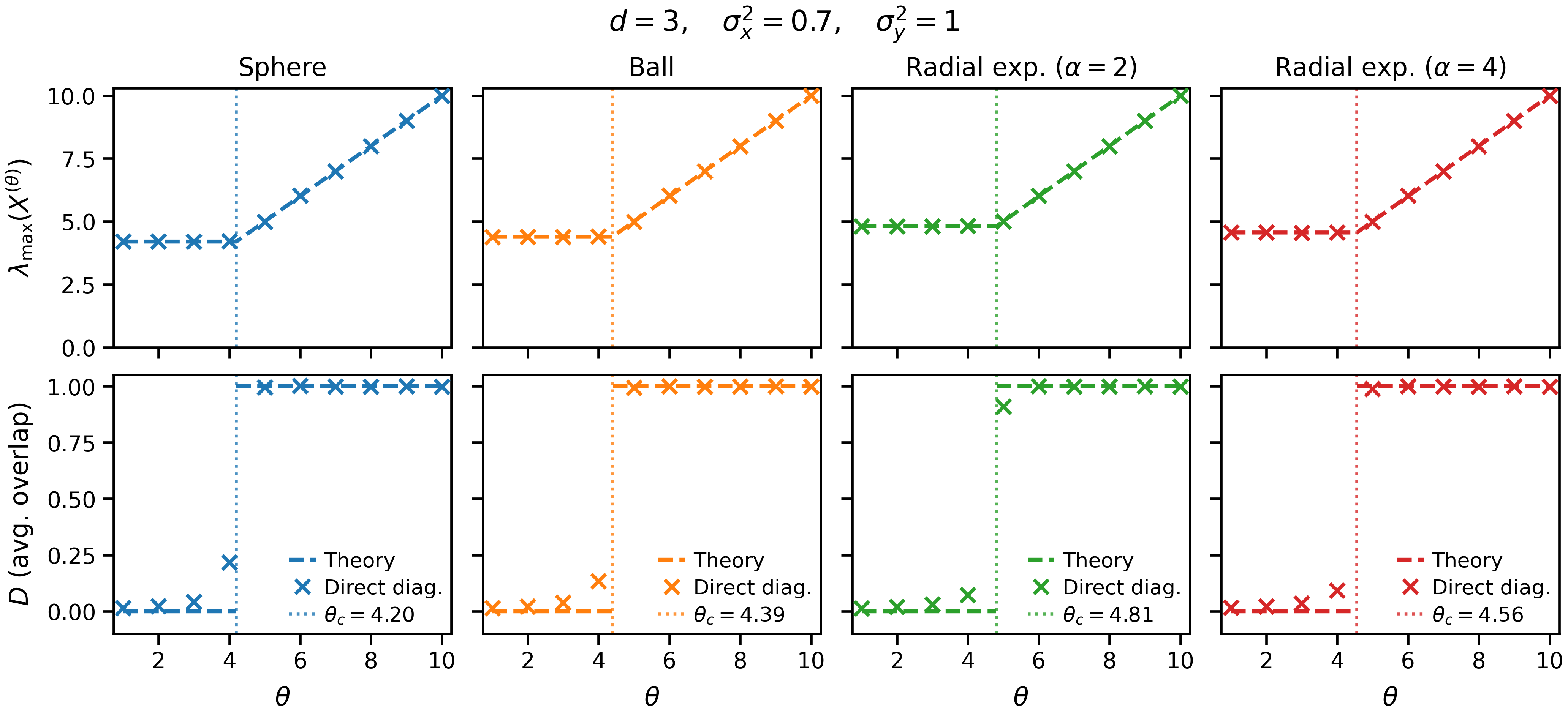}
    \caption{
        Largest eigenvalue and spike overlap for the quadratic
        distance kernel with a centered Gaussian spike.
        Comparison between the theoretical predictions in
        Eqs.~\eqref{eq:summary_quadratic_centered_eigenvalue}
        and~\eqref{eq:summary_quadratic_centered_overlap}
        (dashed lines) and the results obtained via direct
        numerical diagonalisation (crosses), as functions
        of $\theta$.
        The top row shows the average largest eigenvalue,
        while the bottom row shows the average overlap,
        whose limiting value is $D$.
        From left to right, columns correspond to the uniform
        distribution on the sphere, the uniform distribution
        in the ball, and the generalized radial exponential
        with $\alpha=2$ and $\alpha=4$, as defined in
        Table~\ref{tab:isotropic}.
        The case $\alpha=2$ corresponds to the centered
        Gaussian spatial distribution.
        Parameters are fixed to $d=3$ and $\sigma_x^2=0.7$.
        The spike components are independently drawn from a
        Gaussian distribution with $\langle y\rangle_G=0$
        and $\sigma_y^2=1$.
        Vertical dotted lines mark the predicted thresholds
        $\theta_c$ given by
        Eq.~\eqref{eq:summary_quadratic_centered_threshold}.
        Numerical results are obtained for matrices of size
        $N=5000$ and averaged over $100$ independent realisations.
    }
    \label{fig:quadratic_centered}
\end{figure}

In Fig.~\ref{fig:quadratic_centered}, we compare the theoretical
predictions for the average largest eigenvalue and the average
overlap, given by
Eqs.~\eqref{eq:summary_quadratic_centered_eigenvalue}
and~\eqref{eq:summary_quadratic_centered_overlap}
(dashed lines), with the results obtained via direct numerical
diagonalisation (crosses), as functions of $\theta$.
Vertical dotted lines indicate the predicted transition
thresholds $\theta_c$ given by
Eq.~\eqref{eq:summary_quadratic_centered_threshold}.
We fix $d=3$ and $\sigma_x^2=0.7$, and draw the spike
components independently from a Gaussian distribution with
$\langle y\rangle_G=0$ and $\sigma_y^2=1$.
From left to right, the four columns correspond to the uniform
distribution on the sphere, the uniform distribution in the
ball, and the generalized radial exponential with $\alpha=2$
and $\alpha=4$.
The case $\alpha=2$ corresponds to the centered Gaussian
distribution, which also has i.i.d.\ components drawn
from an even marginal distribution.
The top row shows the average largest eigenvalue, while the
bottom row shows the average overlap, whose limiting value
is $D$.
Numerical results are obtained for matrices of size $N=5000$
and averaged over $100$ independent realisations.

In all four cases, the numerical results agree very well
with the analytical predictions and clearly display the
transition at $\theta_c$.
Below the threshold, the largest eigenvalue remains at its
unperturbed value and the overlap is close to zero.
Above the threshold, the largest eigenvalue increases
linearly, and the overlap
becomes $\sigma_y=1$.

\begin{table}[t]
\centering
\renewcommand{\arraystretch}{1.9}
\setlength{\tabcolsep}{14pt}

\begin{tabular}{|l|c|c|c|}
\hline
\rule{0pt}{3.2ex}
& $P(\x)$
& $\sigma_x^2$
& $M$ \\
\hline

\rule{0pt}{4.8ex}
Uniform on the sphere 
& $\displaystyle \frac{\Gamma(d/2)}{2\pi^{d/2}\,r^{d-1}} \delta(r-||\x||)$
& $\displaystyle \frac{r^2}{d}$
& $\displaystyle \frac{r^4}{d(d+2)}$
\\[1.2ex]
\hline

\rule{0pt}{4.8ex}
Uniform in the ball 
& $\displaystyle \frac{\Gamma(d/2)d}{2\pi^{d/2}\,r^d}\,\Theta(r - ||\x||)$
& $\displaystyle \frac{r^2}{d+2}$
& $\displaystyle \frac{r^4}{(d+2)(d+4)}$
\\[1.2ex]
\hline

\rule{0pt}{4.8ex}
Radial exponential
& $\displaystyle 
\frac{\alpha\,b^{d/\alpha}\Gamma(d/2)}{2\pi^{d/2}\Gamma(d/\alpha)}
\,\exp\{-b||\x||^\alpha\}
$
& $\displaystyle 
\frac{b^{-2/\alpha}}{d}\,\frac{\Gamma((d+2)/\alpha)}{\Gamma(d/\alpha)}
$
& $\displaystyle 
\frac{b^{-4/\alpha}}{d(d+2)}\,\frac{\Gamma((d+4)/\alpha)}{\Gamma(d/\alpha)}
$
\\[1.2ex]
\hline
\end{tabular}
\caption{Isotropic distributions in dimension $d$: probability density $P(\x)$ and the corresponding parameters $\sigma_x^2$ and $M$. Here $r$ denotes the radius of the sphere or ball, while $b$ and $\alpha$ are the parameters of the generalized radial exponential. Setting $\alpha=2$ corresponds to the centered Gaussian distribution.}
\label{tab:isotropic}
\end{table}

\FloatBarrier

\subsection{Non-centered spike}\label{sec:summary_quadratic_noncentered}
For a non-centered spike, the conditions $\Lambda_0>0$ and
$\Pi_{E_1}\mathbf1\neq0$ place the quadratic distance kernel
in the regime where the general results of
Sec.~\ref{sec:recap_results} and
Table~\ref{tab:overlap_transition_summary} predict no sharp
transition. Since the perturbation $\theta yy'$ is invariant
under the global sign change $y\mapsto-y$, we assume
$\langle y\rangle_G>0$ without loss of generality. As proved in Appendix~\ref{appendix:quadratic_distance_kernel},
for $\langle y\rangle_G\neq0$ and every $\theta>0$,
the matrix in Eq.~\eqref{eq:summary_quadratic_reduced_system}
has three real and distinct eigenvalues, ordered as
\begin{equation}\label{eq:summary_quadratic_noncentered_root_ordering}
    \lambda_1\leq0<\lambda_2<\theta\sigma_y^2
    <\lambda_{\max}(T_\theta)
    \ .
\end{equation}
The two positive branches therefore never cross, and the
principal eigenvalue is given for every $\theta>0$ by
\begin{equation}\label{eq:summary_quadratic_noncentered_eigenvalue}
    \lambda_{\max}(T_\theta)
    =
    \frac{
        2\langle\lVert\x\rVert^2\rangle_P
        +\theta\langle y^2\rangle_G
    }{3}
    +2\sqrt{-\frac p3}\cos\phi
    \ ,
\end{equation}
where
\begin{equation}\label{eq:summary_quadratic_cubic_angle}
    \phi
    =
    \frac13\arccos\left(
        \frac{3w}{2p}\sqrt{-\frac3p}
    \right)
    \ ,
    \qquad
    0<\phi<\frac{\pi}{3}
    \ .
\end{equation}
\begin{equation}\label{eq:summary_quadratic_cubic_coefficients_p}
\begin{aligned}
    p={}&
    2\theta\langle\lVert\x\rVert^2\rangle_P\sigma_y^2
    -\sigma_{\lVert\x\rVert^2}^2
    -\frac13\left(
        2\langle\lVert\x\rVert^2\rangle_P
        +\theta\langle y^2\rangle_G
    \right)^2<0
    \ ,
\end{aligned}
\end{equation}
\begin{equation}\label{eq:summary_quadratic_cubic_coefficients_w}
\begin{aligned}
    w={}&
    -\frac{2}{27}\left(
        2\langle\lVert\x\rVert^2\rangle_P
        +\theta\langle y^2\rangle_G
    \right)^3
    +\frac13\left(
        2\langle\lVert\x\rVert^2\rangle_P
        +\theta\langle y^2\rangle_G
    \right)\left(
        2\theta\langle\lVert\x\rVert^2\rangle_P\sigma_y^2
        -\sigma_{\lVert\x\rVert^2}^2
    \right)
    +\theta\sigma_{\lVert\x\rVert^2}^2\sigma_y^2
    \ ,
\end{aligned}
\end{equation}
and the radial variance is defined by
\begin{equation}\label{eq:summary_quadratic_radial_variance}
    \sigma_{\lVert\x\rVert^2}^2
    =
    \langle\lVert\x\rVert^4\rangle_P
    -\langle\lVert\x\rVert^2\rangle_P^2 \geq 0
    \ .
\end{equation}
For every $\theta>0$, the principal eigenfunction normalized
according to Eq.~\eqref{eq:spiked_psi_normalisation} has the form
\begin{equation}\label{eq:summary_quadratic_noncentered_eigenfunction}
    \Psi_{\max}(\x,y)
    =
    \sqrt{
        \frac{\lambda_{\max}(T_\theta)}
        {2r_C+\theta r_D^2}
    }
    \left(\lVert\x\rVert^2+r_C+\theta r_D y\right)
    \ ,
\end{equation}
and its limiting average overlap with the spike remains
strictly positive for all $\theta>0$
\begin{equation}\label{eq:summary_quadratic_noncentered_overlap}
    \lim_{N\to\infty}
    \left\langle
        \frac{(\vec y,\vec v^{\, (1)})}{N}
    \right\rangle = D
    =
    r_D\sqrt{
        \frac{\lambda_{\max}(T_\theta)}
        {2r_C+\theta r_D^2}
    }
    >0
    \ ,
\end{equation}
where the ratios are defined by
\begin{equation}\label{eq:summary_quadratic_ratios}
    r_C
    =
    \langle\lVert\x\rVert^2\rangle_P
    +\frac{\sigma_{\lVert\x\rVert^2}^2}
          {\lambda_{\max}(T_\theta)}
    \ ,\qquad
    r_D
    =
    \frac{\lambda_{\max}(T_\theta)\langle y\rangle_G}
         {\lambda_{\max}(T_\theta)-\theta\sigma_y^2}
    \ .
\end{equation}
For the quadratic distance kernel with a non-centered spike,
the overlap remains strictly positive for every $\theta>0$
and no transition occurs, in agreement with the general
criteria established in Sec.~\ref{sec:recap_results} and
summarized in Table~\ref{tab:overlap_transition_summary}.
Moreover, the overlap has finite, positive limits for both weak and
strong perturbations. As $\theta\to0^+$, the principal eigenvalue
approaches the unperturbed value
$\lambda_{\max}(T_f)
=\langle\lVert\x\rVert^2\rangle_P
+\sqrt{\langle\lVert\x\rVert^4\rangle_P}$
given in Eq.~\eqref{eq:summary_quadratic_unperturbed}.
Consequently, Eq.~\eqref{eq:summary_quadratic_ratios} gives
\begin{equation}
    \lim_{\theta\to0^+}r_C
    =
    \sqrt{\langle\lVert\x\rVert^4\rangle_P}
    \ ,\qquad
    \lim_{\theta\to0^+}r_D
    =
    \langle y\rangle_G
    \ ,
\end{equation}
and Eq.~\eqref{eq:summary_quadratic_noncentered_overlap}
yields
\begin{equation}
    \lim_{\theta\to0^+}D
    =
    \langle y\rangle_G
    \sqrt{
        \frac{
            \langle\lVert\x\rVert^2\rangle_P
            +\sqrt{\langle\lVert\x\rVert^4\rangle_P}
        }{
            2\sqrt{\langle\lVert\x\rVert^4\rangle_P}
        }
    }
    >0
    \ .
\end{equation}
For $\theta\to\infty$,
Eq.~\eqref{eq:summary_quadratic_noncentered_root_ordering}
implies that
\begin{equation}
    \lim_{\theta\to\infty}\lambda_{\max}(T_\theta)=+\infty\ .
\end{equation}
Evaluating the leading terms in
Eq.~\eqref{eq:summary_quadratic_noncentered_eigenvalue}
and using Eq.~\eqref{eq:summary_quadratic_ratios} gives
\begin{equation}
\begin{aligned}
    \lim_{\theta\to\infty}
    \frac{\lambda_{\max}(T_\theta)}{\theta}
    &=
    \langle y^2\rangle_G
    \ ,\\
    \lim_{\theta\to\infty}r_C
    &=
    \langle\lVert\x\rVert^2\rangle_P
    \ ,\\
    \lim_{\theta\to\infty}r_D
    &=
    \frac{\langle y^2\rangle_G}{\langle y\rangle_G}
    \ .
\end{aligned}
\end{equation}
Taking the limit in
Eq.~\eqref{eq:summary_quadratic_noncentered_overlap} gives
\begin{equation}
    \lim_{\theta\to\infty}D
    =
    \lim_{\theta\to\infty}
    r_D
    \sqrt{
        \frac{\lambda_{\max}(T_\theta)}
             {2r_C+\theta r_D^2}
    }
    =
    \sqrt{\langle y^2\rangle_G}>0
    \ .
\end{equation}
The overlap therefore remains finite even though the
principal eigenvalue grows linearly with $\theta$.

\begin{figure}[t]
    \centering
    \includegraphics[width=\textwidth]{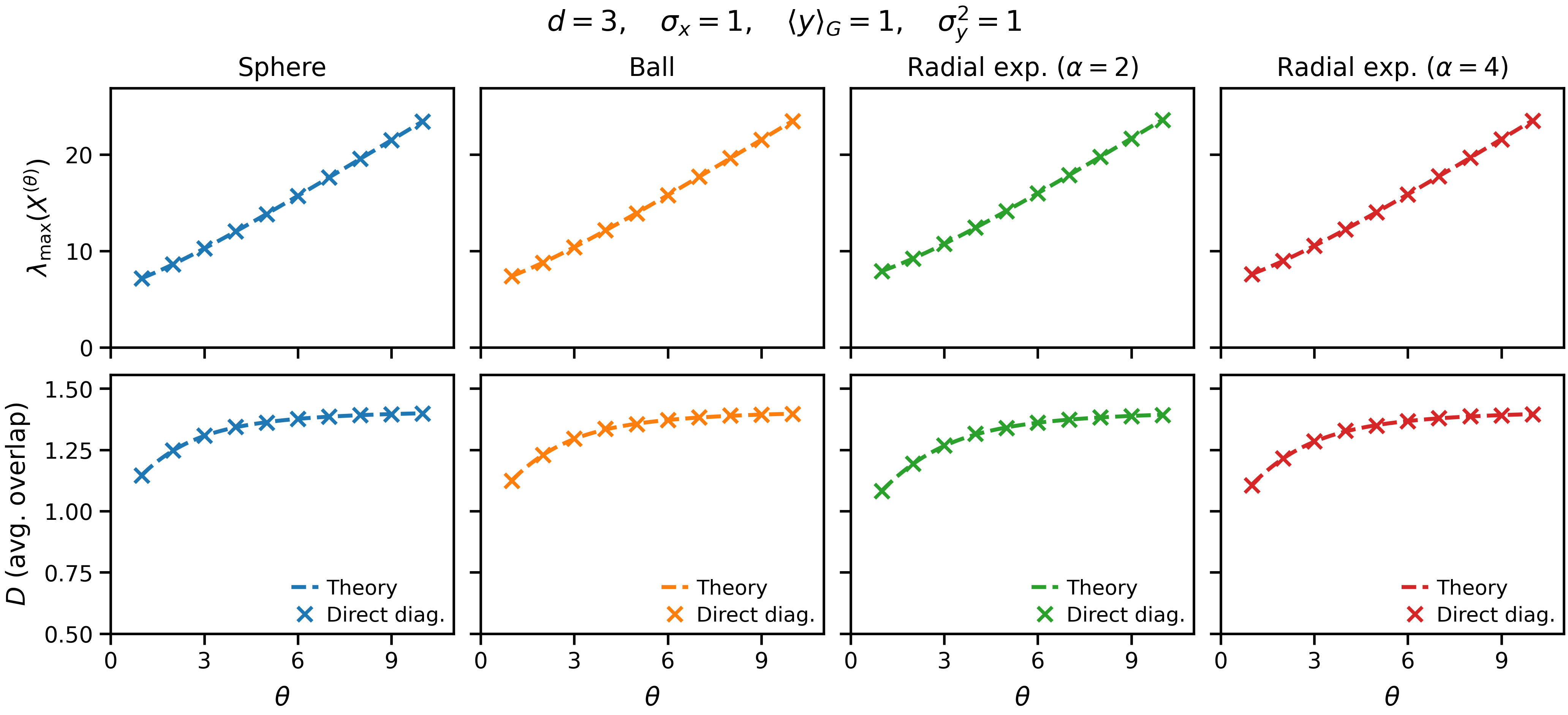}
    \caption{
        Largest eigenvalue and spike overlap for the quadratic
        distance kernel with a non-centered Gaussian spike.
        Comparison between the theoretical predictions in
        Eqs.~\eqref{eq:summary_quadratic_noncentered_eigenvalue}
        and~\eqref{eq:summary_quadratic_noncentered_overlap}
        (dashed lines) and the results obtained via direct
        numerical diagonalisation (crosses), as functions
        of $\theta$.
        The top row shows the average largest eigenvalue,
        while the bottom row shows the average overlap,
        whose limiting value is $D$.
        From left to right, columns correspond to the uniform
        distribution on the sphere, the uniform distribution
        in the ball, and the generalized radial exponential
        with $\alpha=2$ and $\alpha=4$, as defined in
        Table~\ref{tab:isotropic}.
        The case $\alpha=2$ corresponds to the centered
        Gaussian spatial distribution.
        Parameters are fixed to $d=3$ and $\sigma_x^2=1$.
        The spike components are independently drawn from a
        Gaussian distribution with $\langle y\rangle_G=1$
        and $\sigma_y^2=1$.
        Numerical results are obtained for matrices of size
        $N=5000$ and averaged over $100$ independent realisations.
    }
    \label{fig:quadratic_noncentered}
\end{figure}

To test the analytical results for a non-centered spike,
we again consider the three isotropic distributions reported
in Table~\ref{tab:isotropic}.
In Fig.~\ref{fig:quadratic_noncentered}, we compare the
theoretical predictions for the average largest eigenvalue
and the average overlap, given by
Eqs.~\eqref{eq:summary_quadratic_noncentered_eigenvalue}
and~\eqref{eq:summary_quadratic_noncentered_overlap}
(dashed lines), with the results obtained via direct numerical
diagonalisation (crosses), as functions of $\theta$.
We fix $d=3$ and $\sigma_x^2=1$, and draw the spike
components independently from a Gaussian distribution with
$\langle y\rangle_G=1$ and $\sigma_y^2=1$.
From left to right, the four columns correspond to the uniform
distribution on the sphere, the uniform distribution in the
ball, and the generalized radial exponential with $\alpha=2$
and $\alpha=4$.
The top row shows the average largest eigenvalue, while the
bottom row shows the average overlap, whose limiting value
is $D$.
Numerical results are obtained for matrices of size $N=5000$
and averaged over $100$ independent realisations. In all four cases, the numerical results agree very well
with the analytical predictions.
Both the largest eigenvalue and the overlap increase
with $\theta$. The overlap is strictly positive for all positive values of $\theta$, confirming the predicted
absence of a transition.

\FloatBarrier

\section{Results for the rank-two kernel}\label{sec:results_rank2}
In this section we consider the kernel
\begin{equation}\label{eq:results_rank_two_kernel}
    f(\x,\x')=a+b\,h(\x)h(\x')
    \ ,
\end{equation}
where $h:\mathbb R^d\to\mathbb R$ is real-valued and
\begin{equation}\label{eq:results_rank_two_assumptions}
    \langle h\rangle_P=0
    \ ,\qquad
    \langle h^2\rangle_P=1
    \ ,\qquad
    0<a<b
    \ .
\end{equation}
These assumptions ensure Eq.~\eqref{eq:kernel_L2_condition},
so $T_f$ is a Hilbert–Schmidt operator and, by symmetry of
the kernel, is self-adjoint.

As shown in Appendix~\ref{appendix:rank_two_kernel}, the
unperturbed principal eigenvalue and its eigenfunction
normalized according to Eq.~\eqref{eq:psi_normalisation} are
\begin{equation}\label{eq:summary_rank2_unperturbed}
    \Lambda_0=\lambda_{\max}(T_f)=b
    \ ,\qquad
    \psi_{\max}(\x)=b\,h(\x)
    \ .
\end{equation}
This eigenvalue is simple and the corresponding eigenfunction has zero $P$-average,
so the kernel belongs to the case $\Lambda_0>0$ and
$\Pi_{E_1}\mathbf1=0$. The results of
Sec.~\ref{sec:recap_results}, summarized in
Table~\ref{tab:overlap_transition_summary}, therefore predict
a sharp transition at a positive threshold for both centered
and non-centered spikes. The corresponding spiked kernel is
\begin{equation}\label{eq:summary_rank2_spiked_kernel}
    F_\theta\bigl((\x,y),(\x',y')\bigr)
    =
    a+b\,h(\x)h(\x')+\theta yy'
    \ .
\end{equation}
As before, $y$ is independent of $\x$, and its distribution
$G$ has a finite second moment and is non-degenerate, so that
$0<\sigma_y^2<\infty$. Together with the assumptions on $h$,
these conditions ensure Eq.~\eqref{eq:spiked_kernel_L2_condition},
making the spiked spectral problem well defined.

As shown in Appendix~\ref{appendix:rank_two_kernel}, every
eigenfunction associated with a nonzero eigenvalue $\lambda$
of the perturbed operator $T_\theta$ has the form
\begin{equation}\label{eq:summary_rank2_form}
    \Psi(\x,y)=A+B\,h(\x)+\theta Dy
    \ ,\qquad
    D=\frac{q[\Psi]}{\lambda}
    \ .
\end{equation}
In terms of the three coefficients $(A,B,D)$\footnote{Here $(A,B,D)$ denote coefficients specific to the
rank-two kernel, distinct from those used for the quadratic
distance kernel.}, the
self-consistent system in Eq.~\eqref{eq:final_sc_system} becomes
\begin{equation}\label{eq:summary_rank2_system}
    \left\{
    \begin{aligned}
        \lambda A
        &=a\left[A+\theta D\langle y\rangle_G\right]
        \ ,\\
        \lambda B
        &=bB
        \ ,\\
        \lambda D
        &=A\langle y\rangle_G
        +\theta D\langle y^2\rangle_G
        \ .
    \end{aligned}
    \right.
\end{equation}
Diagonalizing the corresponding $3\times3$ matrix and selecting
its largest eigenvalue gives $\lambda_{\max}(T_\theta)$ 
and the coefficients $(A,B,D)$ of the principal eigenfunction.
Normalizing this eigenfunction according to
Eq.~\eqref{eq:spiked_psi_normalisation} imposes the constraint
\begin{equation}\label{eq:summary_rank2_normalization}
    \frac{A^2}{a}+\frac{B^2}{b}+\theta D^2
    =
    \lambda_{\max}(T_\theta)
    \ .
\end{equation}
With this normalization, $D$ again gives the limiting average
overlap between the top eigenvector and the spike, as follows
by comparing Eqs.~\eqref{eq:summary_rank2_form}
and~\eqref{eq:spiked_overlap_main}
\begin{equation}
    \lim_{N\to\infty}
    \left\langle
        \frac{(\vec y,\vec v^{\, (1)})}{N}
    \right\rangle
    =
    \frac{q[\Psi_{\max}]}{\lambda_{\max}(T_\theta)}
    =
    D
    \ .
\end{equation}

\subsection{Centered spike}\label{sec:summary_rank2_centered}
\begin{figure}[t]
    \centering
    \includegraphics[width=\textwidth]{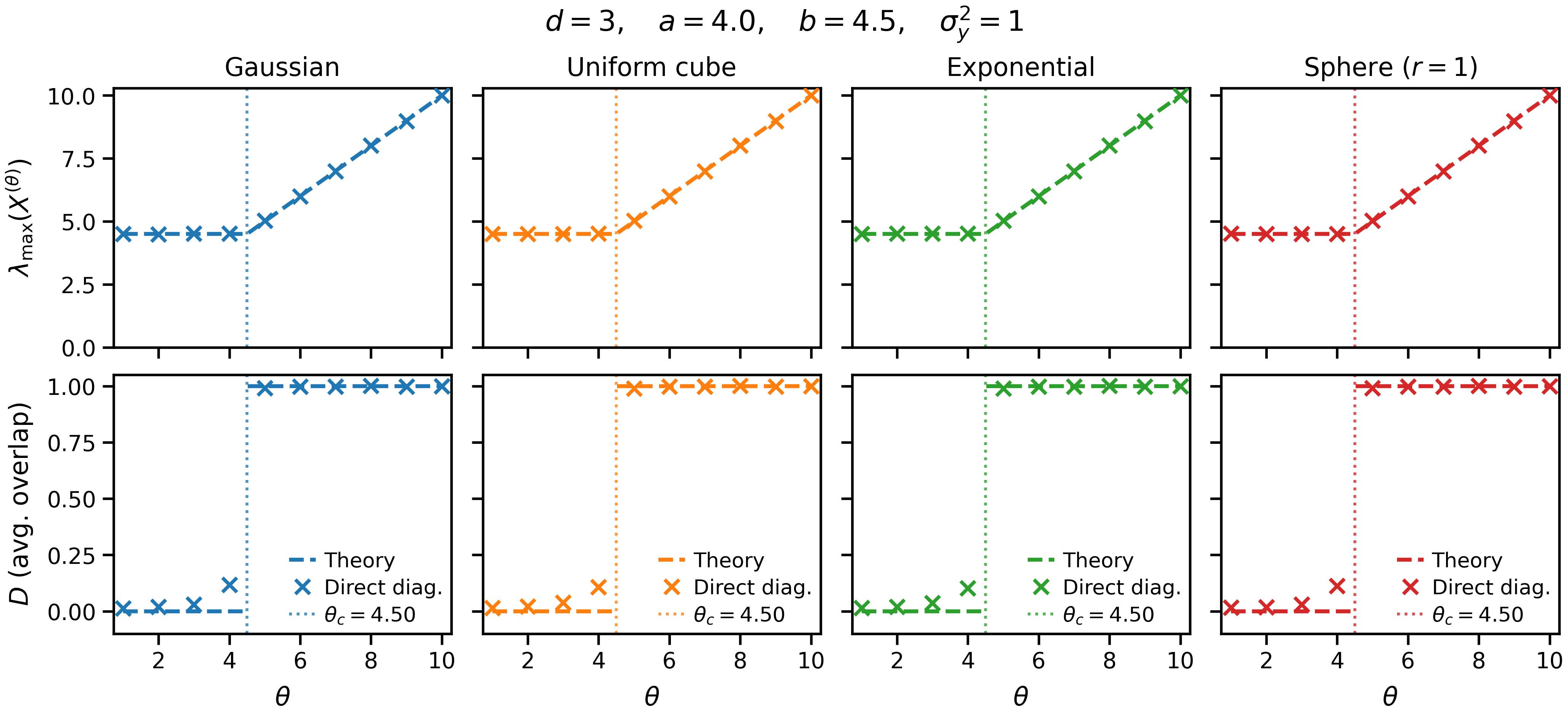}
    \caption{
        Largest eigenvalue and spike overlap for the rank-two
        kernel with a centered Gaussian spike.
        Comparison between the theoretical predictions in
        Eqs.~\eqref{eq:summary_rank2_centered_eigenvalue}
        and~\eqref{eq:summary_rank2_centered_overlap}
        (dashed lines) and the results obtained via direct
        numerical diagonalisation (crosses), as functions
        of $\theta$.
        The top row shows the average largest eigenvalue,
        while the bottom row shows the average overlap,
        whose limiting value is $D$.
        From left to right, columns correspond to the Gaussian,
        uniform cube, independent exponential, and uniform
        unit sphere choices of $(P(\x),h(\x))$ defined in
        Table~\ref{tab:rank2_distributions}.
        Parameters are fixed to $d=3$, $a=4$, and $b=4.5$.
        The spike components are independently drawn from a
        Gaussian distribution with $\langle y\rangle_G=0$
        and $\sigma_y^2=1$.
        Vertical dotted lines mark the common threshold
        $\theta_c=4.5$ given by
        Eq.~\eqref{eq:summary_rank2_centered_threshold}.
        Numerical results are obtained for matrices of size
        $N=5000$ and averaged over $100$ independent realisations.
    }
    \label{fig:rank2_centered}
\end{figure}

For $\langle y\rangle_G=0$, the results of
Sec.~\ref{sec:recap_results}, summarized in
Table~\ref{tab:overlap_transition_summary}, predict a sharp
transition from zero to nonzero overlap at a positive threshold.
Solving the coefficient system in
Eq.~\eqref{eq:summary_rank2_system}, as detailed in
Appendix~\ref{appendix:rank_two_kernel}, gives
\begin{equation}\label{eq:summary_rank2_centered_eigenvalue}
    \lambda_{\max}(T_\theta)
    =
    \begin{cases}
        b\ ,
        & 0<\theta<\theta_c\ ,
        \\[1ex]
        \theta\sigma_y^2\ ,
        & \theta>\theta_c\ ,
    \end{cases}
\end{equation}
and threshold
\begin{equation}\label{eq:summary_rank2_centered_threshold}
    \theta_c
    =
    \frac{b}{\sigma_y^2}
    =
    \frac{\lambda_{\max}(T_f)}{\sigma_y^2}
    \ .
\end{equation}
The branch with eigenvalue $b$ reproduces the unperturbed
value in Eq.~\eqref{eq:summary_rank2_unperturbed} and remains
dominant below $\theta_c$, after which the spike branch
$\theta\sigma_y^2$ takes over.
The principal eigenfunction normalized according to
Eq.~\eqref{eq:summary_rank2_normalization} is
\begin{equation}\label{eq:summary_rank2_centered_eigenfunction}
    \Psi_{\max}(\x,y)
    =
    \begin{cases}
        b\,h(\x)\ ,
        & 0<\theta<\theta_c\ ,
        \\[1ex]
        \theta\sigma_y y\ ,
        & \theta>\theta_c\ ,
    \end{cases}
\end{equation}
and the limiting average overlap is
\begin{equation}\label{eq:summary_rank2_centered_overlap}
    \lim_{N\to\infty}
    \left\langle
        \frac{(\vec y,\vec v^{\, (1)})}{N}
    \right\rangle
    =D
    =
    \begin{cases}
        0\ ,&0<\theta<\theta_c\ ,
        \\[1ex]
        \sigma_y\ ,&\theta>\theta_c\ .
    \end{cases}
\end{equation}
Thus, the principal eigenfunction retains its unperturbed
form below the threshold and becomes proportional to $y$
above it, with the overlap displaying the predicted sharp
transition. At $\theta=\theta_c$,
$\theta_c\sigma_y^2=b=\lambda_{\max}(T_f)$ and the principal
eigenvalue is doubly degenerate. To test the analytical results for the rank-two kernel,
we consider four choices of the spatial distribution $P(\x)$
and the associated function $h(\x)$: a Gaussian distribution
with independent standard normal components, the uniform
distribution in the unit cube $[-1,1]^d$, independent exponential
components with unit rate, and the uniform distribution
on the sphere of radius $r=1$.
The probability densities and the corresponding functions
$h(\x)$ for these four choices are reported in
Table~\ref{tab:rank2_distributions}.
In each case, $h(\x)$ satisfies
$\langle h\rangle_P=0$ and $\langle h^2\rangle_P=1$,
as required by Eq.~\eqref{eq:results_rank_two_assumptions}. In Fig.~\ref{fig:rank2_centered}, we compare the theoretical
predictions for the average largest eigenvalue and the average
overlap, given by
Eqs.~\eqref{eq:summary_rank2_centered_eigenvalue}
and~\eqref{eq:summary_rank2_centered_overlap}
(dashed lines), with the results obtained via direct numerical
diagonalisation (crosses), as functions of $\theta$.
Vertical dotted lines indicate the predicted transition
threshold $\theta_c$ given by
Eq.~\eqref{eq:summary_rank2_centered_threshold}.
We fix $d=3$, $a=4$, and $b=4.5$, and draw the spike
components independently from a Gaussian distribution with
$\langle y\rangle_G=0$ and $\sigma_y^2=1$.
From left to right, the four columns correspond to the
Gaussian, uniform cube, independent exponential, and uniform
unit sphere choices of $(P(\x),h(\x))$ reported in
Table~\ref{tab:rank2_distributions}.
The top row shows the average largest eigenvalue, while the
bottom row shows the average overlap, whose limiting value
is $D$.
Numerical results are obtained for matrices of size $N=5000$
and averaged over $100$ independent realisations.

The theoretical predictions coincide for all four choices
of $(P(\x),h(\x))$, and the numerical results agree very well with
these predictions, clearly displaying the transition at
$\theta_c=4.5$.
Below the threshold, the largest eigenvalue stays at its
unperturbed value $b$ and the overlap is close to zero.
Above the threshold, the largest eigenvalue increases
linearly, and the overlap
becomes $\sigma_y=1$.

\begin{table}[t]
    \centering
    \small
    \renewcommand{\arraystretch}{1.9}
    \setlength{\tabcolsep}{10pt}
    \begin{tabular}{|l|c|c|}
        \hline
        \rule{0pt}{3.2ex}
        Distribution & $P(\x)$ & $h(\x)$ \\
        \hline

        \rule{0pt}{4.8ex}
        Gaussian
        & $\displaystyle
            \frac{1}{(2\pi)^{d/2}}
            \exp\left\{-\frac{\lVert\x\rVert^2}{2}\right\}$
        & $\displaystyle
            \frac{\sum_{\ell=1}^d x_\ell}{\sqrt d}$
        \\[1.2ex]
        \hline

        \rule{0pt}{4.8ex}
        Uniform in the cube
        & $\displaystyle
            \frac{1}{2^d}
            \prod_{\ell=1}^d\Theta(1-|x_\ell|)$
        & $\displaystyle
            \frac{\lVert\x\rVert^2-d/3}{\sqrt{4d/45}}$
        \\[1.2ex]
        \hline

        \rule{0pt}{4.8ex}
        Independent exponential
        & $\displaystyle
            \exp\left\{-\sum_{\ell=1}^d x_\ell\right\}
            \prod_{\ell=1}^d\Theta(x_\ell)$
        & $\displaystyle
            \frac{\sum_{\ell=1}^d x_\ell-d}{\sqrt d}$
        \\[1.2ex]
        \hline

        \rule{0pt}{4.8ex}
        Uniform on the sphere
        & $\displaystyle
            \frac{\Gamma(d/2)}{2\pi^{d/2}r^{d-1}}
            \delta(r-\lVert\x\rVert)$
        & $\displaystyle
            \frac{\sum_{\ell=1}^d x_\ell}{r}$
        \\[1.2ex]
        \hline
    \end{tabular}
    \caption{
        Spatial distributions and associated functions
        for the rank-two kernel.
        Probability density $P(\x)$ and corresponding
        function $h(\x)$ for each of the four choices,
        all satisfying $\langle h\rangle_P=0$ and
        $\langle h^2\rangle_P=1$.
        The Gaussian, cube, and exponential distributions
        have independent components.
        For the uniform distribution on the sphere,
        the radius is fixed to $r=1$.
        Here $\Theta$ denotes the Heaviside step function.
    }
    \label{tab:rank2_distributions}
\end{table}

\FloatBarrier

\subsection{Non-centered spike}\label{sec:summary_rank2_noncentered}
For $\langle y\rangle_G\neq0$, the conditions $\Lambda_0=b>0$
and $\Pi_{E_1}\mathbf1=0$ place the rank-two kernel in the
case where Sec.~\ref{sec:recap_results} and
Table~\ref{tab:overlap_transition_summary} predict a sharp
transition at a positive threshold. Since the perturbation $\theta yy'$ is invariant under the
global sign change $y\mapsto-y$, we again assume
$\langle y\rangle_G>0$ without loss of generality.
The solution obtained in Appendix~\ref{appendix:rank_two_kernel}
gives the principal eigenvalue
\begin{equation}\label{eq:summary_rank2_noncentered_eigenvalue}
    \lambda_{\max}(T_\theta)
    =
    \begin{cases}
        b\ ,
        &0<\theta\leq\theta_c\ ,
        \\[1ex]
        \lambda_c(\theta)\ ,
        &\theta>\theta_c\ ,
    \end{cases}
\end{equation}
where the competing branch is strictly positive for every
$\theta>0$ and is given by
\begin{equation}\label{eq:summary_rank2_lambda_plus}
    \lambda_c(\theta)
    =
    \frac{
        a+\theta\langle y^2\rangle_G
        +\sqrt{
            (a-\theta\langle y^2\rangle_G)^2
            +4a\theta\langle y\rangle_G^2
        }
    }{2}
    \ ,
\end{equation}
and the crossing occurs at
\begin{equation}\label{eq:summary_rank2_noncentered_threshold}
    \theta_c
    =
    \left[
        \frac{\langle y\rangle_G^2}{b-a}
        +\frac{\sigma_y^2}{b}
    \right]^{-1}
    >0
    \ .
\end{equation}
At $\theta=\theta_c$, the equality $\lambda_c(\theta_c)=b$
produces a doubly degenerate principal eigenvalue.
With $\Lambda_0=b$, the threshold can equivalently be written
as $\theta_c=[F(b^+)]^{-1}$, in agreement with
Eqs.~\eqref{eq:recap_F_definition} and~\eqref{eq:recap_F_edge}.
The evaluation of this limit is given in
Appendix~\ref{appendix:rank_two_kernel}. The principal eigenfunction normalized according to
Eq.~\eqref{eq:summary_rank2_normalization} is
\begin{equation}\label{eq:summary_rank2_noncentered_eigenfunction}
    \Psi_{\max}(\x,y)
    =
    \begin{cases}
        b\,h(\x)\ ,
        &0<\theta<\theta_c\ ,
        \\[1ex]
        D\bigl(r_A+\theta y\bigr)\ ,
        &\theta>\theta_c\ ,
    \end{cases}
\end{equation}
and the limiting average overlap takes the form
\begin{equation}\label{eq:summary_rank2_noncentered_overlap}
    \lim_{N\to\infty}
    \left\langle
        \frac{(\vec y,\vec v^{\, (1)})}{N}
    \right\rangle
    =D
    =
    \begin{cases}
        0\ ,
        &0<\theta<\theta_c\ ,
        \\[2ex]
        \displaystyle
        \sqrt{\frac{a\lambda_c(\theta)}{a\theta+r_A^2}}\ ,
        &\theta>\theta_c\ ,
    \end{cases}
\end{equation}
with 
\begin{equation}\label{eq:summary_rank2_rA}
    r_A
    =
    \frac{a\theta\langle y\rangle_G}{\lambda_c(\theta)-a}>0
    \ .
\end{equation}
The overlap therefore vanishes below $\theta_c$ and is
strictly positive above it. To establish that this transition
is discontinuous, we compute the right limit of $D$ at
$\theta_c$.
Eqs.~\eqref{eq:summary_rank2_lambda_plus}
and~\eqref{eq:summary_rank2_noncentered_threshold} give
\begin{equation}
    \lim_{\theta\to\theta_c^+}\lambda_c(\theta)=b\ .
\end{equation}
Equation~\eqref{eq:summary_rank2_rA} therefore implies
\begin{equation}
    \lim_{\theta\to\theta_c^+}r_A
    =
    \frac{a\theta_c\langle y\rangle_G}{b-a}
    \ .
\end{equation}
Substituting these limits into
Eq.~\eqref{eq:summary_rank2_noncentered_overlap} yields
\begin{align}
    \lim_{\theta\to\theta_c^+}D
    &=
    \frac{
        b\langle y\rangle_G^2+(b-a)\sigma_y^2
    }{
        \sqrt{
            b^2\langle y\rangle_G^2
            +(b-a)^2\sigma_y^2
        }
    }
    >0
    \ ,
\end{align}
where we used
Eq.~\eqref{eq:summary_rank2_noncentered_threshold}.
Since the overlap vanishes below $\theta_c$, this strictly
positive right limit proves that the transition is
discontinuous.
To characterize the overlap far above the threshold,
we now consider the limit $\theta\to\infty$.
Eqs.~\eqref{eq:summary_rank2_lambda_plus}
and~\eqref{eq:summary_rank2_rA} give
\begin{equation}
\begin{aligned}
    \lim_{\theta\to\infty}
    \frac{\lambda_c(\theta)}{\theta}
    &=
    \langle y^2\rangle_G
    \ ,\\
    \lim_{\theta\to\infty}r_A
    &=
    \frac{a\langle y\rangle_G}{\langle y^2\rangle_G}
    \ .
\end{aligned}
\end{equation}
Since $r_A$ has a finite limit,
Eq.~\eqref{eq:summary_rank2_noncentered_overlap} then yields
\begin{equation}
    \lim_{\theta\to\infty}D
    =
    \sqrt{\langle y^2\rangle_G}
    \ .
\end{equation}

\begin{figure}[t]
    \centering
    \includegraphics[width=\textwidth]{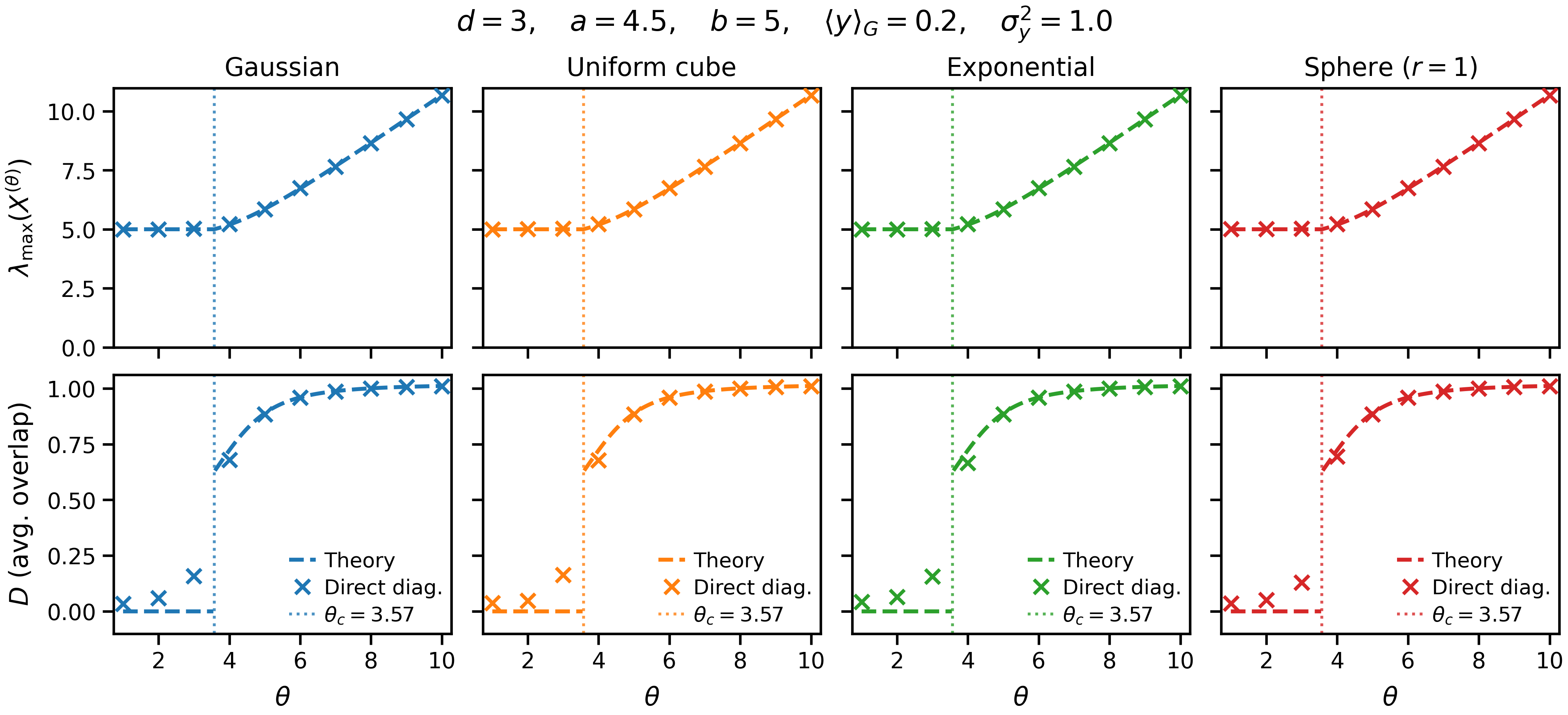}
    \caption{
        Largest eigenvalue and spike overlap for the rank-two
        kernel with a non-centered Gaussian spike.
        Comparison between the theoretical predictions in
        Eqs.~\eqref{eq:summary_rank2_noncentered_eigenvalue}
        and~\eqref{eq:summary_rank2_noncentered_overlap}
        (dashed lines) and the results obtained via direct
        numerical diagonalisation (crosses), as functions
        of $\theta$.
        The top row shows the average largest eigenvalue,
        while the bottom row shows the average overlap,
        whose limiting value is $D$.
        From left to right, columns correspond to the Gaussian,
        uniform cube, independent exponential, and uniform
        unit sphere distributions, with the associated
        functions $h(\x)$ defined in
        Table~\ref{tab:rank2_distributions}.
        Parameters are fixed to $d=3$, $a=4.5$, and $b=5$.
        The spike components are independently drawn from a
        Gaussian distribution with $\langle y\rangle_G=0.2$
        and $\sigma_y^2=1$.
        Vertical dotted lines mark the common threshold
        $\theta_c=3.57$ given by
        Eq.~\eqref{eq:summary_rank2_noncentered_threshold}.
        Numerical results are obtained for matrices of size
        $N=5000$ and averaged over $100$ independent realisations.
    }
    \label{fig:rank2_noncentered}
\end{figure}

In Fig.~\ref{fig:rank2_noncentered}, we compare the theoretical
predictions for the average largest eigenvalue and the average
overlap, given by
Eqs.~\eqref{eq:summary_rank2_noncentered_eigenvalue}
and~\eqref{eq:summary_rank2_noncentered_overlap}
(dashed lines), with the results obtained via direct numerical
diagonalisation (crosses), as functions of $\theta$, using
the four choices of $P(\x)$ and $h(\x)$ reported in
Table~\ref{tab:rank2_distributions}. Vertical dotted lines indicate the predicted transition
threshold $\theta_c$ given by
Eq.~\eqref{eq:summary_rank2_noncentered_threshold}.
We fix $d=3$, $a=4.5$, and $b=5$, and draw the spike
components independently from a Gaussian distribution with
$\langle y\rangle_G=0.2$ and $\sigma_y^2=1$.
From left to right, the four columns correspond to the
Gaussian, uniform cube, independent exponential, and uniform
unit sphere distributions, with their associated functions
$h(\x)$.
The top row shows the average largest eigenvalue, while the
bottom row shows the average overlap, whose limiting value
is $D$.
Numerical results are obtained for matrices of size $N=5000$
and averaged over $100$ independent realisations.

The theoretical predictions coincide for all four choices
of $P(\x)$ and $h(\x)$, and the numerical results agree
very well with these predictions.
A clear transition is observed at the common threshold
$\theta_c$, confirming that a transition also
occurs for a non-centered spike in this model.
Below the threshold, the largest eigenvalue stays at its
unperturbed value $b=5$ and the overlap is zero.
Above the threshold, the largest eigenvalue follows the
branch $\lambda_c(\theta)$ in
Eq.~\eqref{eq:summary_rank2_lambda_plus}, and the overlap
becomes nonzero.

\FloatBarrier

\section{Conclusion}\label{sec:conclusion}

In this work, we investigated signal recovery through standard PCA for symmetric Euclidean random matrices with an independent additive rank-one perturbation. By treating the spike component as an additional coordinate of each random point, we showed how the replica framework of Ref.~\cite{casaburi2026} can be extended to study the perturbed problem. The resulting Fredholm formulation determines the average largest eigenvalue, the density of the top eigenvector's components, and the average overlap with the signal in the large-$N$ limit, while accounting for the geometric correlations without imposing rotational invariance on the background noise. Its self-consistency equations depend on the unperturbed operator and the first two moments of the spike distribution. We also established that the perturbed principal eigenvalue is strictly positive for every $\theta>0$, ensuring that the formulation remains applicable even when the unperturbed upper spectral edge is nonpositive.

Our main result is a general analytical classification of recovery transitions, with criteria and threshold formulas derived without specifying the kernel or the point and spike distributions. Writing $\Lambda_0$ for the unperturbed upper spectral edge, we found that the overlap is nonzero for every $\theta>0$ when $\Lambda_0\leq0$. For $\Lambda_0>0$, a centered spike always produces a positive threshold $\theta_c=\Lambda_0/\sigma_y^2$, where $\sigma_y^2$ is the spike variance. For a non-centered spike, a positive threshold persists precisely when every eigenfunction in the unperturbed principal eigenspace has zero average under the point distribution. If this eigenspace contains even one eigenfunction with nonzero average, recovery occurs for every positive signal strength. A nonzero spike mean can therefore remove or preserve a sharp recovery threshold according to its coupling to the principal modes of the Euclidean background.

The most striking feature of these transitions is their interpretation in terms of a reordering of principal directions. In the centered case, the signal direction is an eigenfunction of the perturbed operator for every positive signal strength, but remains subleading below the threshold. The transition occurs when its eigenvalue overtakes the largest eigenvalue of the Euclidean background, causing the leading eigenvector to switch abruptly from a noise direction to the signal-bearing direction. For non-centered spikes, an analogous crossing survives when the principal eigenspace of the background is orthogonal to the constant function, although the competing signal-bearing eigenfunction generally mixes spatial and signal coordinates. Thus, the discontinuity reflects a change in which eigendirection is selected by the leading spectral estimator, rather than the sudden appearance of the signal in the spectrum. Unlike transitions governed by the edge of a continuous spectral bulk~\cite{Bocchi2026}, this mechanism involves competing isolated eigenvalues of a compact operator. Related crossings occur in sparse symmetric and sparse Wishart models~\cite{Adomaityte2025,Forer2026}, but there the signal eigenvalue can first detach from a continuous bulk before overtaking a structural background outlier, whereas our transition takes place entirely within the discrete positive spectrum of the limiting operator. 

We illustrated the general classification using the quadratic distance kernel and a rank-two kernel, obtaining explicit solutions through finite-dimensional reductions of the Fredholm problem. For the quadratic distance kernel, a nonzero spike mean removes the positive threshold present for centered spikes, whereas the rank-two kernel retains a threshold for both centered and non-centered signals. These complementary examples make explicit the role of the spatial averages of the background principal eigenfunctions. We tested the theoretical predictions through direct numerical diagonalisation of large matrices, finding excellent agreement for the average largest eigenvalue and overlap across the distributions and parameter regimes considered.

Several natural extensions of this work can be considered. A first question concerns finite-size fluctuations near the transition, including how they resolve the degeneracy and determine the overlap at the crossing. This analysis would also clarify how the discontinuous limiting transition is approached at finite matrix size. Extending the framework to perturbations of higher but fixed rank would allow several signal directions to compete with the geometric background. It would also be interesting to relax the independence condition between the signal and the spatial variables, and determine how correlations with the underlying point configuration modify the recovery criteria.

\section{Acknowledgments}
P.V. acknowledges support from UKRI FLF Scheme (No. MR/X023028/1). 

\newpage

\appendix
\section{Proof for spectral edges}\label{appendix:spectral_edges}
In this Appendix, we compare the upper spectral edges of
the perturbed and unperturbed operators
\begin{equation}
    \Lambda_\theta
    \coloneqq
    \sup\operatorname{Spec}(T_\theta)\ ,
    \qquad\qquad
    \Lambda_0
    \coloneqq
    \sup\operatorname{Spec}(T_f)
    \ ,
\end{equation}
and we will prove that
\begin{equation}
    \Lambda_\theta\geq\Lambda_0
    \qquad\text{for all }\theta\geq0
    \ ,
\end{equation}
\begin{equation}
    \Lambda_\theta>0
    \qquad\text{for all }\theta>0
    \ .
\end{equation}
We consider the spaces $L^2(P)$ and $L^2(P\otimes G)$, equipped with the scalar products
\begin{align}
    (\phi,\psi)_P
    &\coloneqq
    \int \dd\x\,P(\x)\,
    \phi(\x)\psi(\x)
    \ ,
    \label{eq:appendix_scalar_product_P}
    \\
    (\Phi,\Psi)_{P\otimes G}
    &\coloneqq
    \int \dd\x\,\dd y\,
    P(\x)G(y)\,
    \Phi(\x,y)\Psi(\x,y)
    \ .
    \label{eq:appendix_scalar_product_PG}
\end{align}
We recall that the unperturbed and perturbed operators act on $L^2(P)$ and $L^2(P\otimes G)$, respectively, as
\begin{align}
    (T_f\psi)(\x)
    &=
    \int \dd\x'\,P(\x')\,
    f(\x,\x')\psi(\x')
    \ ,
    \\
    (T_\theta\Psi)(\x,y)
    &=
    \int \dd\x'\,\dd y'\,
    P(\x')G(y')\,
    \left[f(\x,\x')+\theta yy'\right]
    \Psi(\x',y')
    \ .
\end{align}
We assume that the unperturbed kernel $f$ is symmetric and satisfies
\begin{equation}
    \int \dd\x\,\dd\x'\,
    P(\x)P(\x')\,
    \left|f(\x,\x')\right|^2
    <
    \infty
    \ ,
\end{equation}
so that $T_f$ is a Hilbert–Schmidt operator on $L^2(P)$ and is self-adjoint with respect to the scalar product in Eq.~\eqref{eq:appendix_scalar_product_P}.
We further assume that the distribution $G$ has a finite second moment and is non-degenerate, so that $0<\sigma_y^2<\infty$.
As discussed in Sec.~\ref{sec:spiked model}, these assumptions ensure that also the symmetric kernel
$F_\theta((\x,y),(\x',y'))=f(\x,\x')+\theta yy'$ satisfies
\begin{align}
    &\int \dd\x\,\dd y\,\dd\x'\,\dd y'\,
    P(\x)G(y)P(\x')G(y')\,
    \left|F_\theta\bigl((\x,y),(\x',y')\bigr)\right|^2
    \nonumber\\
    &\qquad\leq
    2\int \dd\x\,\dd\x'\,
    P(\x)P(\x')\left|f(\x,\x')\right|^2
    +2\theta^2\left\langle y^2\right\rangle_G^2
    <
    \infty
    \ .
\end{align}
Hence, $T_\theta$ is also a Hilbert–Schmidt operator on $L^2(P\otimes G)$ and is self-adjoint with respect to the scalar product in Eq.~\eqref{eq:appendix_scalar_product_PG}. For these bounded self-adjoint operators, the variational
characterization of the upper spectral edge gives
\begin{equation}
    \Lambda_0
    =
    \sup_{\lVert\psi\rVert_{L^2(P)}=1}
    (\psi,T_f\psi)_P
    \ ,
    \qquad
    \Lambda_\theta
    =
    \sup_{\lVert\Psi\rVert_{L^2(P\otimes G)}=1}
    (\Psi,T_\theta\Psi)_{P\otimes G}
    \ .
\end{equation}
To compare these suprema, we consider each $\psi\in L^2(P)$ as a function in $L^2(P\otimes G)$ independent of $y$, by setting $\Psi(\x,y)=\psi(\x)$. This identification preserves the norm because $G$ is a probability distribution
\begin{equation}
    \lVert\Psi\rVert_{L^2(P\otimes G)}^2
    =(\Psi,\Psi)_{P\otimes G}=
    \int \dd\x\,P(\x)\psi(\x)^2
    \int \dd y\,G(y)
    =(\psi,\psi)_P=
    \lVert\psi\rVert_{L^2(P)}^2
    \ .
\end{equation}
For any such function with $\lVert\psi\rVert_{L^2(P)}=1$, we compute
\begin{align}
    (\Psi,T_\theta\Psi)_{P\otimes G}
    &=
    \int \dd\x\,\dd y\,\dd\x'\,\dd y'\,
    P(\x)G(y)P(\x')G(y')\,
    \psi(\x)
    \left[f(\x,\x')+\theta yy'\right]
    \psi(\x')
    \nonumber\\
    &=
    \int \dd\x\,\dd\x'\,
    P(\x)P(\x')\,
    \psi(\x)f(\x,\x')\psi(\x')
    +
    \theta
    \left[
        \int \dd\x\,\dd y\,
        P(\x)G(y)y\psi(\x)
    \right]^2
    \nonumber\\
    &=
    (\psi,T_f\psi)_P
    +
    \theta\langle y\rangle_G^2
    \langle\psi\rangle_P^2
    \nonumber\\
    &\geq
    (\psi,T_f\psi)_P
    \ ,
\end{align}
where the last inequality holds for $\theta\geq0$.
Since the supremum defining $\Lambda_\theta$ includes all normalized $y$-independent functions, it is at least as large as the supremum over this subset:
\begin{equation}
    \Lambda_\theta
    \geq
    \sup_{\lVert\psi\rVert_{L^2(P)}=1}
    \left[
        (\psi,T_f\psi)_P
        +
        \theta\langle y\rangle_G^2
        \langle\psi\rangle_P^2
    \right]
    \geq
    \Lambda_0
    \ .
\end{equation}
To prove the strict positivity of $\Lambda_{\theta}$ for $\theta>0$, consider the normalized function
\begin{equation}
    \Psi_c(\x,y)
    \coloneqq
    \frac{y-\langle y\rangle_G}{\sigma_y}
    \ ,
    \qquad
    \lVert\Psi_c\rVert_{L^2(P\otimes G)}^2
    =
    \frac{\left\langle
        (y-\langle y\rangle_G)^2
    \right\rangle_G}{\sigma_y^2}
    =
    1
    \ .
\end{equation}
Since $\Psi_c$ is independent of $\x$ and has zero $G$-average, we obtain
\begin{align}
    (\Psi_c,T_\theta\Psi_c)_{P\otimes G}
    &=
    \frac{1}{\sigma_y^2}
    \int \dd\x\,\dd y\,\dd\x'\,\dd y'\,
    P(\x)G(y)P(\x')G(y')\,
    \left(y-\langle y\rangle_G\right)
    \left[f(\x,\x')+\theta yy'\right]
    \left(y'-\langle y\rangle_G\right)
    \nonumber\\
    &=
    \frac{1}{\sigma_y^2}
    \left[
        \int \dd\x\,\dd\x'\,
        P(\x)P(\x')f(\x,\x')
    \right]
    \left[
        \int \dd y\,G(y)
        \left(y-\langle y\rangle_G\right)
    \right]^2
    +
    \frac{\theta}{\sigma_y^2}
    \left[
        \int \dd y\,G(y)y
        \left(y-\langle y\rangle_G\right)
    \right]^2
    \nonumber\\
    &=
    \frac{\theta}{\sigma_y^2}
    \left[
        \langle y^2\rangle_G-\langle y\rangle_G^2
    \right]^2
    =
    \theta\sigma_y^2
    \ ,
\end{align}
where the $f$ term vanishes by centering, and the integrals over $\x$ and $\x'$ in the spike term equal one.
Since $\Psi_c$ is normalized, the variational principle gives
\begin{equation}
    \Lambda_\theta
    \geq
    \theta\sigma_y^2
    >
    0
    \ .
\end{equation}

\section{Replica derivation of the principal eigenpair statistics for the spiked model}\label{appendix_replica_der}
In this Appendix, we use the replica method to derive the large-$N$
average largest eigenvalue, the density of the top eigenvector's components,
and the average overlap between the top eigenvector and the spike for the
matrix $X^{(\theta)}$ defined in Eq.~\eqref{eq:spiked_matrix}. For the first two observables, after introducing
the augmented variables and the corresponding kernel $F_\theta$,
the replica calculation closely follows that developed in
Ref.~\cite{casaburi2026} for generic symmetric Euclidean random matrices.
We therefore follow the same construction, referring to
Ref.~\cite{casaburi2026} for the detailed replica computations and retaining
here only the main steps needed to make explicit the modifications introduced
by the spiked model. The calculation of
the average overlap is specific to the present setting, but follows closely
the derivation of the density of the top eigenvector's components, with a
suitable modification of the source term in the auxiliary partition
function.

\subsection{Largest eigenvalue}
In this section we derive the average largest eigenvalue of the spiked matrix
$X^{(\theta)}$ directly from the replica formalism. For a fixed realisation of the disorder, we introduce the partition function
\begin{equation}
    Z_{N,\theta}(\beta)
    \coloneqq
    \int \dd\vec v\,
    \exp\left[
        \frac{\beta}{2}
        \left(
            \vec v,
            X^{(\theta)}\vec v
        \right)
    \right]
    \delta\left(
        \lVert\vec v\rVert^2-N
    \right)
    \ .
\end{equation}
To evaluate the average largest eigenvalue through
Eq.~\eqref{eq:general_eigenvalue_replica}, we first compute the replicated
partition function averaged over the disorder
\begin{align}
    \left\langle
        \left[
            Z_{N,\theta}(\beta)
        \right]^n
    \right\rangle
    &=
    \int
    \prod_{c=1}^{n}
    \left[
        \dd\vec v_c\,
        \delta\left(
            \lVert\vec v_c\rVert^2-N
        \right)
    \right]
    \prod_{i=1}^{N}
    \left[
        \dd\x_i\,P(\x_i)\,
        \dd y_i\,G(y_i)
    \right]
    \exp\left\{
        \frac{\beta}{2N}
        \sum_{c=1}^{n}
        \sum_{i,j=1}^{N}
        \left[
            f(\x_i,\x_j)
            +
            \theta y_i y_j
        \right]
        v_{ic}v_{jc}
    \right\}
    \ ,
    \label{eq:spiked_replicated_partition_xy}
\end{align}
where we used that the vectors $\{\x_i\}_{i=1}^N$ are i.i.d.\ with density $P$, the variables $\{y_i\}_{i=1}^N$ are i.i.d.\ with density $G$, and the two families are independent. As in Sec.~\ref{sec:spiked model} of the main text, we now introduce the augmented random
vectors
\begin{equation}
    \z_i
    \coloneqq
    (\x_i,y_i)
    \in\mathbb R^{d+1}
    \ ,
\end{equation}
whose first $d$ components coincide with $\x_i$ and the last one with
$y_i$. We then define the corresponding probability density as
\begin{equation}\label{eq:augmented_probability_density}
    \mathcal P(\z)
    \coloneqq
    P(\x)G(y)
    \ ,
    \qquad
    \z=(\x,y)
    \ ,
\end{equation}
and use the shorthand $\dd\z\equiv\dd\x\,\dd y$.
Using the augmented kernel $F_\theta$ defined in
Eq.~\eqref{eq:spiked_kernel} of the main text, Eq.~\eqref{eq:spiked_replicated_partition_xy}
becomes
\begin{align}
    \left\langle
        \left[
            Z_{N,\theta}(\beta)
        \right]^n
    \right\rangle
    &=
    \int
    \prod_{c=1}^{n}
    \left[
        \dd\vec v_c\,
        \delta\left(
            \lVert\vec v_c\rVert^2-N
        \right)
    \right]
    \prod_{i=1}^{N}
    \left[
        \dd\z_i\,\mathcal P(\z_i)
    \right]
    \exp\left\{
        \frac{\beta}{2N}
        \sum_{c=1}^{n}
        \sum_{i,j=1}^{N}
        F_\theta(\z_i,\z_j)
        v_{ic}v_{jc}
    \right\}
    \ .
    \label{eq:spiked_replicated_partition_z}
\end{align}
Equation~\eqref{eq:spiked_replicated_partition_z} has exactly the same
functional form as Eq.~(8) of Ref.~\cite{casaburi2026}, corresponding to the
replicated partition function of the generic symmetric Euclidean random
matrix. The only replacements are
\begin{equation}
    \x
    \longrightarrow
    \z
    \ ,
    \qquad
    P(\x)
    \longrightarrow
    \mathcal P(\z)
    \ ,
    \qquad
    f(\x,\x')
    \longrightarrow
    F_\theta(\z,\z')
    \ ,
\end{equation}
and the integrations are now performed over $\mathbb R^{d+1}$ rather than
$\mathbb R^d$. The remainder of the replica calculation can therefore be carried out
without modification. We briefly reproduce the steps needed to identify the
corresponding spectral problem. We first rewrite the product of Dirac delta
functions through their Fourier representation
\begin{equation}\label{eq:spiked_fourier_representation}
\begin{aligned}
    \prod_{c=1}^{n}
    \delta\left(
        \lVert\Vec{v}_c\rVert^2-N
    \right)
    &=
    \left(
        \frac{\beta}{4\pi}
    \right)^n
    \int
    \dd\boldsymbol{\eta}\,
    \exp\left\{
        -\ii\frac{\beta}{2}
        \sum_{c=1}^{n}
        \sum_{i=1}^{N}
        \eta_c v_{ic}^2
        +
        \ii\frac{\beta}{2}N
        \sum_{c=1}^{n}
        \eta_c
    \right\}
    \ ,
\end{aligned}
\end{equation}
where $\dd\boldsymbol{\eta} = \prod_{c=1}^{n}\dd\eta_c$. For each replica, we then introduce the density
\begin{equation}\label{eq:spiked_replica_density}
    \rho_c(\z)
    \coloneqq
    \frac{1}{N}
    \sum_{i=1}^{N}
    \delta(\z-\z_i)v_{ic}
    \ ,
    \qquad
    c \in \{1,\ldots,n\}
    \ ,
\end{equation}
which allow us to write
\begin{equation}\label{eq:interaction_term_density}
    \frac{1}{N}
    \sum_{i,j=1}^{N}
    F_\theta(\z_i,\z_j)v_{ic}v_{jc}
    =
    N
    \int
    \dd\z\,\dd\z'\,
    \rho_c(\z)
    \rho_c(\z')
    F_\theta(\z,\z')
    \ .
\end{equation}
We enforce the definition of the densities in
Eq.~\eqref{eq:spiked_replica_density} through functional deltas, introducing the conjugate densities $\hat\rho_c(\z)$ as
\begin{equation}\label{eq:spiked_functional_delta}
\begin{aligned}
    1
    &=
    \int
    \{\mathcal D\rho_c\}
    \{\mathcal D\hat\rho_c\}
    \exp\left\{
        \ii N
        \sum_{c=1}^{n}
        \int
        \dd\z\,
        \rho_c(\z)\hat\rho_c(\z)
        -
        \ii
        \sum_{c=1}^{n}
        \sum_{i=1}^{N}
        \hat\rho_c(\z_i)v_{ic}
    \right\}
    \ ,
\end{aligned}
\end{equation}
where we used the notation $\{\mathcal{D}\rho_c\}=\prod_c \mathcal{D}\rho_c$, $\{\mathcal{D}\hrho_c\}=\prod_c \mathcal{D}\hrho_c$.
Using Eqs.~\eqref{eq:spiked_fourier_representation},
\eqref{eq:interaction_term_density}, and
\eqref{eq:spiked_functional_delta}, the averaged replicated partition
function in Eq.~\eqref{eq:spiked_replicated_partition_z} can then be written
in a form suitable for a saddle-point evaluation in the large-$N$ limit (see Eqs.~(15)–(19) of Ref.~\cite{casaburi2026})
\begin{equation}\label{eq:amenable_saddle}
    \left\langle
        \left[
            Z_{N,\theta}(\beta)
        \right]^n
    \right\rangle
    \propto
    \int
    \dd\boldsymbol{\eta}\,
    \{\mathcal D\rho_c\}
    \{\mathcal D\hat\rho_c\}
    \exp\left\{
        N\mathcal S_\theta
        [
            \boldsymbol{\eta},
            \{\rho_c\},
            \{\hat\rho_c\}
        ]
    \right\}
    \ ,
\end{equation}
where the action can be written as $\mathcal S_\theta[\boldsymbol{\eta},\{\rho_c\},\{\hat\rho_c\}]=\mathcal S_\theta^{(1)}[\boldsymbol{\eta}]+\mathcal S_\theta^{(2)}[\{\rho_c\}]+\mathcal S_\theta^{(3)}[\{\rho_c\},\{\hat\rho_c\}]+\mathcal S_\theta^{(4)}[\boldsymbol{\eta},\{\hat\rho_c\}]$,
with
\begin{equation}\label{eq:spiked_replica_action_1}
    \mathcal S_\theta^{(1)}[\boldsymbol{\eta}]
    =
    \frac{\ii\beta}{2}
    \sum_{c=1}^{n}\eta_c
    \ ,
\end{equation}
\begin{equation}\label{eq:spiked_replica_action_2}
    \mathcal S_\theta^{(2)}[\{\rho_c\}]
    =
    \frac{\beta}{2}
    \sum_{c=1}^{n}
    \int
    \dd\z\,\dd\z'\,
    \rho_c(\z)
    \rho_c(\z')
    F_\theta(\z,\z')
    \ ,
\end{equation}
\begin{equation}\label{eq:spiked_replica_action_3}
    \mathcal S_\theta^{(3)}[\{\rho_c\},\{\hat\rho_c\}]
    =
    \ii
    \sum_{c=1}^{n}
    \int
    \dd\z\,
    \rho_c(\z)
    \hat\rho_c(\z)
    \ ,
\end{equation}
\begin{equation}\label{eq:spiked_replica_action_4}
\begin{aligned}
    \mathcal S_\theta^{(4)}[\boldsymbol{\eta},\{\hat\rho_c\}]
    &=
    \log\left[
        \int
        \dd\z\,
        \mathcal P(\z)
        \prod_{c=1}^{n}
        \int
        \dd v_c\,
        \exp\left(
            -\ii\frac{\beta}{2}
            \eta_c v_c^2
            -
            \ii\hat\rho_c(\z)v_c
        \right)
    \right]
    \ .
\end{aligned}
\end{equation}
At this stage, we assume that in
Eq.~\eqref{eq:general_eigenvalue_replica} the limit $N\to\infty$ can be
taken before the zero-temperature and replica limits. Therefore the saddle-point evaluation of the
functional integral in Eq. \eqref{eq:amenable_saddle} yields
\begin{equation}\label{eq:spiked_replica_saddle_approximation}
    \left\langle
        \left[
            Z_{N,\theta}(\beta)
        \right]^n
    \right\rangle
    \underset{N\to\infty}{\approx}
    \exp\left\{
        N\,
        \mathcal S_\theta
        \left[
            \boldsymbol{\eta}^*,
            \{\rho_c^*\},
            \{\hat\rho_c^*\}
        \right]
    \right\}
    \ ,
\end{equation}
where $[\boldsymbol{\eta^*},\{\rho^*_c\}, \{\hat \rho^*_c\}]$ maximise the action $\mathcal S_\theta$. Assuming replica symmetry
\begin{equation}
    \eta_c=\eta
    \ ,
    \qquad
    \rho_c=\rho
    \ ,
    \qquad
    \hat\rho_c=\hat\rho
    \ ,
    \qquad
    \forall c=1,\ldots,n
    \ ,
\end{equation}
and expanding the action to first order in $n$, the saddle-point equations
take the same form as in the unperturbed calculation of Ref.~\cite{casaburi2026}. Introducing
\begin{equation}\label{eq:Psi_spiked_replica}
    \Psi(\z)
    \coloneqq
    \int
    \dd\z'\,
    \rho^*(\z')
    F_\theta(\z,\z')\ ,
\end{equation}
\begin{equation}\label{eq:s_theta_replica}
    s_\theta
    \coloneqq
    \int
    \dd\z\,
    \mathcal P(\z)
    \Psi(\z)^2
    \ ,
\end{equation}
the solutions of the saddle-point equations that maximize the functional
$\mathcal S_\theta[\eta,\rho,\hat\rho]$ are given by
(see Eqs.~(29)–(31) of Ref.~\cite{casaburi2026})
\begin{equation}\label{eq:spiked_rhohat_saddle}
    \hat\rho^*(\z)
    =
    \ii\beta\,
    \Psi(\z)
    \ ,
\end{equation}
\begin{equation}\label{eq:spiked_rho_saddle}
    \rho^*(\z)
    =
    \frac{\mathcal P(\z)}
    {\ii\eta^*}
    \Psi(\z)
    \ ,
\end{equation}
\begin{equation}\label{eq:spiked_eta_saddle}
    (\eta^*)^2
    +
    \frac{\ii}{\beta}\eta^*
    +
    s_\theta
    =
    0
    \quad\Longleftrightarrow\quad
    \eta^*
    =
    \frac{\ii}{2\beta}
    \left[
        -1
        -
        \sqrt{1+4\beta^2s_\theta}
    \right]
    =
    -\ii\sqrt{s_\theta}
    +
    O\left(\frac{1}{\beta}\right)
    \ .
\end{equation}
In Eq.~\eqref{eq:spiked_eta_saddle}, we selected the negative root for
$\eta^*$ according to the prescription discussed in
Ref.~\cite{casaburi2026}.
Inserting the saddle-point solution into the replica expression
Eq.~\eqref{eq:general_eigenvalue_replica} and taking the zero-temperature
limit therefore gives
\begin{equation}\label{eq:replica_lambda_spiked_s}
    \lim_{N\to\infty}\left\langle
        \lambda_{\max}
        \left(
            X^{(\theta)}
        \right)
    \right\rangle
    =
    \sqrt{s_\theta}
    \ .
\end{equation}

It remains to determine $s_\theta$ self-consistently. Combining
Eqs.~\eqref{eq:Psi_spiked_replica} and
\eqref{eq:spiked_rho_saddle} gives
\begin{equation}
    \Psi(\z)
    =
    \frac{1}{\ii\eta^*}
    \int
    \dd\z'\,
    \mathcal P(\z')
    F_\theta(\z,\z')
    \Psi(\z')
    \ .
\end{equation}
Taking the limit $\beta\to\infty$ and using
Eq.~\eqref{eq:spiked_eta_saddle}, we obtain
\begin{equation}\label{eq:spiked_replica_fredholm_z}
    \sqrt{s_\theta}\,
    \Psi(\z)
    =
    \int
    \dd\z'\,
    \mathcal P(\z')
    F_\theta(\z,\z')
    \Psi(\z')
    \ .
\end{equation}
Returning explicitly to $\z=(\x,y)$, this becomes
\begin{align}
    \sqrt{s_\theta}\,
    \Psi(\x,y)
    &=
    \int
    \dd\x'\,\dd y'\,
    P(\x')G(y')
    F_\theta
    \bigl(
        (\x,y),
        (\x',y')
    \bigr)
    \Psi(\x',y')
    =
    (T_\theta\Psi)(\x,y)
    \ .
    \label{eq:spiked_replica_fredholm_xy}
\end{align}
Thus, the replica saddle-point equations reduce precisely to the Fredholm
spectral problem of the operator $T_\theta$ introduced in
Eq.~\eqref{eq:spiked_Fredholm_operator}. As proved in Appendix~\ref{appendix:spectral_edges}, for $\theta>0$ the upper spectral
edge of $T_\theta$ is strictly positive. Since $T_\theta$ is compact and
self-adjoint, this upper edge is therefore an eigenvalue. The saddle
corresponding to the largest eigenvalue consequently yields
\begin{equation}\label{eq:spiked_replica_s_lamda}
    \sqrt{s_\theta}
    =
    \lambda_{\max}(T_\theta)
    \ ,
\end{equation}
and Eq.~\eqref{eq:replica_lambda_spiked_s} becomes
\begin{equation}\label{eq:spiked_replica_final_lambda}
    \lim_{N\to\infty}\left\langle
        \lambda_{\max}
        \left(
            X^{(\theta)}
        \right)
    \right\rangle
    =
    \lambda_{\max}(T_\theta)
    \ .
\end{equation}
This reproduces directly from the replica calculation the result stated in
Sec.~\ref{sec:spiked model}. Finally, denoting by $\Psi_{\max}$ the
eigenfunction associated with the principal eigenvalue
$\lambda_{\max}(T_\theta)$, Eqs.~\eqref{eq:s_theta_replica} and
\eqref{eq:spiked_replica_s_lamda} give
\begin{equation}
    1
    =
    \int
    \dd\x\,\dd y\,
    P(\x)G(y)
    \left[
        \frac{
            \Psi_{\max}(\x,y)
        }{
            \lambda_{\max}(T_\theta)
        }
    \right]^2
    \ ,
\end{equation}
which is exactly the normalization given in
Eq.~\eqref{eq:spiked_psi_normalisation}.

\subsection{Density of the top eigenvector's components}
In this section we compute the density of the top eigenvector's components $T(u)$.
According to the general replica representation in
Eq.~\eqref{eq:general_eigenvector_replica}, we introduce the auxiliary
partition function
\begin{equation}\label{eq:spiked_auxiliary_partition}
\begin{aligned}
    Z_{\epsilon,\beta}
    \left(
        t,X^{(\theta)};u
    \right)
    \coloneqq
    \int
    \dd\vec v\,
    \exp\left\{
        \frac{\beta}{2}
        \left(
            \vec v,
            X^{(\theta)}\vec v
        \right)
        +
        \beta t
        \sum_{i=1}^{N}
        \delta_\epsilon(u-v_i)
    \right\}
    \delta\left(
        \lVert\vec v\rVert^2-N
    \right)
    \ ,
\end{aligned}
\end{equation}
where $\delta_\epsilon$ denotes a smooth regularization of the Dirac delta
function. Replicating Eq.~\eqref{eq:spiked_auxiliary_partition} and averaging
over the disorder yields
\begin{align}
    \left\langle
        \left[
            Z_{\epsilon,\beta}
            \left(
                t,X^{(\theta)};u
            \right)
        \right]^n
    \right\rangle
    &=
    \int
    \prod_{c=1}^{n}
    \left[
        \dd\vec v_c\,
        \delta\left(
            \lVert\vec v_c\rVert^2-N
        \right)
    \right]
    \prod_{i=1}^{N}
    \left[
        \dd\x_i\,P(\x_i)\,
        \dd y_i\,G(y_i)
    \right]
    \nonumber\\
    &\quad\times
    \exp\left\{
        \frac{\beta}{2N}
        \sum_{c=1}^{n}
        \sum_{i,j=1}^{N}
        \left[
            f(\x_i,\x_j)
            +
            \theta y_i y_j
        \right]
        v_{ic}v_{jc}
        +
        \beta t
        \sum_{c=1}^{n}
        \sum_{i=1}^{N}
        \delta_\epsilon(u-v_{ic})
    \right\}
    \ .
    \label{eq:spiked_auxiliary_replicated_partition_xy}
\end{align}
Introducing again the augmented variables $\z_i=(\x_i,y_i)$, together with
$\mathcal P(\z)=P(\x)G(y)$ and the kernel $F_\theta$ defined in
Eq.~\eqref{eq:spiked_kernel}, the previous equation can be rewritten as
\begin{align}
    \left\langle
        \left[
            Z_{\epsilon,\beta}
            \left(
                t,X^{(\theta)};u
            \right)
        \right]^n
    \right\rangle
    &=
    \int
    \prod_{c=1}^{n}
    \left[
        \dd\vec v_c\,
        \delta\left(
            \lVert\vec v_c\rVert^2-N
        \right)
    \right]
    \prod_{i=1}^{N}
    \left[
        \dd\z_i\,\mathcal P(\z_i)
    \right]
    \nonumber\\
    &\quad\times
    \exp\left\{
        \frac{\beta}{2N}
        \sum_{c=1}^{n}
        \sum_{i,j=1}^{N}
        F_\theta(\z_i,\z_j)
        v_{ic}v_{jc}
        +
        \beta t
        \sum_{c=1}^{n}
        \sum_{i=1}^{N}
        \delta_\epsilon(u-v_{ic})
    \right\}
    \ ,
    \label{eq:spiked_auxiliary_replicated_partition_z}
\end{align}
that has exactly the same functional form as Eq.(59) of Ref.~\cite{casaburi2026}, with the same replacements introduced in the previous section:
$\x\to\z$, $P(\x)\to\mathcal P(\z)$, and
$f(\x,\x')\to F_\theta(\z,\z')$.

Using the Fourier representation and the densities introduced in
Eqs.~\eqref{eq:spiked_fourier_representation} and
\eqref{eq:spiked_replica_density}, and carrying out the same computations as
in the previous section and in Ref.~\cite{casaburi2026}, the replicated
auxiliary partition function can be written as
\begin{equation}\label{eq:spiked_eigenvector_action}
    \left\langle
        \left[
            Z_{\epsilon,\beta}
            \left(
                t,X^{(\theta)};u
            \right)
        \right]^n
    \right\rangle
    \propto
    \int
    \dd\boldsymbol{\eta}\,
    \{\mathcal D\rho_c\}
    \{\mathcal D\hat\rho_c\}
    \exp\left\{
        N\mathcal S_\theta
        [
            \boldsymbol{\eta},
            \{\rho_c\},
            \{\hat\rho_c\};
            \epsilon,t,u
        ]
    \right\}
    \ ,
\end{equation}
which is the analogue of Eq.~(61) of Ref.~\cite{casaburi2026}. The action can again be decomposed into four contributions, $\mathcal S_\theta[\boldsymbol{\eta},\{\rho_c\},\{\hat\rho_c\};\epsilon,t,u]=\mathcal S_\theta^{(1)}[\boldsymbol{\eta}]+\mathcal S_\theta^{(2)}[\{\rho_c\}]+\mathcal S_\theta^{(3)}[\{\rho_c\},\{\hat\rho_c\}]+\mathcal S_\theta^{(4)}[\boldsymbol{\eta},\{\hat\rho_c\};\epsilon,t,u]$. The first three terms coincide with Eqs.~\eqref{eq:spiked_replica_action_1}, \eqref{eq:spiked_replica_action_2}, and \eqref{eq:spiked_replica_action_3}, while the fourth becomes
\begin{equation}\label{eq:spiked_eigenvector_action_4}
\begin{aligned}
    \mathcal S_\theta^{(4)}
    [
        \boldsymbol{\eta},
        \{\hat\rho_c\};
        \epsilon,t,u
    ]
    &=
    \log\left[
        \int
        \dd\z\,
        \mathcal P(\z)
        \prod_{c=1}^{n}
        \int
        \dd v_c\,
        \exp\left(
            -\ii\frac{\beta}{2}\eta_c v_c^2
            -
            \ii\hat\rho_c(\z)v_c
            +
            \beta t\delta_\epsilon(u-v_c)
        \right)
    \right]
    \ ,
\end{aligned}
\end{equation}
in direct correspondence with Eq.~(62) of
Ref.~\cite{casaburi2026}. We assume that, in Eq.~\eqref{eq:general_eigenvector_replica}, the
large-$N$ limit can be taken before the zero-temperature, regularization,
and replica limits. Equation~\eqref{eq:spiked_eigenvector_action} can then
be evaluated by the saddle-point method, yielding
\begin{equation}\label{eq:spiked_eigenvector_saddle}
    \left\langle
        \left[
            Z_{\epsilon,\beta}
            \left(
                t,X^{(\theta)};u
            \right)
        \right]^n
    \right\rangle
    \underset{N\to\infty}{\approx}
    \exp\left\{
        N\mathcal S_\theta
        [
            \boldsymbol{\eta}^*,
            \{\rho_c^*\},
            \{\hat\rho_c^*\};
            \epsilon,t,u
        ]
    \right\}
    \ ,
\end{equation}
where quantities $\boldsymbol{\eta}^*$, $\{\rho_c^*\}$, and
$\{\hat\rho_c^*\}$ denote the saddle-point values that maximize the action
$\mathcal S_\theta$.
Since the derivative with respect to $t$ in
Eq.~\eqref{eq:general_eigenvector_replica} acts only on the explicit
$t$-dependence of the action, the saddle-point equations can be evaluated at
$t=0$. They then reduce exactly to those obtained in the previous
section. Hence, assuming replica symmetry, $\hat\rho^*$, $\rho^*$, and
$\eta^*$ are still given by
Eqs.~\eqref{eq:spiked_rhohat_saddle},
\eqref{eq:spiked_rho_saddle}, and
\eqref{eq:spiked_eta_saddle}. To first order in $n$, the only contribution that explicitly depends on
$t$ is therefore
\begin{equation}\label{eq:spiked_eigenvector_action_4_RS}
\begin{aligned}
    \mathcal S_\theta^{(4)}
    [
        \eta^*,
        \hat\rho^*;
        \epsilon,t,u
    ]
    &=
    n
    \int
    \dd\z\,
    \mathcal P(\z)
    \log\left[
        \int
        \dd v\,
        \exp\left(
            -\ii\frac{\beta}{2}\eta^*v^2
            -
            \ii\hat\rho^*(\z)v
            +
            \beta t\delta_\epsilon(u-v)
        \right)
    \right]
    +
    o(n)
    \ .
\end{aligned}
\end{equation}
Inserting Eqs.~\eqref{eq:spiked_eigenvector_saddle} and
\eqref{eq:spiked_eigenvector_action_4_RS} into
Eq.~\eqref{eq:general_eigenvector_replica} then yields
\begin{align}
    T(u)
    &=\lim_{\beta \to \infty}\frac{1}{\beta}\lim_{n \to 0}\frac{\partial}{\partial t}\frac{1}{n}\mathcal S_\theta^{(4)}[\eta^*,\hat\rho^*;\epsilon,t,u]\Big|_{\epsilon=t=0}\\
    &=\lim_{\beta\to\infty}
    \int
    \dd\z\,
    \mathcal P(\z)
    \frac{
        \exp\left[
            -\ii\frac{\beta}{2}\eta^*u^2
            -
            \ii\hat\rho^*(\z)u
        \right]
    }{
        \displaystyle
        \int
        \dd v\,
        \exp\left[
            -\ii\frac{\beta}{2}\eta^*v^2
            -
            \ii\hat\rho^*(\z)v
        \right]
    }
    \nonumber\\
    &=
    \int
    \dd\z\,
    \mathcal P(\z)
    \delta\left(
        u-
        \frac{\Psi(\z)}{\sqrt{s_\theta}}
    \right)
    \ ,
    \label{eq:spiked_top_eigenvector_density_z}
\end{align}
where in the last equality we used
Eqs.~\eqref{eq:spiked_rhohat_saddle} and
\eqref{eq:spiked_eta_saddle}. This is the direct analogue of Eq.~(65) of
Ref.~\cite{casaburi2026}. Finally, using $\z=(\x,y)$ and
$\mathcal P(\z)\dd\z=P(\x)G(y)\dd\x\,\dd y$, we can rewrite $T(u)$ as
\begin{equation}\label{eq:spiked_top_eigenvector_density_xy}
    T(u)
    =
    \int
    \dd\x\,\dd y\,
    P(\x)G(y)
    \delta\left(
        u-
        \frac{
            \Psi_{\max}(\x,y)
        }{
            \lambda_{\max}(T_\theta)
        }
    \right)
    \ ,
\end{equation}
where we used Eq.~\eqref{eq:spiked_replica_s_lamda} and denoted by
$\bigl(\lambda_{\max}(T_\theta),\Psi_{\max}\bigr)$ the principal eigenpair
of $T_\theta$, corresponding to the solution of the saddle-point equations
obtained in the previous section.  This reproduces
Eq.~\eqref{eq:spiked_top_eigenvector_distribution} of the main text.

\subsection{Average overlap between the top eigenvector and the spike}

In this section, we consider the overlap between the top eigenvector
$\vec v^{\, (1)}$ of $X^{(\theta)}$ and the spike vector
$\vec y=(y_1,\ldots,y_N)$. To this end, we first introduce the overlap density
\begin{equation}\label{eq:spiked_overlap_density_definition}
    T_{\mathrm{ov}}(u)
    =\lim_{N\to\infty}
    \left\langle
        \frac{1}{N}
        \sum_{i=1}^{N}
        \delta\left(
            u-y_i v_i^{\, (1)}
        \right)
    \right\rangle
    \ .
\end{equation}
The average overlap is then obtained from the first moment of
$T_{\mathrm{ov}}(u)$.

According to the general replica representation in
Eq.~\eqref{eq:general_eigenvector_replica}, the density $T_{\mathrm{ov}}(u)$
can be accessed in the large-$N$ limit by introducing the auxiliary partition function
\begin{equation}\label{eq:spiked_overlap_auxiliary_partition}
\begin{aligned}
    Z_{\epsilon,\beta}^{\mathrm{ov}}
    \left(
        t,X^{(\theta)};u
    \right)
    \coloneqq
    \int
    \dd\vec v\,
    \exp\left\{
        \frac{\beta}{2}
        \left(
            \vec v,
            X^{(\theta)}\vec v
        \right)
        +
        \beta t
        \sum_{i=1}^{N}
        \delta_\epsilon(u-y_i v_i)
    \right\}
    \delta\left(
        \lVert\vec v\rVert^2-N
    \right)
    \ .
\end{aligned}
\end{equation}
This auxiliary partition function differs from
Eq.~\eqref{eq:spiked_auxiliary_partition} only in the source term, namely
$\delta_\epsilon(u-v_i)$ is replaced by $\delta_\epsilon(u-y_i v_i)$. 
The replica calculation therefore proceeds exactly as in the previous
section, and the average replicated partition function can again be
written in a form amenable to a saddle-point evaluation 
\begin{equation}\label{eq:spiked_overlap_action}
    \left\langle
        \left[
            Z_{\epsilon,\beta}^{\mathrm{ov}}
            \left(
                t,X^{(\theta)};u
            \right)
        \right]^n
    \right\rangle
    \propto
    \int
    \dd\boldsymbol{\eta}\,
    \{\mathcal D\rho_c\}
    \{\mathcal D\hat\rho_c\}
    \exp\left\{
        N\mathcal S_{\theta,\mathrm{ov}}
        [
            \boldsymbol{\eta},
            \{\rho_c\},
            \{\hat\rho_c\};
            \epsilon,t,u
        ]
    \right\}
    \ .
\end{equation}
The action can again be decomposed into four contributions, $\mathcal S_{\theta,\mathrm{ov}}[\boldsymbol{\eta},\{\rho_c\},\{\hat\rho_c\};\epsilon,t,u]=\mathcal S_\theta^{(1)}[\boldsymbol{\eta}]+\mathcal S_\theta^{(2)}[\{\rho_c\}]+\mathcal S_\theta^{(3)}[\{\rho_c\},\{\hat\rho_c\}]+\mathcal S_{\theta,\mathrm{ov}}^{(4)}[\boldsymbol{\eta},\{\hat\rho_c\};\epsilon,t,u]$. The first three terms coincide with Eqs.~\eqref{eq:spiked_replica_action_1}, \eqref{eq:spiked_replica_action_2}, and \eqref{eq:spiked_replica_action_3}, while the fourth is given by (see Eq.~\eqref{eq:spiked_eigenvector_action_4})
\begin{equation}\label{eq:spiked_overlap_action_4}
\begin{aligned}
    \mathcal S_{\theta,\mathrm{ov}}^{(4)}
    [
        \boldsymbol{\eta},
        \{\hat\rho_c\};
        \epsilon,t,u
    ]
    &=
    \log\left[
        \int
        \dd\z\,
        \mathcal P(\z)
        \prod_{c=1}^{n}
        \int
        \dd v_c\,
        \exp\left(
            -\ii\frac{\beta}{2}\eta_c v_c^2
            -
            \ii\hat\rho_c(\z)v_c
            +
            \beta t
            \delta_\epsilon(u-z_{d+1}v_c)
        \right)
    \right]
    \ ,
\end{aligned}
\end{equation}
where $z_{d+1}=y$ denotes the last component of the augmented vector
$\z=(\x,y)$ introduced above. Since the derivative with respect to $t$ acts only on the explicit
$t$-dependence of the action, the saddle-point equations can again be
evaluated at $t=0$. They therefore coincide with those of the previous
sections and, assuming replica symmetry, the saddle-point variables
$\hat\rho^*$, $\rho^*$, and
$\eta^*$ are still given by
Eqs.~\eqref{eq:spiked_rhohat_saddle},
\eqref{eq:spiked_rho_saddle}, and
\eqref{eq:spiked_eta_saddle}. To first order in $n$, the only contribution that explicitly depends on
$t$ is therefore
\begin{equation}\label{eq:spiked_overlap_action_4_RS}
\begin{aligned}
    \mathcal S_{\theta,\mathrm{ov}}^{(4)}
    [
        \eta^*,
        \hat\rho^*;
        \epsilon,t,u
    ]
    &=
    n
    \int
    \dd\z\,
    \mathcal P(\z)
    \log\left[
        \int
        \dd v\,
        \exp\left(
            -\ii\frac{\beta}{2}\eta^*v^2
            -
            \ii\hat\rho^*(\z)v
            +
            \beta t\delta_\epsilon(u-z_{d+1}v)
        \right)
    \right]
    +
    o(n)
    \ .
\end{aligned}
\end{equation}
This is the analogue of Eq.~\eqref{eq:spiked_eigenvector_action_4_RS},
with the modified source term introduced above. Inserting Eq.~\eqref{eq:spiked_overlap_action_4_RS} into the general
replica representation in
Eq.~\eqref{eq:general_eigenvector_replica} and proceeding as in the previous section, we
obtain
\begin{equation}\label{eq:spiked_overlap_density_z}
\begin{aligned}
    T_{\mathrm{ov}}(u)
    &=\lim_{\beta \to \infty}\frac{1}{\beta}\lim_{n \to 0}\frac{\partial}{\partial t}\frac{1}{n}\mathcal  S_{\theta,\mathrm{ov}}^{(4)}
    [
        \eta^*,
        \hat\rho^*;
        \epsilon,t,u
    ]\Big|_{\epsilon=t=0}\\
    &=\int
    \dd\z\,
    \mathcal P(\z)
    \delta\left(
        u-
        z_{d+1}\,
        \frac{\Psi(\z)}{\sqrt{s_\theta}}
    \right)
    \ ,
\end{aligned}
\end{equation}
where we used
Eqs.~\eqref{eq:spiked_rhohat_saddle} and
\eqref{eq:spiked_eta_saddle}. Finally, using $\z=(\x,y)$ and
$\mathcal P(\z)\dd\z=P(\x)G(y)\dd\x\,\dd y$, this becomes
\begin{equation}\label{eq:spiked_overlap_density_xy}
    T_{\mathrm{ov}}(u)
    =
    \int
    \dd\x\,\dd y\,
    P(\x)G(y)
    \delta\left(
        u-
        y\,
        \frac{
            \Psi_{\max}(\x,y)
        }{
            \lambda_{\max}(T_\theta)
        }
    \right)
    \ ,
\end{equation}
where we used Eq.~\eqref{eq:spiked_replica_s_lamda} and denoted by
$\bigl(\lambda_{\max}(T_\theta),\Psi_{\max}\bigr)$ the principal eigenpair
of $T_\theta$, corresponding to the solution of the saddle-point equations
obtained above.
The average overlap between the top eigenvector and the spike is then
given by the first moment of $T_{\mathrm{ov}}(u)$
\begin{align}
    \lim_{N\to\infty}
    \left\langle
        \frac{
            (\vec y,\vec v^{\, (1)})
        }{N}
    \right\rangle
    &=
    \int
    \dd u\,
    u\,T_{\mathrm{ov}}(u)
    \nonumber\\
    &=
    \frac{1}{
        \lambda_{\max}(T_\theta)
    }
    \int
    \dd\x\,\dd y\,
    P(\x)G(y)\,
    y\,\Psi_{\max}(\x,y)
    \nonumber\\
    &=
    \frac{
        q[\Psi_{\max}]
    }{
        \lambda_{\max}(T_\theta)
    }
    \ ,
    \label{eq:spiked_average_overlap}
\end{align}

where $q[\Psi_{\max}]$ is the functional defined in Eq.~\eqref{eq:q_psi} of the
main text, evaluated on the principal eigenfunction $\Psi_{\max}$ of the operator
$T_\theta$. This therefore reproduces the expression for the average overlap
given in the main text.

\section{Derivation of the transition thresholds}\label{appendix:transition_thresholds}
In this appendix, the spectral problem is solved
to determine when the principal eigenpair of $T_\theta$ changes
from the branch $q[\Psi]=0$ to the branch $q[\Psi]\neq0$ and
to compute the corresponding transition threshold.
The analysis recovers the criteria and thresholds summarized
in Sec.~\ref{sec:recap_results} and
Tables~\ref{tab:overlap_transition_summary}
and~\ref{tab:transition_observables} of the main text,
treating centered and non-centered spikes separately. The starting point is the self-consistent system in
Eq.~\eqref{eq:final_sc_system} of the main text
\begin{equation}\label{eq:thresholds_sc_system}
    \begin{cases}
        \displaystyle
        \lambda\bar{\Psi}(\x)
        =
        (T_f\bar{\Psi})(\x)
        +\theta\langle y\rangle_G\,q[\Psi]
        \ ,
        \\[2mm]
        \displaystyle
        q[\Psi]\left(
            1-\frac{\theta\sigma_y^2}{\lambda}
        \right)
        =
        \langle y\rangle_G\langle\bar{\Psi}\rangle_P
        \ ,
    \end{cases}
\end{equation}
which determines every eigenpair of $T_\theta$ with a nonzero
eigenvalue $\lambda$. The corresponding eigenfunction has the general form given in
Eq.~\eqref{eq:psi_complete_parametrisation} of the main text:
\begin{equation}\label{eq:thresholds_psi_reconstruction}
    \Psi(\x,y)
    =
    \bar{\Psi}(\x)
    +
    \frac{\theta}{\lambda}
    q[\Psi]
    \left(
        y-\langle y\rangle_G
    \right)
    \ ,
\end{equation}
where $\bar{\Psi}$ and $q[\Psi]$ are defined in
Eqs.~\eqref{eq:barpsi} and~\eqref{eq:q_psi} of the main text.
We are interested in the principal eigenpair
$\bigl(\lambda_{\max}(T_\theta),\Psi_{\max}\bigr)$, 
with the principal eigenfunction normalized according to
Eq.~\eqref{eq:spiked_psi_normalisation} of the main text 
\begin{equation}\label{eq:thresholds_normalisation}
    1
    =
    \int \dd\x\,\dd y\,P(\x)G(y)
    \left(
        \frac{\Psi_{\max}(\x,y)}
             {\lambda_{\max}(T_\theta)}
    \right)^2
    \ .
\end{equation}
The limiting average overlap between the top eigenvector and the spike
is then given by Eq.~\eqref{eq:spiked_overlap_main}
of the main text:
\begin{equation}\label{eq:thresholds_overlap}
    \lim_{N\to\infty}
    \left\langle
        \frac{(\vec y,\vec v^{\, (1)})}{N}
    \right\rangle
    =
    \frac{q[\Psi_{\max}]}{\lambda_{\max}(T_\theta)}
    \ .
\end{equation}
Since $\lambda_{\max}(T_\theta)>0$ for every $\theta>0$,
as established in Eq.~\eqref{eq:lambda_max_positive} of the
main text, this overlap vanishes precisely when
$q[\Psi_{\max}]=0$. The analysis below examines the solutions of
Eq.~\eqref{eq:thresholds_sc_system}, focusing on the principal
eigenpair and the corresponding overlap.

\subsection{Centered spike}\label{sec:spiked model_centered}
For $\langle y\rangle_G=0$, the system in
Eq.~\eqref{eq:thresholds_sc_system} reduces to
\begin{equation}\label{eq:centered_sc_system}
    \begin{cases}
        \displaystyle
        \lambda\bar{\Psi}(\x)
        =
        (T_f\bar{\Psi})(\x)
        \ ,
        \\[2mm]
        \displaystyle
        q[\Psi]\left(
            1-\frac{\theta\sigma_y^2}{\lambda}
        \right)
        =
        0
        \ .
    \end{cases}
\end{equation}
The solutions separate into two branches. We determine the possible eigenpairs associated with each branch, compare their eigenvalues, and select the largest one. 

We first consider the case in which the upper spectral edge
of the unperturbed operator is strictly positive,
$\Lambda_0>0$. As discussed in
Sec.~\ref{sec:recap_results} of the main text, this guarantees
that $T_f$ admits a principal eigenvalue
$\lambda_{\max}(T_f)=\Lambda_0>0$ and a corresponding
non-trivial eigenfunction $\psi_{\max}(\x)$.

\begin{description}
    \item[1. \boldmath{$q[\Psi]=0$}]
    Equation~\eqref{eq:thresholds_psi_reconstruction} shows that $\Psi$ is independent of $y$ and coincides with $\bar{\Psi}$. The first equation in Eq.~\eqref{eq:centered_sc_system} then reduces to the unperturbed spectral problem for $T_f$. Consequently, the principal eigenpair in this branch is the principal eigenpair of the unperturbed problem, denoted by
    \begin{equation}
        \bigl(\lambda_{\max}(T_f),\psi_{\max}\bigr)
        \ .
    \end{equation}

    \item[2. \boldmath{$q[\Psi]\neq0$}]
    The second equation in
    Eq.~\eqref{eq:centered_sc_system} fixes the eigenvalue to
    \begin{equation}
        \lambda
        =
        \theta\sigma_y^2
        =
        \theta\langle y^2\rangle_G
        \ .
    \end{equation}
    A self-consistent solution of the first equation is
    obtained by choosing $\bar{\Psi}=0$\footnote{
    If $\theta\sigma_y^2$ is also an eigenvalue of $T_f$,
    the first equation in Eq.~\eqref{eq:centered_sc_system}
    also admits solutions with $\bar{\Psi}\neq0$.
    This possible degeneracy does not affect the eigenvalue
    associated with this branch.}.
    Equation~\eqref{eq:thresholds_psi_reconstruction} then gives
    \begin{equation}
        \Psi(\x,y)
        =
        \frac{\theta}{\lambda}q[\Psi]\,y
        =
        \frac{q[\Psi]}{\sigma_y^2}\,y
        \ .
    \end{equation}
    For this solution, imposing the normalization in
    Eq.~\eqref{eq:thresholds_normalisation} gives
    \begin{equation}
        1
        =
        \frac{q[\Psi]^2}{\lambda^2\sigma_y^4}
        \langle y^2\rangle_G
        =
        \frac{q[\Psi]^2}{\lambda^2\sigma_y^2}
        \quad\Longleftrightarrow\quad
        q[\Psi]=\pm\theta\sigma_y^3
        \ .
    \end{equation}
    Choosing the overall sign of the eigenfunction such that $q[\Psi]>0$, we obtain that the normalized eigenfunction associated with this branch is
    \begin{equation}
        \Psi(\x,y)
        =
        \theta\sigma_y\,y
        \ .
    \end{equation}
\end{description}
Comparing the eigenvalues obtained in the two branches, we conclude that when $\Lambda_0=\lambda_{\max}(T_f)>0$
\begin{equation}\label{eq:centered_largest_eigenvalue}
    \lambda_{\max}(T_\theta)
    =
    \max\left\{
        \lambda_{\max}(T_f),
        \theta\sigma_y^2
    \right\}
    \ .
\end{equation}
Therefore, the transition between the two branches occurs at the strictly positive threshold
\begin{equation}\label{eq:centered_theta_threshold}
    \theta_c
    =
    \frac{\lambda_{\max}(T_f)}{\sigma_y^2}>0
    \ .
\end{equation}
For $0<\theta<\theta_c$, the largest eigenvalue and the corresponding principal eigenfunction coincide with those of the unperturbed operator. In this regime
\begin{equation}
    q[\Psi_{\max}]
    =
    0
    \ ,
\end{equation}
meaning that the principal eigenfunction has zero overlap with the perturbation direction. For $\theta>\theta_c$, the perturbation branch dominates, and
\begin{equation}\label{eq:centered_recap_branch_qnonzero}
    \lambda_{\max}(T_\theta)
    =
    \theta\sigma_y^2
    \ ,
    \qquad
    \Psi_{\max}(\x,y)
    =
    \theta\sigma_y\,y
    \ ,
    \qquad
    q[\Psi_{\max}]
    =
    \theta\sigma_y^3
    >
    0
    \ .
\end{equation}
By Eq.~\eqref{eq:thresholds_overlap}, the limiting average
overlap is therefore zero for $0<\theta<\theta_c$
and equals $\sigma_y>0$ for $\theta>\theta_c$.
At $\theta=\theta_c$, the two branches have the
same eigenvalue, so the principal eigenvalue of $T_\theta$
is degenerate and the overlap is not uniquely defined.

When $\Lambda_0\leq0$, the replica results are not applicable
to the unperturbed problem, since they require a strictly
positive principal eigenvalue. Nevertheless, the
analysis above still shows that any eigenvalue in the branch
$q[\Psi]=0$ is bounded above by $\Lambda_0$, whereas the
branch $q[\Psi]\neq0$ has eigenvalue $\theta\sigma_y^2>0$.
The latter therefore dominates for every $\theta>0$, giving
the same results of Eq.~\eqref{eq:centered_recap_branch_qnonzero}.
Thus, the perturbed problem always admits a strictly positive
principal eigenvalue and a corresponding normalized
principal eigenfunction, even when $\Lambda_0=0$ is not
an eigenvalue of $T_f$. The replica results consequently
apply to the perturbed problem, and
Eq.~\eqref{eq:thresholds_overlap} gives a limiting average
overlap equal to $\sigma_y>0$ for every $\theta>0$,
with no transition at a strictly positive threshold
when $\Lambda_0\leq0$. These results recover the centered-spike transition criterion
and threshold summarized in Sec.~\ref{sec:recap_results}
and Tables~\ref{tab:overlap_transition_summary}
and~\ref{tab:transition_observables} of the main text.

\subsection{Non-centered spike}\label{sec:spiked model_non_centered}
We now solve the self-consistent system in Eq.~\eqref{eq:thresholds_sc_system} when the perturbation has non-zero mean, $\langle y\rangle_G\neq0$
\begin{equation}\label{eq:non_centered_sc_system}
    \begin{cases}
        \displaystyle
        \lambda\bar{\Psi}(\x)
        =
        (T_f\bar{\Psi})(\x)
        +
        \theta\langle y\rangle_G\,q[\Psi]
        \ ,
        \\[2mm]
        \displaystyle
        q[\Psi]
        \left(
            1-\frac{\theta\sigma_y^2}{\lambda}
        \right)
        =
        \langle y\rangle_G
        \left\langle\bar{\Psi}(\x)\right\rangle_P
        \ .
    \end{cases}
\end{equation}
Unlike in the centered case, the two equations are coupled through the non-zero mean $\langle y\rangle_G$. Suppose first that $\lambda\notin\operatorname{Spec}(T_f)$, so that the resolvent
\begin{equation}\label{eq:Tf_resolvent}
    R_\lambda
    \coloneqq
    \left(\lambda I-T_f\right)^{-1}
\end{equation}
is well defined. Applying $R_\lambda$ to the first equation in \eqref{eq:non_centered_sc_system} gives
\begin{equation}\label{eq:non_centered_barPsi_resolvent}
    \bar{\Psi}
    =
    \theta\langle y\rangle_G\,q[\Psi]\,
    R_\lambda\mathbf{1}
    \ ,
\end{equation}
where $\mathbf{1}(\x)=1$ denotes the constant function. Taking the average with respect to $P$, we obtain
\begin{equation}\label{eq:non_centered_barPsi_average}
    \left\langle\bar{\Psi}(\x)\right\rangle_P
    =
    (\mathbf{1},\bar{\Psi})_P
    =
    \theta\langle y\rangle_G\,q[\Psi]\,
    (\mathbf{1},R_\lambda\mathbf{1})_P
    \ .
\end{equation}
Inserting Eq.~\eqref{eq:non_centered_barPsi_average} into the second equation of \eqref{eq:non_centered_sc_system} yields
\begin{equation}\label{eq:non_centered_q_condition}
    q[\Psi]
    \left[
        1
        -
        \frac{\theta\sigma_y^2}{\lambda}
        -
        \theta\langle y\rangle_G^2
        (\mathbf{1},R_\lambda\mathbf{1})_P
    \right]
    =
    0
    \ .
\end{equation}
Equation~\eqref{eq:non_centered_q_condition} identifies two distinct branches, corresponding to $q[\Psi]=0$ and $q[\Psi]\neq0$. We analyze them separately, determine the possible eigenpairs associated with each branch, and finally compare the resulting eigenvalues to identify the largest one.

\begin{description}
    \item[1. \boldmath{$q[\Psi]=0$}]
    In this case, the original system~\eqref{eq:non_centered_sc_system} reduces to
    \begin{equation}\label{eq:non_centered_q_zero_system}
        \begin{cases}
            \displaystyle
            (T_f\bar{\Psi})(\x)
            =
            \lambda\bar{\Psi}(\x)
            \ ,
            \\[2mm]
            \displaystyle
            \left\langle\bar{\Psi}\right\rangle_P
            =
            0
            \ ,
        \end{cases}
    \end{equation}
    where the second equation follows from $\langle y\rangle_G\neq0$. Therefore, any eigenpair arising from this branch must also be an eigenpair of $T_f$, with an eigenfunction that is independent of $y$ and has zero $P$-average. We now determine whether the branch $q[\Psi]=0$ can contain the principal eigenpair of $T_\theta$. Recall from the discussion in Sec.~\ref{sec:spiked model} that the upper spectral edges are defined by
    \begin{equation}
        \Lambda_\theta
        \coloneqq
        \sup\operatorname{Spec}(T_\theta)
        \qquad\text{and}\qquad
        \Lambda_0
        \coloneqq
        \sup\operatorname{Spec}(T_f)
        \ ,
    \end{equation}
    and, as proved in Appendix~\ref{appendix:spectral_edges}, satisfy
    \begin{equation}\label{eq:recap_appendix_A}
        \Lambda_\theta
        \geq
        \Lambda_0
        \qquad\text{and}\qquad
        \Lambda_\theta>0
        \quad
        \forall \theta>0
        \ .
    \end{equation}
    We therefore distinguish the following cases.
    
    \begin{description}
        \item[\boldmath{(a) $\Lambda_0\leq0$}]
        Any eigenvalue arising from the branch $q[\Psi]=0$ belongs to $\operatorname{Spec}(T_f)$ and therefore satisfies
        \begin{equation}
            \lambda
            \leq
            \Lambda_0
            \leq
            0
            \ .
        \end{equation}
        On the other hand, Eq.~\eqref{eq:recap_appendix_A} guarantees that
        \begin{equation}
            \Lambda_\theta>0
            \qquad
            \forall \theta>0
            \ .
        \end{equation}
        Therefore, the principal eigenpair of $T_\theta$ cannot belong to the branch $q[\Psi]=0$ in this case.

        \item[\boldmath{(b) $\Lambda_0>0$}] In this case, compactness of $T_f$ guarantees that $\Lambda_0$ is an eigenvalue of $T_f$ with finite multiplicity. We denote its eigenspace by
        \begin{equation}
            E_1
            \coloneqq
            \ker\!\left(\Lambda_0 I-T_f\right)
        \end{equation}
        and by $\Pi_{E_1}$ the orthogonal projector from $L^2(P)$ onto $E_1$. The relevant distinction for the existence of a transition is whether the constant function has a non-zero component along the top eigenspace
        \begin{equation}
            \Pi_{E_1}\mathbf{1}=0
            \quad\Longleftrightarrow\quad
            (\mathbf{1},\psi)_P=\langle\psi\rangle_P=0
            \quad
            \forall\psi\in E_1\ ,
        \end{equation}
        whereas $\Pi_{E_1}\mathbf{1}\neq0$ means that $E_1$ contains at least one eigenfunction with non-zero $P$-average.
        \begin{description}
            \item[\boldmath{(b1) $\Lambda_0>0$} and \boldmath{$\Pi_{E_1}\mathbf{1}\neq0$}]
            Since $\Pi_{E_1}\mathbf{1}\neq0$, there exists a normalized eigenfunction $\psi\in E_1$ such that
            \begin{equation}
                \left\langle\psi\right\rangle_P
                =
                (\mathbf{1},\psi)_P
                \neq
                0
                \ .
            \end{equation}
            Regarding $\psi$ as a $y$-independent function in $L^2(P\otimes G)$, the variational principle gives
            \begin{align}
                \Lambda_\theta
                &\geq
                (\psi,T_\theta\psi)_{P\otimes G}
                \nonumber\\
                &=
                \Lambda_0
                +
                \theta
                \langle y\rangle_G^2
                \left\langle\psi\right\rangle_P^2
                \nonumber\\
                &>
                \Lambda_0
                \qquad
                \forall\theta>0
                \ ,
            \end{align}
            where the strict inequality follows from
            $\langle y\rangle_G\neq0$ and
            $\langle\psi\rangle_P\neq0$. Since every eigenvalue arising from the branch $q[\Psi]=0$ is at most $\Lambda_0$, the principal eigenpair of $T_\theta$ cannot belong to this branch.
    
            Thus, the existence of a single eigenfunction in $E_1$
            with non-zero $P$-average rules out a sharp overlap
            transition, since the principal eigenpair of $T_\theta$
            then belongs to the branch $q[\Psi]\neq0$ for every
            $\theta>0$. When $\Lambda_0$ is degenerate, it is enough
            that one linear combination of its eigenfunctions
            have non-zero $P$-average.     
            
            \item[\boldmath{(b2) $\Lambda_0>0$} and \boldmath{$\Pi_{E_1}\mathbf{1}=0$}]
            In this case
            \begin{equation}
                \left\langle\psi\right\rangle_P
                =
                (\mathbf{1},\psi)_P
                =
                0
                \qquad
                \forall\psi\in E_1
                \ .
            \end{equation}
            Therefore, every normalized $\psi\in E_1$, regarded as a $y$-independent function in $L^2(P\otimes G)$, satisfies the system~\eqref{eq:non_centered_q_zero_system}. Hence, the principal eigenpair $(\Lambda_0,\psi)$ of $T_f$ is also an eigenpair of $T_\theta$ belonging to the branch $q[\Psi]=0$ and is a candidate for the principal eigenpair of $T_\theta$.
    
            However, this candidate must still be compared with the eigenvalues arising from the branch $q[\Psi]\neq0$, which we analyze below. Denoting the largest eigenvalue arising from that branch by $\lambda_c(\theta)$, the principal eigenvalue of $T_\theta$ in this case is given by
            \begin{equation}
                \Lambda_\theta
                =
                \max\left\{
                    \Lambda_0,
                    \lambda_c(\theta)
                \right\}
                \ .
            \end{equation}
            Consequently, there may exist a critical value $\theta_c$ at which the principal eigenpair transitions from the branch $q[\Psi]=0$ to the branch $q[\Psi]\neq0$. Determining whether such a transition occurs requires the analysis of the latter branch, which we carry out below.
        \end{description}
    \end{description}
    \item[2. \boldmath{$q[\Psi]\neq0$}] 
    In this case, Eq.~\eqref{eq:non_centered_q_condition} directly yields the equation
    \begin{equation}\label{eq:non_centered_secular_equation}
        1
        =
        \theta
        \left[
            \langle y\rangle_G^2\,\mathcal G(\lambda)
            +
            \frac{\sigma_y^2}{\lambda}
        \right],
    \end{equation}
    where we have defined
    \begin{equation}\label{eq:G_resolvent}
        \mathcal G(\lambda)
        \coloneqq
        \left(
            \mathbf 1,
            (\lambda I-T_f)^{-1}\mathbf 1
        \right)_P=\left(
            \mathbf 1,
            R_{\lambda}\mathbf 1
        \right)_P\ .
    \end{equation}
    
    Since $T_f$ is compact and self-adjoint, the spectral theorem guarantees the existence of a complete orthonormal basis $\{e_k\}_k$ of $L^2(P)$ consisting of eigenfunctions of $T_f$. We write
    \begin{equation}\label{eq:appendix_spectral_decomposition}
        T_fe_k=\mu_k e_k\ ,
        \qquad
        (e_k,e_\ell)_P=\delta_{k\ell}\ ,
    \end{equation}
    where the basis includes an orthonormal basis of $\ker T_f$, whenever the kernel is non-trivial. Defining
    \begin{equation}\label{eq:coeffs_ck}
        c_k
        \coloneqq
        (\mathbf 1,e_k)_P
        =
        \langle e_k\rangle_P\ ,
    \end{equation}
    the constant function admits the expansion
    \begin{equation}\label{eq:constant_spectral_decomposition}
        \mathbf 1
        =
        \sum_k
        (\mathbf 1,e_k)_P e_k
        =
        \sum_k c_k e_k\ .
    \end{equation}
    Moreover the coefficients $c_k$ satisfy
    \begin{equation}\label{eq:ck_normalization}
        \sum_k c_k^2
        =
        \|\mathbf 1\|_P^2
        =
        (\mathbf 1,\mathbf 1)_P
        =
        \int \mathrm d \x~P(\x) 
        =
        1
        \ ,
    \end{equation}
    where we used the fact that $P$ is a probability measure. Applying the resolvent yields
    \begin{equation}\label{eq:appendix_R_inbasis}
        R_{\lambda}\mathbf 1=(\lambda I-T_f)^{-1}\mathbf 1
        =
        \sum_k
        \frac{c_k}{\lambda-\mu_k}\,e_k\ .
    \end{equation}
    Taking the inner product with $\mathbf 1$, we therefore obtain the spectral representation
    \begin{equation}\label{eq:G_spectral_decomposition}
        \mathcal G(\lambda)
        =
        \sum_k
        \frac{c_k^2}{\lambda-\mu_k}\ .
    \end{equation}
    Defining
    \begin{equation}\label{eq:F_definition}
        F(\lambda)
        \coloneqq
        \langle y\rangle_G^2
        \sum_k
        \frac{c_k^2}{\lambda-\mu_k}
        +
        \frac{\sigma_y^2}{\lambda}
        \ ,
    \end{equation}
    Eq.~\eqref{eq:non_centered_secular_equation} can be equivalently rewritten as
    \begin{equation}\label{eq:F_secular_equation}
        F(\lambda)
        =
        \frac{1}{\theta}
        \ .
    \end{equation}
    We now analyze separately the cases $\Lambda_0\leq0$ and $\Lambda_0>0$ to determine whether Eq.~\eqref{eq:F_secular_equation} admits a solution corresponding to the principal eigenpair of $T_\theta$.
    \begin{description}
        \item[\boldmath{(a) $\Lambda_0\leq0$}]
        From the results of Appendix~\ref{appendix:spectral_edges}, we already know that 
        \begin{equation}
            \Lambda_\theta>0
            \qquad
            \forall\theta>0
            \ .
        \end{equation}
        We therefore seek a solution of Eq.~\eqref{eq:F_secular_equation} in the region
        \begin{equation}
            \lambda>0\geq\Lambda_0\geq\mu_k
            \qquad
            \forall k
            \ .
        \end{equation}
        It follows directly from Eq.~\eqref{eq:F_definition} that
        \begin{equation}
            F(\lambda)>0
            \qquad
            \forall\lambda>0
            \ ,
        \end{equation}
        while its derivative is
        \begin{equation}
            F'(\lambda)
            =
            -
            \langle y\rangle_G^2
            \sum_k
            \frac{c_k^2}{(\lambda-\mu_k)^2}
            -
            \frac{\sigma_y^2}{\lambda^2}
            <
            0
            \qquad
            \forall\lambda>0
            \ .
        \end{equation}
        Moreover,
        \begin{equation}
            \lim_{\lambda\to0^+}F(\lambda)
            =
            +\infty,
            \qquad
            \lim_{\lambda\to+\infty}F(\lambda)
            =
            0
            \ .
        \end{equation}
        Hence, $F$ is continuous and strictly decreasing on $(0,+\infty)$. Therefore, for every $\theta>0$, there exists a unique solution $\lambda_c(\theta)>0$ such that
        \begin{equation}
            F\!\left(\lambda_c(\theta)\right)
            =
            \frac{1}{\theta}
            \ .
        \end{equation}
        Since every eigenvalue arising from the branch $q[\Psi]=0$
        is non-positive, as established above, the unique positive
        solution $\lambda_c(\theta)$ is the principal eigenvalue
        of $T_\theta$, consistently with the strict positivity
        of $\Lambda_\theta$. Thus
        \begin{equation}
            \Lambda_\theta
            =
            \lambda_c(\theta)
            >
            0
            \qquad
            \forall\theta>0
            \ .
        \end{equation}
    
        \item[\boldmath{(b) $\Lambda_0>0$}] If we now consider the case $\Lambda_0>0$, since $T_f$ is a compact and self-adjoint operator, $\Lambda_0$ is its principal eigenvalue and has finite multiplicity. Therefore, for $\lambda>\Lambda_0$, Eq.~\eqref{eq:F_definition} can  be written as
            \begin{align}\label{eq:F_decomposition_top_eigenspace}
                F(\lambda)
                &=
                \langle y\rangle_G^2
                \left[
                    \frac{\displaystyle\sum_{\mu_k= \Lambda_0}c_k^2}
                         {\lambda-\Lambda_0}
                    +
                    \sum_{\mu_k < \Lambda_0}
                    \frac{c_k^2}{\lambda-\mu_k}
                \right]
                +
                \frac{\sigma_y^2}{\lambda}
                \nonumber\\
                &=
                \langle y\rangle_G^2
                \left[
                    \frac{\lVert\Pi_{E_1}\mathbf 1\rVert_P^2}
                         {\lambda-\Lambda_0}
                    +
                    \sum_{\mu_k < \Lambda_0}
                    \frac{c_k^2}{\lambda-\mu_k}
                \right]
                +
                \frac{\sigma_y^2}{\lambda},
            \end{align}
            where $\Pi_{E_1}\mathbf 1$ is the orthogonal projection of the constant function onto the top eigenspace $E_1$. In fact, since $\{e_k\}_{\mu_k = \Lambda_0}$ is an orthonormal basis of $E_1$
            \begin{equation}
                \Pi_{E_1}\mathbf 1
                =
                \sum_{\mu_k = \Lambda_0}
                (\mathbf 1,e_k)_P e_k
                =
                \sum_{\mu_k = \Lambda_0}c_k e_k\ ,
            \end{equation}
            and hence
            \begin{equation}
                \lVert\Pi_{E_1}\mathbf 1\rVert_P^2
                =
                \sum_{\mu_k = \Lambda_0}c_k^2
                \ .
            \end{equation}
            From Eq.~\eqref{eq:F_decomposition_top_eigenspace}, it still follows that
        \begin{equation}
            F(\lambda)>0
            \qquad
            \forall\lambda>\Lambda_0
            \ ,
        \end{equation}
        and
        \begin{equation}\label{eq:appendix_F_prime}
            F'(\lambda)
            =
            -
            \langle y\rangle_G^2
            \sum_k
            \frac{c_k^2}{(\lambda-\mu_k)^2}
            -
            \frac{\sigma_y^2}{\lambda^2}
            <
            0
            \qquad
            \forall\lambda>\Lambda_0
            \ .
        \end{equation}
        Moreover,
        \begin{equation}\label{eq:appendix_F_to_zero}
            \lim_{\lambda\to+\infty}F(\lambda)
            =
            0
            \ .
        \end{equation}
        To determine whether the principal eigenvalue of $T_\theta$ belongs to the branch $q[\Psi]\neq0$, we must check whether Eq.~\eqref{eq:F_secular_equation} admits a solution with $\lambda>\Lambda_0$. Since $F$ is strictly decreasing on $(\Lambda_0,+\infty)$, this requires examining the limit
        \begin{equation}
            \lim_{\lambda\to\Lambda_0^+}F(\lambda)
            \ .
        \end{equation}
        \begin{description}    
            \item[\boldmath{(b1) $\Lambda_0>0$ and $\Pi_{E_1}\mathbf 1\neq0$}]
            In this case $\lVert\Pi_{E_1}\mathbf 1\rVert_P^2>0$ and the first term in
            Eq.~\eqref{eq:F_decomposition_top_eigenspace} has a pole at
            $\lambda=\Lambda_0$, therefore
            \begin{equation}
                \lim_{\lambda\to\Lambda_0^+}F(\lambda)
                =
                +\infty
                \ .
            \end{equation}
            Since $F$ is continuous and strictly decreasing on $(\Lambda_0,+\infty)$, the equation
            \begin{equation}
                F(\lambda)
                =
                \frac{1}{\theta}
            \end{equation}
            admits a unique solution 
            \begin{equation}
                \lambda_c(\theta)>\Lambda_0
                \qquad
                \forall\theta>0
                \ .
            \end{equation}
            Hence, the branch $q[\Psi]\neq0$ always yields the principal
            eigenpair of $T_\theta$, and no transition between the two branches
            occurs.
        
            \item[\boldmath{(b2) $\Lambda_0>0$ and $\Pi_{E_1}\mathbf 1 = 0$}] 
            In this case $\lVert\Pi_{E_1}\mathbf 1\rVert_P^2=0$, there is no pole at $\Lambda_0$ and
            \begin{equation}\label{eq:F_edge_finite}
                F(\Lambda_0^+) \coloneqq \lim_{\lambda\to\Lambda_0^+}F(\lambda)
                =
                \langle y\rangle_G^2
                \sum_{\mu_k<\Lambda_0}
                \frac{c_k^2}{\Lambda_0-\mu_k}
                +
                \frac{\sigma_y^2}{\Lambda_0}        
                \ .
            \end{equation}
            This quantity is finite and strictly positive. We therefore define
            the critical value
            \begin{equation}\label{eq:theta_critical}
                \theta_c
                \coloneqq
                \frac{1}{F(\Lambda_0^+)}
                =
                \left[
                \langle y\rangle_G^2
                \sum_{\mu_k<\Lambda_0}
                \frac{c_k^2}{\Lambda_0-\mu_k}
                +
                \frac{\sigma_y^2}{\Lambda_0}
                \right]^{-1}
                \ .
            \end{equation}
            Since $F$ is strictly decreasing on $(\Lambda_0,+\infty)$, Eq. \eqref{eq:F_secular_equation} admits a solution with $\lambda>\Lambda_0$ if and
            only if
            \begin{equation}
                \frac{1}{\theta}
                <
                F(\Lambda_0^+)\ ,
            \end{equation}
            or equivalently if $\theta>\theta_c$. Thus
            \begin{equation}\label{eq:appendix_recap_noncentered_transition}
                \begin{cases}
                    0<\theta<\theta_c:
                    &
                    \Lambda_\theta=\Lambda_0,
                    \quad
                    \text{and the principal eigenpair belongs to the branch }
                    q[\Psi]=0,
                    \\[1mm]
                    \theta>\theta_c:
                    &
                    \Lambda_\theta=\lambda_c(\theta)>\Lambda_0,
                    \quad
                    \text{and the principal eigenpair belongs to the branch }
                    q[\Psi]\neq0.
                \end{cases}
            \end{equation}
            At $\theta=\theta_c$, the $q[\Psi]\neq0$ branch reaches
            $\lambda=\Lambda_0$ and meets the $q[\Psi]=0$ branch, so that the
            principal eigenspace is degenerate.
        \end{description}
    \end{description}
\end{description}
This analysis recovers the transition criteria
summarized in Sec.~\ref{sec:recap_results} and
Tables~\ref{tab:overlap_transition_summary}
and~\ref{tab:transition_observables} of the main text.

We now prove that every transition identified above is
discontinuous by showing that the limiting average overlap
in Eq.~\eqref{eq:thresholds_overlap} approaches a strictly
positive value as $\theta\to\theta_c^+$.
In the present non-centered setting, we therefore focus on
the transition regime $\Lambda_0>0$ and
$\Pi_{E_1}\mathbf1=0$.\footnote{Notice that in the centered
case, Eq.~\eqref{eq:centered_recap_branch_qnonzero} already
gives a constant overlap $\sigma_y>0$ above the threshold, while
the overlap vanishes below it, giving a discontinuous transition.}
Above the threshold, Eq.~\eqref{eq:appendix_recap_noncentered_transition}
gives the principal eigenvalue
$\lambda_{\max}(T_\theta)=\lambda_c(\theta)>\Lambda_0>0$.
Combining Eqs.~\eqref{eq:non_centered_barPsi_resolvent}
and~\eqref{eq:thresholds_psi_reconstruction} gives for the principal eigenfunction
\begin{equation}
    \Psi_{\max}(\x,y)
    =
    \theta q[\Psi_{\max}]
    \left[
        \langle y\rangle_G
        (R_{\lambda_c(\theta)}\mathbf1)(\x)
        +\frac{y-\langle y\rangle_G}{\lambda_c(\theta)}
    \right]
    \ .
\end{equation}
Inserting this expression into the normalization condition
in Eq.~\eqref{eq:thresholds_normalisation}, and using the
resolvent expansion in Eq.~\eqref{eq:appendix_R_inbasis}
together with the orthonormality of the eigenfunctions
$e_k$ in Eq.~\eqref{eq:appendix_spectral_decomposition}, yields
\begin{align}
    1
    &=
    \int \dd\x\,\dd y\,P(\x)G(y)
    \left(
        \frac{\Psi_{\max}(\x,y)}{\lambda_c(\theta)}
    \right)^2
    \nonumber\\
    &=
    \theta^2
    \left(
        \frac{q[\Psi_{\max}]}{\lambda_c(\theta)}
    \right)^2
    \left[
        \langle y\rangle_G^2
        \sum_k
        \frac{c_k^2}{(\lambda_c(\theta)-\mu_k)^2}
        +\frac{\sigma_y^2}{\lambda_c(\theta)^2}
    \right]
    \nonumber\\
    &=
    -\theta^2
    \left(
        \frac{q[\Psi_{\max}]}{\lambda_c(\theta)}
    \right)^2
    F'(\lambda_c(\theta))
    \ ,
    \label{eq:thresholds_overlap_normalization}
\end{align}
where the last equality follows from
Eq.~\eqref{eq:appendix_F_prime}.
Choosing the positive solution therefore gives the overlap
\begin{equation}\label{eq:appendix_general_overlap_noncentered}
     \lim_{N\to\infty}
    \left\langle
        \frac{(\vec y,\vec v^{\, (1)})}{N}
    \right\rangle
    =\frac{q[\Psi_{\max}]}{\lambda_c(\theta)}
    =
    \frac{1}{\theta\sqrt{-F'(\lambda_c(\theta))}}
    \ ,
\end{equation}
where we used Eq.~\eqref{eq:thresholds_overlap}. Eqs.~\eqref{eq:F_secular_equation}
and~\eqref{eq:theta_critical}, together with the strict
monotonicity of $F$, imply that
$\lambda_c(\theta)\to\Lambda_0^+$ as
$\theta\to\theta_c^+$.
By the overlap formula above, it therefore suffices to show
that $0<-F'(\Lambda_0^+)<\infty$ to prove that the transition
is discontinuous.

Since $\Pi_{E_1}\mathbf1=0$,
Eq.~\eqref{eq:coeffs_ck} gives $c_k=0$ whenever
$\mu_k=\Lambda_0$.
Moreover, compactness and self-adjointness of $T_f$ ensure
that its positive principal eigenvalue $\Lambda_0$ is
isolated. The remaining denominators are therefore bounded
away from zero, and Eq.~\eqref{eq:ck_normalization}
ensures convergence of the sum.
Taking the right limit in Eq.~\eqref{eq:appendix_F_prime}
thus gives
\begin{equation}
    0<
    -F'(\Lambda_0^+)
    =
    \langle y\rangle_G^2
    \sum_{\,\mu_k<\Lambda_0}
    \frac{c_k^2}{(\Lambda_0-\mu_k)^2}
    +
    \frac{\sigma_y^2}{\Lambda_0^2}
    <\infty
    \ ,
\end{equation}
where strict positivity follows from $\sigma_y^2>0$.
Hence
\begin{equation}\label{eq:thresholds_overlap_jump}
    \lim_{\theta\to\theta_c^+}
    \frac{q[\Psi_{\max}]}{\lambda_c(\theta)}
    =
    \frac{1}{\theta_c\sqrt{-F'(\Lambda_0^+)}}
    >0
    \ .
\end{equation}
Since Eq.~\eqref{eq:appendix_recap_noncentered_transition}
gives $q[\Psi_{\max}]=0$ for $0<\theta<\theta_c$,
the limiting average overlap vanishes below the threshold.
This strictly positive right limit therefore proves
that the transition is discontinuous.

We finally consider the limit $\theta\to\infty$. As $\theta$ increases, the right-hand side of
Eq.~\eqref{eq:F_secular_equation} decreases to zero.
Since $F$ is strictly decreasing and tends to zero at
infinity by Eqs.~\eqref{eq:appendix_F_prime}
and~\eqref{eq:appendix_F_to_zero}, the principal eigenvalue
$\lambda_c(\theta)$ increases with $\theta$ and satisfies
\begin{equation}
    \lim_{\theta\to\infty}\lambda_c(\theta)=+\infty
\end{equation}
Using Eq.~\eqref{eq:ck_normalization}
in Eqs.~\eqref{eq:F_definition} and~\eqref{eq:appendix_F_prime}
therefore gives
\begin{equation}
    \lim_{\lambda\to+\infty}\lambda F(\lambda)
    =
    \lim_{\lambda\to+\infty}\bigl[-\lambda^2F'(\lambda)\bigr]
    =
    \langle y\rangle_G^2+\sigma_y^2
    =
    \langle y^2\rangle_G
    \ .
\end{equation}
The secular equation~\eqref{eq:F_secular_equation} then yields
\begin{equation}
    \lim_{\theta\to\infty}\frac{\lambda_c(\theta)}{\theta}
    =
    \lim_{\theta\to\infty}
    \lambda_c(\theta)F(\lambda_c(\theta))
    =
    \langle y^2\rangle_G
    \ .
\end{equation}
Combining the limits gives
\begin{equation}
    \lim_{\theta\to\infty}
    \bigl[-\theta^2F'(\lambda_c(\theta))\bigr]
    =
    \lim_{\theta\to\infty}
    \frac{-\lambda_c(\theta)^2F'(\lambda_c(\theta))}
         {(\lambda_c(\theta)/\theta)^2}
    =
    \frac{1}{\langle y^2\rangle_G}
    \ .
\end{equation}
Thus, Eq.~\eqref{eq:appendix_general_overlap_noncentered}
gives
\begin{equation}
    \lim_{\theta\to\infty}
    \frac{q[\Psi_{\max}]}{\lambda_c(\theta)}
    =
    \sqrt{\langle y^2\rangle_G}
    \ .
\end{equation}
For a centered spike, the eigenvector aligns fully with the
spike as soon as $\theta$ exceeds $\theta_c$, with overlap
$\sigma_y=\sqrt{\langle y^2\rangle_G}$ by Eq.~\eqref{eq:centered_recap_branch_qnonzero}.
For a non-centered spike, full alignment may not be immediate,
but it is always reached in the limit $\theta\to\infty$, when
the overlap equals its maximum.

\section{Quadratic distance kernel: detailed computations}\label{appendix:quadratic_distance_kernel}
In this Appendix, we derive the results summarized in Sec.~\ref{sec:results_quadratic_distance} for the quadratic distance kernel
\begin{equation}
    f(\x,\x')=\lVert\x-\x'\rVert^2
    \ .
\end{equation}
We assume that the distribution $P$ of $\x$ has a finite fourth moment, ensuring that Eq.~\eqref{eq:kernel_L2_condition} holds.
Thus, $T_f$ is a Hilbert–Schmidt operator on $L^2(P)$ and, by symmetry of the kernel, is self-adjoint.
We also assume that $P$ is not supported on a single point, which ensures that the upper spectral edge $\Lambda_0$ of $T_f$ is strictly positive.
To prove this, we use the variational characterization of the upper spectral edge for a bounded self-adjoint operator
\begin{equation}
    \Lambda_0
    =
    \sup_{\lVert\psi\rVert_{L^2(P)}=1}
    (\psi,T_f\psi)_P
    \ .
\end{equation}
Since the constant function $\mathbf1$ has unit norm in $L^2(P)$, it follows that
\begin{align}
    \Lambda_0
    &\geq
    (\mathbf1,T_f\mathbf1)_P
    \nonumber\\
    &=
    \int \dd\x\,\dd\x'\,
    P(\x)P(\x')\,
    \lVert\x-\x'\rVert^2
    \nonumber\\
    &=
    2\left(
        \langle\lVert\x\rVert^2\rangle_P
        -
        \lVert\langle\x\rangle_P\rVert^2
    \right)
    >
    0
    \ ,
\end{align}
where the last inequality holds because $P$ is not supported on a single point.
Compactness and self-adjointness therefore imply that
\begin{equation}
    \Lambda_0=\lambda_{\max}(T_f)>0
    \ .
\end{equation}

The form of the kernel also ensures that this principal eigenvalue is simple and that its eigenfunction has nonzero $P$-average.
Indeed, for any $\psi\in L^2(P)$ with $\langle\psi\rangle_P=0$, expanding the squared distance gives
\begin{align}
    (\psi,T_f\psi)_P
    &=
    \int \dd\x\,\dd\x'\,
    P(\x)P(\x')\,
    \psi(\x)
    \left[
        \lVert\x\rVert^2+\lVert\x'\rVert^2
        -2\x\cdot\x'
    \right]
    \psi(\x')
    \nonumber\\
    &=
    2\langle\psi\rangle_P
    \left\langle\lVert\x\rVert^2\psi\right\rangle_P
    -
    2\left\lVert\langle\x\psi\rangle_P\right\rVert^2
    \nonumber\\
    &=
    -2\left\lVert\langle\x\psi\rangle_P\right\rVert^2
    \leq0
    \ .
\end{align}
A nonzero eigenfunction associated with $\Lambda_0>0$ cannot therefore have zero $P$-average, since it satisfies
\begin{equation}
    (\psi,T_f\psi)_P
    =
    \Lambda_0\lVert\psi\rVert_{L^2(P)}^2
    >
    0
    \ .
\end{equation}
Moreover, if the principal eigenspace had dimension greater than one, it would contain a nonzero linear combination of eigenfunctions with zero $P$-average, contradicting the same inequality.
Thus, $\Lambda_0$ is simple and its normalized eigenfunction $\psi_{\max}$ satisfies
$\langle\psi_{\max}\rangle_P\neq0$. Therefore, projecting the constant function onto the principal eigenspace $E_1$ gives
\begin{equation}
    \Pi_{E_1}\mathbf1
    =
    \frac{(\psi_{\max},\mathbf1)_P}{\lVert\psi_{\max}\rVert_{L^2(P)}^2}\,\psi_{\max}
    =
    \frac{\langle\psi_{\max}\rangle_P}{\lVert\psi_{\max}\rVert_{L^2(P)}^2}\,\psi_{\max}
    \neq0
    \ .
\end{equation}
The quadratic distance kernel therefore belongs to the case
$\Lambda_0>0$ and $\Pi_{E_1}\mathbf1\neq0$ examined in Appendix~\ref{appendix:transition_thresholds}.
For the corresponding spiked kernel
\begin{equation}
    F_\theta\bigl((\x,y),(\x',y')\bigr)
    =
    \lVert\x-\x'\rVert^2+\theta yy'
    \ ,
\end{equation}
we assume that the distribution $G$ of the spike components $y$
has a finite second moment and is non-degenerate, so that
$0<\sigma_y^2<\infty$. Under the above assumptions on $P$, the
finite second moment ensures
Eq.~\eqref{eq:spiked_kernel_L2_condition}, making the spiked
spectral problem well defined. The general results of Appendix~\ref{appendix:transition_thresholds} therefore
predict a transition, for centered spikes, from the dominant
branch $q[\Psi]=0$ to the branch $q[\Psi]\neq0$ at the positive
threshold given in Eq.~\eqref{eq:centered_theta_threshold}.
For non-centered spikes, no such transition occurs: the unperturbed principal eigenvalue is simple and the corresponding eigenfunction has a
non-zero $P$-average. As established in
Sec.~\ref{sec:spiked model_non_centered}, this places the principal
eigenpair of $T_\theta$ on the branch $q[\Psi]\neq0$ for every
$\theta>0$.
For this choice of the kernel,  the eigenvalue equation reads
\begin{align}
    \lambda\Psi(\x,y)
    =(T_\theta \Psi)(\x,y)&=
    \int \dd\x'\,\dd y'\,
    P(\x')G(y')
    \left[
        \lVert\x\rVert^2+\lVert\x'\rVert^2
        -2\x\cdot\x'+\theta yy'
    \right]\Psi(\x',y')
    \nonumber\\
    &=
    \lVert\x\rVert^2
        \left\langle\bar{\Psi}(\x')\right\rangle_P
    -2\x\cdot
        \left\langle\x'\bar{\Psi}(\x')\right\rangle_P
    +\left\langle
        \lVert\x'\rVert^2\bar{\Psi}(\x')
    \right\rangle_P
    +\theta yq[\Psi]
    \ ,
\end{align}
where we used the definitions of $\bar{\Psi}$ and $q[\Psi]$
in Eqs.~\eqref{eq:barpsi} and~\eqref{eq:q_psi} of the main text.
Every eigenfunction associated with a nonzero eigenvalue
$\lambda$ therefore has the form
\begin{equation}\label{eq:results_quadratic_Psi}
    \Psi(\x,y)
    =
    A\lVert\x\rVert^2+\B\cdot\x+C+\theta Dy
    \ ,
\end{equation}
where the $d+3$ coefficients are defined by
\begin{equation}\label{eq:results_quadratic_coefficients}
\begin{aligned}
    A
    &\coloneqq
    \frac{1}{\lambda}
    \left\langle\bar{\Psi}(\x)\right\rangle_P
    \ ,\\
    B_\ell
    &\coloneqq
    -\frac{2}{\lambda}
    \left\langle x_\ell\bar{\Psi}(\x)\right\rangle_P
    \ ,
    \qquad \ell=1,\ldots,d
    \ ,\\
    C
    &\coloneqq
    \frac{1}{\lambda}
    \left\langle
        \lVert\x\rVert^2\bar{\Psi}(\x)
    \right\rangle_P
    \ ,\\
    D
    &\coloneqq
    \frac{q[\Psi]}{\lambda}
    \ .
\end{aligned}
\end{equation}
Averaging Eq.~\eqref{eq:results_quadratic_Psi} over $y$ yields
\begin{equation}\label{eq:results_quadratic_barPsi}
    \bar{\Psi}(\x)
    =
    A\lVert\x\rVert^2+\B\cdot\x+C
    +\theta D\langle y\rangle_G
    \ .
\end{equation}
Inserting this expression into the definitions of $A$, $\B$,
and $C$ in Eq.~\eqref{eq:results_quadratic_coefficients},
and inserting Eqs.~\eqref{eq:q_psi} and \eqref{eq:results_quadratic_Psi}
into the definition of $D$, yields
\begin{equation}\label{eq:results_quadratic_system}
\left\{
\begin{aligned}
    \lambda A
    &=
    A\left\langle\lVert\x\rVert^2\right\rangle_P
    +\B\cdot\left\langle\x\right\rangle_P
    +C+\theta D\langle y\rangle_G
    \ ,\\
    \lambda B_\ell
    &=
    -2\left[
        A\left\langle x_\ell\lVert\x\rVert^2\right\rangle_P
        +\B\cdot\left\langle x_\ell\x\right\rangle_P
        +(C+\theta D\langle y\rangle_G)
            \left\langle x_\ell\right\rangle_P
    \right]
    \ ,
    \qquad \ell=1,\ldots,d
    \ ,\\
    \lambda C
    &=
    A\left\langle\lVert\x\rVert^4\right\rangle_P
    +\B\cdot\left\langle\x\lVert\x\rVert^2\right\rangle_P
    +(C+\theta D\langle y\rangle_G)
        \left\langle\lVert\x\rVert^2\right\rangle_P
    \ ,\\
    \lambda D
    &=
    \langle y\rangle_G
    \left[
        A\left\langle\lVert\x\rVert^2\right\rangle_P
        +\B\cdot\left\langle\x\right\rangle_P+C
    \right]
    +\theta D\langle y^2\rangle_G
    \ .
\end{aligned}
\right.
\end{equation}
This system holds for every nonzero eigenvalue of $T_\theta$ and
is the general self-consistent system in
Eq.~\eqref{eq:final_sc_system} of the main text expressed in terms of the coefficients
$(A,\B,C,D)$. It takes the form of an eigenvalue problem for a
$(d+3)\times(d+3)$ matrix whose entries depend on moments and mixed
correlators of the components of $\x$ up to fourth order, and on
the first two moments of $y$.

To obtain the principal eigenpair $\bigl(\lambda_{\max}(T_\theta),\Psi_{\max}\bigr)$, we seek the largest value of
$\lambda$ for which the system in Eq.~\eqref{eq:results_quadratic_system} admits a
non-trivial solution, while imposing the normalization condition
in Eq.~\eqref{eq:spiked_psi_normalisation} on the corresponding
coefficients. For the principal eigenfunction $\Psi_{\max}$, using the
normalization in Eq.~\eqref{eq:spiked_psi_normalisation},
the representation in Eq.~\eqref{eq:results_quadratic_Psi},
and the coefficient relations in
Eq.~\eqref{eq:results_quadratic_coefficients}, we obtain
\begin{equation}\label{eq:results_quadratic_norm_identity}
\begin{aligned}
    \lambda_{\max}(T_\theta)^2
    &=
    A\left\langle\lVert\x\rVert^2\Psi_{\max}\right\rangle_{P\otimes G}
    +\B\cdot\left\langle\x\Psi_{\max}\right\rangle_{P\otimes G}
    +C\left\langle\Psi_{\max}\right\rangle_{P\otimes G}
    +\theta D\left\langle y\Psi_{\max}\right\rangle_{P\otimes G}
    \\
    &=
    \lambda_{\max}(T_\theta)\left(
        2AC-\frac12\lVert\B\rVert^2+\theta D^2
    \right)
    \ .
\end{aligned}
\end{equation}
The normalization constraint for the principal eigenfunction
therefore reduces to
\begin{equation}\label{eq:results_quadratic_normalization}
    2AC-\frac12\lVert\B\rVert^2+\theta D^2
    =
    \lambda_{\max}(T_\theta)
    \ .
\end{equation}

To solve Eq.~\eqref{eq:results_quadratic_system}, we find
the values of $\lambda$ for which it admits a non-trivial solution
and determine the corresponding coefficients $(A,\B,C,D)$.
The largest such value is $\lambda_{\max}(T_\theta)$, and its
coefficients, normalized according to
Eq.~\eqref{eq:results_quadratic_normalization}, determine the
principal eigenfunction $\Psi_{\max}$ through
Eq.~\eqref{eq:results_quadratic_Psi}. The coefficient $D$ associated with this normalized principal
eigenfunction also has a direct interpretation: as shown in
Appendix~\ref{appendix_replica_der}, it is precisely the limiting average overlap between
the top eigenvector and the spike
\begin{equation}\label{eq:results_quadratic_overlap}
    D
    =
    \frac{q[\Psi_{\max}]}{\lambda_{\max}(T_\theta)}
    =
    \lim_{N\to\infty}
    \left\langle
        \frac{(\vec y,\vec v^{\, (1)})}{N}
    \right\rangle
    \ .
\end{equation}
The system in Eq.~\eqref{eq:results_quadratic_system} and the normalization constraint leave an overall sign
ambiguity, since replacing $(A,\B,C,D)$ by $(-A,-\B,-C,-D)$
preserves the validity of the solution and its normalization. Following the convention adopted in the main text, we fix
the overall sign of the eigenfunction so that $D>0$
whenever $D\neq0$, corresponding to a positive overlap
between the top eigenvector and the spike according to
Eq.~\eqref{eq:results_quadratic_overlap}. When $D=0$, we instead
choose $A>0$, following the convention of Ref.~\cite{casaburi2026}.

In the remainder of this Appendix, we consider two broad classes of
distributions for which the system in
Eq.~\eqref{eq:results_quadratic_system} simplifies, namely isotropic
distributions, $P(\x)=P(\lVert\x\rVert)$, and distributions with
independent and identically distributed components drawn from an
even marginal distribution,
$P(\x)=\prod_{\ell=1}^d P(x_\ell)$ with $P(x_\ell)=P(-x_\ell)$.
In both cases, their symmetries imply
\begin{equation}\label{eq:results_quadratic_symmetry_moments}
    \langle x_\ell\rangle_P
    =
    \left\langle x_\ell\lVert\x\rVert^2\right\rangle_P
    =
    0
    \ ,
    \qquad
    \langle x_\ell x_m\rangle_P
    =
    \frac{\langle\lVert\x\rVert^2\rangle_P}{d}\,
    \delta_{\ell m}
    \ .
\end{equation}
Consequently, the equation for $B_\ell$ in
Eq.~\eqref{eq:results_quadratic_system} reduces to
\begin{equation}
    \left(
        \lambda+\frac{2}{d}\langle\lVert\x\rVert^2\rangle_P
    \right)B_\ell=0
    \ ,
    \qquad \forall \ell=1,\ldots,d
    \ .
\end{equation}
Since $\lambda_{\max}(T_\theta)>0$, the principal solution therefore
has $\B=0$ in both cases, leaving only the coefficients $(A,C,D)$
to determine. The remaining equations depend on $P(\x)$ only through
$\langle\lVert\x\rVert^2\rangle_P$ and
$\langle\lVert\x\rVert^4\rangle_P$, so that both classes of
distributions are described by the reduced system
\begin{equation}\label{eq:results_quadratic_reduced_system}
\left\{
\begin{aligned}
    \lambda A
    &=
    A\langle\lVert\x\rVert^2\rangle_P
    +C+\theta D\langle y\rangle_G
    \ ,\\
    \lambda C
    &=
    A\langle\lVert\x\rVert^4\rangle_P
    +(C+\theta D\langle y\rangle_G)
        \langle\lVert\x\rVert^2\rangle_P
    \ ,\\
    \lambda D
    &=
    \langle y\rangle_G
    \left[
        A\langle\lVert\x\rVert^2\rangle_P+C
    \right]
    +\theta D\langle y^2\rangle_G
    \ .
\end{aligned}
\right.
\quad\iff\quad
\lambda
\begin{pmatrix}
    A\\ C\\ D
\end{pmatrix}
=
\begin{pmatrix}
    \langle\lVert\x\rVert^2\rangle_P
    & 1
    & \theta\langle y\rangle_G
    \\
    \langle\lVert\x\rVert^4\rangle_P
    & \langle\lVert\x\rVert^2\rangle_P
    & \theta\langle y\rangle_G\langle\lVert\x\rVert^2\rangle_P
    \\
    \langle y\rangle_G\langle\lVert\x\rVert^2\rangle_P
    & \langle y\rangle_G
    & \theta\langle y^2\rangle_G
\end{pmatrix}
\begin{pmatrix}
    A\\ C\\ D
\end{pmatrix}
\ .
\end{equation}
The reduced system in Eq.~\eqref{eq:results_quadratic_reduced_system}
is a $3\times3$ eigenvalue problem. Denoting its matrix by $M$,
a non-trivial solution exists if and only if
$\det(\lambda I-M)=0$. Expanding the determinant gives the
characteristic equation
\begin{equation}\label{eq:results_quadratic_characteristic}
    \lambda^3
    -\left(
        2\langle\lVert\x\rVert^2\rangle_P
        +\theta\langle y^2\rangle_G
    \right)\lambda^2
    +\left(
        2\theta\langle\lVert\x\rVert^2\rangle_P\sigma_y^2
        -\sigma_{\lVert\x\rVert^2}^2
    \right)\lambda
    +\theta\sigma_{\lVert\x\rVert^2}^2\sigma_y^2
    =0
    \ ,
\end{equation}
where we have defined the variances
\begin{equation}\label{eq:results_quadratic_variances}
    \sigma_y^2
    =
    \langle y^2\rangle_G-\langle y\rangle_G^2
    \ ,
    \qquad
    \sigma_{\lVert\x\rVert^2}^2
    =
    \langle\lVert\x\rVert^4\rangle_P
    -\langle\lVert\x\rVert^2\rangle_P^2
    \ .
\end{equation}
The characteristic polynomial in
Eq.~\eqref{eq:results_quadratic_characteristic} coincides with that
of the real symmetric matrix
\begin{equation}
    H=
    \begin{pmatrix}
        2\langle\lVert\x\rVert^2\rangle_P
        +\theta\langle y\rangle_G^2
        & \sigma_{\lVert\x\rVert^2}
        & \theta\langle y\rangle_G\sigma_y
        \\
        \sigma_{\lVert\x\rVert^2} & 0 & 0
        \\
        \theta\langle y\rangle_G\sigma_y
        & 0 & \theta\sigma_y^2
    \end{pmatrix}
    \ ,
    \label{eq:results_quadratic_symmetric_matrix}
\end{equation}
and therefore has three real roots, counted with multiplicity.
Denoting these roots by $(\lambda_1,\lambda_2,\lambda_3)$,
Vieta's relations give
\begin{equation}\label{eq:results_quadratic_vieta}
\begin{aligned}
    \lambda_1+\lambda_2+\lambda_3
    &=
    2\langle\lVert\x\rVert^2\rangle_P
    +\theta\langle y^2\rangle_G
    \ ,\\
    \lambda_1\lambda_2\lambda_3
    &=
    -\theta\sigma_{\lVert\x\rVert^2}^2\sigma_y^2
    \ .
\end{aligned}
\end{equation}
Since $\theta>0$ and $0<\sigma_y^2<\infty$, the sign of the
product depends on $\sigma_{\lVert\x\rVert^2}^2$.
If $\sigma_{\lVert\x\rVert^2}^2>0$, the product is negative,
implying that either one or three roots are negative. Since
their sum is positive, the latter possibility is excluded.
There is therefore exactly one negative root and two positive
roots.
If instead $\sigma_{\lVert\x\rVert^2}^2=0$, Eq.~\eqref{eq:results_quadratic_characteristic}
reduces to
\begin{equation}
    \lambda\left[
        \lambda^2
        -\left(
            2\langle\lVert\x\rVert^2\rangle_P
            +\theta\langle y^2\rangle_G
        \right)\lambda
        +2\theta\langle\lVert\x\rVert^2\rangle_P\sigma_y^2
    \right]=0
    \ .
\end{equation}
One root is therefore zero, while the remaining two have
positive sum and product and hence are both positive.
Combining the two cases, we order the roots as
\begin{equation}\label{eq:results_quadratic_root_ordering}
    \lambda_1\leq0<\lambda_2\leq\lambda_3
    \ ,
\end{equation}
where $\lambda_1=0$ if and only if
$\sigma_{\lVert\x\rVert^2}^2=0$, and
$\lambda_{\max}(T_\theta)=\lambda_3$.
To solve Eq.~\eqref{eq:results_quadratic_reduced_system}, we
separate the two cases of centered spikes, $\langle y\rangle_G=0$,
and non-centered spikes, $\langle y\rangle_G\neq0$.

\subsection{Centered spike}\label{sec:res_quadratic_centered}
When $\langle y\rangle_G=0$, the equation for $D$ decouples from
those for $(A,C)$, and the system in
Eq.~\eqref{eq:results_quadratic_reduced_system} takes the form
\begin{equation}\label{eq:results_quadratic_centered_system}
\left\{
\begin{aligned}
    \lambda A
    &=A\langle\lVert\x\rVert^2\rangle_P+C
    \ ,\\
    \lambda C
    &=A\langle\lVert\x\rVert^4\rangle_P
    +C\langle\lVert\x\rVert^2\rangle_P
    \ ,\\
    \lambda D
    &=\theta\sigma_y^2D
    \ .
\end{aligned}
\right.
\quad\iff\quad
\lambda
\begin{pmatrix}
    A\\ C\\ D
\end{pmatrix}
=
\begin{pmatrix}
    \langle\lVert\x\rVert^2\rangle_P & 1 & 0\\
    \langle\lVert\x\rVert^4\rangle_P
    & \langle\lVert\x\rVert^2\rangle_P & 0\\
    0 & 0 & \theta\sigma_y^2
\end{pmatrix}
\begin{pmatrix}
    A\\ C\\ D
\end{pmatrix}
\ .
\end{equation}
The matrix is block diagonal. Its upper $2\times2$ block is precisely
the system of equations for $(A,C)$ obtained by setting $\theta=0$
in Eq.~\eqref{eq:results_quadratic_reduced_system}, thereby removing
the rank-one perturbation and reducing $T_\theta$ to $T_f$.
The remaining scalar block gives the spike eigenvalue
$\theta\sigma_y^2$.
Accordingly, Eq.~\eqref{eq:results_quadratic_characteristic}
factorizes as
\begin{equation}\label{eq:results_quadratic_centered_characteristic}
    \left(\lambda-\theta\sigma_y^2\right)
    \left[
        \left(\lambda-\langle\lVert\x\rVert^2\rangle_P\right)^2
        -\langle\lVert\x\rVert^4\rangle_P
    \right]
    =0
    \ .
\end{equation}
The upper block therefore gives the two eigenvalues
$\langle\lVert\x\rVert^2\rangle_P
\pm\sqrt{\langle\lVert\x\rVert^4\rangle_P}$, the larger of which is
the principal eigenvalue of $T_f$
\begin{equation}\label{eq:results_quadratic_unperturbed_largest}
    \lambda_{\max}(T_f)
    =
    \langle\lVert\x\rVert^2\rangle_P
    +\sqrt{\langle\lVert\x\rVert^4\rangle_P}
    \ .
\end{equation}
This reproduces Eqs.~(74) and (80) of Ref.~\cite{casaburi2026}
for isotropic distributions and distributions with symmetric
i.i.d.\ components, respectively. The remaining scalar block
gives the spike eigenvalue $\theta\sigma_y^2$. With the ordering given in Eq.~\eqref{eq:results_quadratic_root_ordering},
the three eigenvalues are therefore
\begin{equation}\label{eq:results_quadratic_centered_roots}
\begin{aligned}
    &\lambda_1
    =
    \langle\lVert\x\rVert^2\rangle_P
    -\sqrt{\langle\lVert\x\rVert^4\rangle_P}
    \ ,\\
    &\lambda_2
    =
    \min\left\{
        \langle\lVert\x\rVert^2\rangle_P
        +\sqrt{\langle\lVert\x\rVert^4\rangle_P},
        \theta\sigma_y^2
    \right\}
    =\min\left\{
        \lambda_{\max}(T_f),
        \theta\sigma_y^2
    \right\}
    \ ,\\
    &\lambda_{\max}(T_\theta)
    =
    \max\left\{
        \langle\lVert\x\rVert^2\rangle_P
        +\sqrt{\langle\lVert\x\rVert^4\rangle_P},
        \theta\sigma_y^2
    \right\}
    =\max\left\{
        \lambda_{\max}(T_f),
        \theta\sigma_y^2
    \right\}
    \ ,
\end{aligned}
\end{equation}
where $\lambda_1=0$ if and only if
$\sigma_{\lVert\x\rVert^2}^2=0$, as anticipated above. The principal solution therefore changes at
\begin{equation}
    \theta_c
    =
    \frac{
        \langle\lVert\x\rVert^2\rangle_P
        +\sqrt{\langle\lVert\x\rVert^4\rangle_P}
    }{\sigma_y^2} =
    \frac{\lambda_{\max}(T_f)}{\sigma_y^2}
    \ ,
\end{equation}
where the two positive eigenvalue branches meet. These expressions for $\lambda_{\max}(T_\theta)$ and $\theta_c$
recover exactly the general results derived in
Sec.~\ref{sec:spiked model_centered}, given in
Eqs.~\eqref{eq:centered_largest_eigenvalue} and
\eqref{eq:centered_theta_threshold}, respectively. To determine the principal eigenfunction and the average overlap
between the top eigenvector and the spike, we solve
Eq.~\eqref{eq:results_quadratic_centered_system} for $(A,C,D)$
together with the normalization constraint in
Eq.~\eqref{eq:results_quadratic_normalization}, which reduces to
\begin{equation}\label{eq:results_quadratic_centered_normalization}
    2AC+\theta D^2=\lambda_{\max}(T_\theta)
    \ ,
\end{equation}
after setting $\B=0$.

For $0<\theta<\theta_c$, the principal eigenvalue is
$\lambda_{\max}(T_\theta)=\lambda_{\max}(T_f)$.
The last equation of
Eq.~\eqref{eq:results_quadratic_centered_system} then requires
$D=0$, while the first gives
$C=A\sqrt{\langle\lVert\x\rVert^4\rangle_P}$.
Imposing the normalization constraint $2AC=\lambda_{\max}(T_f)$ and choosing $A>0$, we obtain
\begin{equation}\label{eq:results_quadratic_centered_below}
    A
    =
    \sqrt{
        \frac{\lambda_{\max}(T_f)}
        {2\sqrt{\langle\lVert\x\rVert^4\rangle_P}}
    }=
    \sqrt{
        \frac{\langle\lVert\x\rVert^2\rangle_P
    +\sqrt{\langle\lVert\x\rVert^4\rangle_P}}
        {2\sqrt{\langle\lVert\x\rVert^4\rangle_P}}
    }
    \ ,
    \qquad
    C=A\sqrt{\langle\lVert\x\rVert^4\rangle_P}
    \ ,
    \qquad
    D=0
    \ .
\end{equation}
The principal eigenfunction is therefore unchanged by the spike,
and its coefficients give zero average overlap according to
Eq.~\eqref{eq:results_quadratic_overlap}.

For $\theta>\theta_c$, the principal eigenvalue is
$\lambda_{\max}(T_\theta)=\theta\sigma_y^2$.
Since this value lies above both eigenvalues of the upper block,
the first two equations of
Eq.~\eqref{eq:results_quadratic_centered_system} require $A=C=0$.
The normalization constraint becomes
$\theta D^2=\theta\sigma_y^2$, so our sign convention gives
\begin{equation}\label{eq:results_quadratic_centered_above}
    A=0
    \ ,
    \qquad
    C=0
    \ ,
    \qquad
    D=\sigma_y
    \ .
\end{equation}
Thus, recalling Eqs.~\eqref{eq:results_quadratic_Psi} and \eqref{eq:results_quadratic_overlap}, the principal eigenfunction and the limiting
average overlap are
\begin{equation}\label{eq:results_quadratic_centered_eigenfunction}
    \Psi_{\max}(\x,y)
    =
    \begin{cases}
        \displaystyle
        \sqrt{
            \frac{\langle\lVert\x\rVert^2\rangle_P
    +\sqrt{\langle\lVert\x\rVert^4\rangle_P}}
            {2\sqrt{\langle\lVert\x\rVert^4\rangle_P}}
        }
        \left(
            \lVert\x\rVert^2
            +\sqrt{\langle\lVert\x\rVert^4\rangle_P}
        \right)\ ,
        & 0<\theta<\theta_c\ ,
        \\[3ex]
        \theta\sigma_y y\ ,
        & \theta>\theta_c\ ,
    \end{cases}
\end{equation}

\begin{equation}\label{eq:results_quadratic_centered_overlap}
    \lim_{N\to\infty}
    \left\langle
        \frac{(\vec y,\vec v^{\, (1)})}{N}
    \right\rangle
    =
    D
    =
    \frac{q[\Psi_{\max}]}{\lambda_{\max}(T_\theta)}
    =
    \begin{cases}
        0, & 0<\theta<\theta_c,\\
        \sigma_y, & \theta>\theta_c.
    \end{cases}
\end{equation}
These expressions recover Eqs.~\eqref{eq:summary_quadratic_centered_eigenvalue}, \eqref{eq:summary_quadratic_centered_eigenfunction} and  \eqref{eq:summary_quadratic_centered_overlap} of the main text and the general results derived in
Sec.~\ref{sec:spiked model_centered}: below the threshold, the
dominant branch has zero overlap ($q[\Psi_{\max}]=0$) and the
principal eigenfunction is independent of the spike; above the
threshold, the dominant branch has non-zero overlap
($q[\Psi_{\max}]\neq0$), with the principal eigenfunction
proportional to $y$.

At $\theta=\theta_c$, the two positive eigenvalue branches coincide,
so that
$\lambda_{\max}(T_{\theta_c})=\lambda_{\max}(T_f)=\theta_c\sigma_y^2=\langle\lVert\x\rVert^2\rangle_P
        +\sqrt{\langle\lVert\x\rVert^4\rangle_P}$
is a doubly degenerate eigenvalue. The first equation of
Eq.~\eqref{eq:results_quadratic_centered_system} gives
$C=A\sqrt{\langle\lVert\x\rVert^4\rangle_P}$, which also satisfies
the second equation, while the third becomes an identity and places
no constraint on $D$. The normalization condition therefore reduces to
\begin{equation}\label{eq:results_quadratic_centered_critical_normalization}
    2\sqrt{\langle\lVert\x\rVert^4\rangle_P}\,A^2
    +\theta_cD^2
    =
    \theta_c\sigma_y^2
    \ .
\end{equation}
Consequently, the normalized principal eigenfunctions form a family
with coefficients
\begin{equation}\label{eq:results_quadratic_centered_critical_coefficients}
\begin{aligned}
    A
    &=
    \pm\sqrt{
        \frac{\theta_c\sigma_y^2}
        {2\sqrt{\langle\lVert\x\rVert^4\rangle_P}}
        \left(1-\frac{D^2}{\sigma_y^2}\right)
    }
    \ ,\\
    C
    &=A\sqrt{\langle\lVert\x\rVert^4\rangle_P}
    \ ,\\
    0&\leq D\leq\sigma_y
    \ .
\end{aligned}
\end{equation}
Here the range of $D$ follows from the normalization and our sign
convention, which also selects $A>0$ when $D=0$.
The endpoints $D=0$ and $D=\sigma_y$ correspond, respectively, to
the unperturbed principal eigenfunction and the eigenfunction
proportional to $y$; intermediate values describe normalized
linear combinations of these two eigenfunctions.
Thus, at the threshold, the limiting operator admits no unique
normalized principal eigenfunction, and the equations above do
not select a unique overlap. Determining the limiting average
overlap of the finite-size top eigenvector at this point would require
an analysis of how finite-size fluctuations resolve the degeneracy.

\subsection{Non-centered spike}\label{sec:res_quadratic_noncentered}
We now consider non-centered spikes, $\langle y\rangle_G\neq0$.
For $\theta>0$, the three roots of
Eq.~\eqref{eq:results_quadratic_characteristic} are distinct.
To see this, we rewrite the characteristic polynomial as
\begin{equation}\label{eq:results_quadratic_noncentered_polynomial}
\begin{aligned}
    P_{\theta}(\lambda)
    &=
    \lambda^3
    -\left(
        2\langle\lVert\x\rVert^2\rangle_P
        +\theta\langle y^2\rangle_G
    \right)\lambda^2
    +\left(
        2\theta\langle\lVert\x\rVert^2\rangle_P\sigma_y^2
        -\sigma_{\lVert\x\rVert^2}^2
    \right)\lambda
    +\theta\sigma_{\lVert\x\rVert^2}^2\sigma_y^2\\
    &=
    \left(\lambda-\theta\sigma_y^2\right)
    \left[
        \lambda^2
        -2\langle\lVert\x\rVert^2\rangle_P\lambda
        -\sigma_{\lVert\x\rVert^2}^2
    \right]
    -\theta\langle y\rangle_G^2\lambda^2
    \ .
\end{aligned}
\end{equation}
Recalling that $\theta>0$ and $\sigma_y^2>0$, this expression shows that
\begin{equation}\label{eq:results_quadratic_noncentered_polynomial_signs}
\begin{aligned}
    &P_{\theta}(\lambda)\xrightarrow{\lambda\to-\infty}-\infty
    \ ,\\
    &P_{\theta}(0)=\theta\sigma_y^2\sigma_{\lVert\x\rVert^2}^2\geq0
    \ ,\\
    &P_{\theta}(\theta\sigma_y^2)
    =-\theta^3\langle y\rangle_G^2\sigma_y^4<0
    \ ,\\
    &P_{\theta}(\lambda)\xrightarrow{\lambda\to+\infty}+\infty
    \ ,
\end{aligned}
\end{equation}

If $\sigma_{\lVert\x\rVert^2}^2>0$, then $P_{\theta}(0)>0$.
By continuity, there is a root in each of the three disjoint
intervals $(-\infty,0)$, $(0,\theta\sigma_y^2)$, and
$(\theta\sigma_y^2,+\infty)$. Since the polynomial is cubic,
these are its only roots, and they are all distinct.

If instead $\sigma_{\lVert\x\rVert^2}^2=0$, one root is zero
and $P_{\theta}(\lambda)=\lambda Q_{\theta}(\lambda)$, where
\begin{equation}
    Q_{\theta}(\lambda)
    =
    \left(\lambda-\theta\sigma_y^2\right)
    \left(\lambda-2\langle\lVert\x\rVert^2\rangle_P\right)
    -\theta\langle y\rangle_G^2\lambda
    \ .
\end{equation}
This polynomial satisfies
\begin{equation}
\begin{aligned}
    &Q_{\theta}(0)=2\theta\langle\lVert\x\rVert^2\rangle_P\sigma_y^2>0
    \ ,\\
    &Q_{\theta}(\theta\sigma_y^2)
    =-\theta^2\langle y\rangle_G^2\sigma_y^2<0
    \ ,\\
    &Q_{\theta}(\lambda)\xrightarrow{\lambda\to+\infty}+\infty
    \ .
\end{aligned}
\end{equation}
By continuity, the remaining two roots lie in
$(0,\theta\sigma_y^2)$ and $(\theta\sigma_y^2,+\infty)$,
respectively.

In both cases, the three roots are therefore distinct and satisfy
\begin{equation}\label{eq:results_quadratic_noncentered_root_ordering}
    \lambda_1\leq0<\lambda_2<\theta\sigma_y^2
    <\lambda_3=\lambda_{\max}(T_\theta)
    \ ,
\end{equation}
with $\lambda_1=0$ if and only if
$\sigma_{\lVert\x\rVert^2}^2=0$.
The two positive eigenvalue branches therefore cannot cross
for $\theta>0$. For a centered spike, instead,
$P_{\theta}(\theta\sigma_y^2)=0$, allowing these branches to coincide
at $\theta=\theta_c$.

To solve Eq.~\eqref{eq:results_quadratic_characteristic} explicitly,
we introduce the shift
\begin{equation}\label{eq:results_quadratic_cubic_shift}
    \lambda
    =
    z+\frac{
        2\langle\lVert\x\rVert^2\rangle_P
        +\theta\langle y^2\rangle_G
    }{3}
    \ ,
\end{equation}
which removes the quadratic term and gives the depressed cubic
$z^3+pz+w=0$, where
\begin{equation}\label{eq:results_quadratic_depressed_coefficients}
\begin{aligned}
    p={}&
    2\theta\langle\lVert\x\rVert^2\rangle_P\sigma_y^2
    -\sigma_{\lVert\x\rVert^2}^2
    -\frac13\left(
        2\langle\lVert\x\rVert^2\rangle_P
        +\theta\langle y^2\rangle_G
    \right)^2
    \ ,\\
    w={}&
    -\frac{2}{27}\left(
        2\langle\lVert\x\rVert^2\rangle_P
        +\theta\langle y^2\rangle_G
    \right)^3
    +\frac13\left(
        2\langle\lVert\x\rVert^2\rangle_P
        +\theta\langle y^2\rangle_G
    \right)\left(
        2\theta\langle\lVert\x\rVert^2\rangle_P\sigma_y^2
        -\sigma_{\lVert\x\rVert^2}^2
    \right)
    +\theta\sigma_{\lVert\x\rVert^2}^2\sigma_y^2
    \ .
\end{aligned}
\end{equation}
Since the three roots are real and distinct, we have
$p<0$ and $\Delta=-4p^3-27w^2>0$. The trigonometric solution
can therefore be expressed in terms of the angle
\begin{equation}\label{eq:results_quadratic_cubic_angle}
    \phi
    =
    \frac13\arccos\left(
        \frac{3w}{2p}\sqrt{-\frac3p}
    \right)
    \ ,
    \qquad
    0<\phi<\frac{\pi}{3}
    \ .
\end{equation}
With the ordering given in
Eq.~\eqref{eq:results_quadratic_noncentered_root_ordering},
the roots are
\begin{equation}\label{eq:results_quadratic_explicit_roots}
\begin{aligned}
    &\lambda_1
    =
    \frac{
        2\langle\lVert\x\rVert^2\rangle_P
        +\theta\langle y^2\rangle_G
    }{3}
    +2\sqrt{-\frac p3}
        \cos\left(\phi+\frac{2\pi}{3}\right)
    \ ,\\
    &\lambda_2
    =
    \frac{
        2\langle\lVert\x\rVert^2\rangle_P
        +\theta\langle y^2\rangle_G
    }{3}
    +2\sqrt{-\frac p3}
        \cos\left(\phi-\frac{2\pi}{3}\right)
    \ ,\\
    &\lambda_{\max}(T_\theta)
    =
    \frac{
        2\langle\lVert\x\rVert^2\rangle_P
        +\theta\langle y^2\rangle_G
    }{3}
    +2\sqrt{-\frac p3}\cos\phi
    \ .
\end{aligned}
\end{equation}
Since the perturbation $\theta yy'$ is invariant under the global
sign change $y\mapsto-y$, we assume $\langle y\rangle_G>0$
without loss of generality. To determine the principal eigenfunction
and the average overlap, we now solve
Eq.~\eqref{eq:results_quadratic_reduced_system} at
$\lambda=\lambda_{\max}(T_\theta)$ and impose the normalization
constraint in Eq.~\eqref{eq:results_quadratic_normalization}
with $\B=0$.

The first equation of
Eq.~\eqref{eq:results_quadratic_reduced_system} gives
$A\langle\lVert\x\rVert^2\rangle_P+C
=\lambda A-\theta D\langle y\rangle_G$.
Inserting this relation into the third equation yields
\begin{equation}\label{eq:results_quadratic_noncentered_D_relation}
    \left(\lambda-\theta\sigma_y^2\right)D
    =
    \lambda\langle y\rangle_G A
    \ .
\end{equation}
Since $\lambda_{\max}(T_\theta)>\theta\sigma_y^2$ by
Eq.~\eqref{eq:results_quadratic_noncentered_root_ordering},
we obtain
\begin{equation}\label{eq:results_quadratic_noncentered_rD}
    D=r_D A
    \ ,
    \qquad
    r_D
    =
    \frac{
        \lambda_{\max}(T_\theta)\langle y\rangle_G
    }{
        \lambda_{\max}(T_\theta)-\theta\sigma_y^2
    }
    >0
    \ .
\end{equation}
Similarly, subtracting
$\langle\lVert\x\rVert^2\rangle_P$ times the first equation
from the second gives
$\lambda(C-A\langle\lVert\x\rVert^2\rangle_P)
=A\sigma_{\lVert\x\rVert^2}^2$, hence
\begin{equation}\label{eq:results_quadratic_noncentered_rC}
    C=r_C A
    \ ,
    \qquad
    r_C
    =
    \langle\lVert\x\rVert^2\rangle_P
    +\frac{
        \sigma_{\lVert\x\rVert^2}^2
    }{
        \lambda_{\max}(T_\theta)
    }
    >0
    \ .
\end{equation}
The normalization constraint therefore yields
\begin{equation}\label{eq:results_quadratic_noncentered_normalization}
    2AC+\theta D^2
    =
    A^2\left(2r_C+\theta r_D^2\right)
    =
    \lambda_{\max}(T_\theta)
    \ .
\end{equation}
Since $r_D>0$, our sign convention $D>0$ selects $A>0$.
The normalized coefficients are therefore
\begin{equation}\label{eq:results_quadratic_noncentered_coefficients}
\begin{aligned}
    &A
    =
    \sqrt{
        \frac{\lambda_{\max}(T_\theta)}
        {2r_C+\theta r_D^2}
    }
    \ ,
    \qquad
    C=r_C A=
    \left[\langle\lVert\x\rVert^2\rangle_P
    +\frac{\sigma_{\lVert\x\rVert^2}^2}
    {\lambda_{\max}(T_\theta)} \right] A
    \ ,
    \qquad
    D=r_D A = 
    \left[\frac{\lambda_{\max}(T_\theta)\langle y\rangle_G}
    {\lambda_{\max}(T_\theta)-\theta\sigma_y^2}\right] A
    \ ,
\end{aligned}
\end{equation}
where $\lambda_{\max}(T_\theta)$ is given in
Eq.~\eqref{eq:results_quadratic_explicit_roots}. 
The coefficients in
Eq.~\eqref{eq:results_quadratic_noncentered_coefficients}
determine the principal eigenfunction through
Eq.~\eqref{eq:results_quadratic_Psi}, giving
\begin{equation}\label{eq:results_quadratic_noncentered_eigenfunction}
    \Psi_{\max}(\x,y)
    =
    \sqrt{
        \frac{\lambda_{\max}(T_\theta)}
        {2r_C+\theta r_D^2}
    }
    \left(\lVert\x\rVert^2+r_C+\theta r_D y\right)
    \ .
\end{equation}
Moreover, recalling Eq.~\eqref{eq:results_quadratic_overlap},
the limiting average overlap is
\begin{equation}\label{eq:results_quadratic_noncentered_overlap}
    \lim_{N\to\infty}
    \left\langle
        \frac{(\vec y,\vec v^{\, (1)})}{N}
    \right\rangle
    =
    \frac{q[\Psi_{\max}]}{\lambda_{\max}(T_\theta)}
    =
    D
    =
    r_D
    \sqrt{
        \frac{\lambda_{\max}(T_\theta)}
        {2r_C+\theta r_D^2}
    }
    >0
    \ .
\end{equation}
Since $\lambda_{\max}(T_\theta)>0$, $r_C>0$, and $r_D>0$,
this overlap is strictly positive for every $\theta>0$. Thus, unlike the centered case in
Sec.~\ref{sec:res_quadratic_centered}, there is no transition
from a zero-overlap branch to a non-zero-overlap branch.

These results recover
Eqs.~\eqref{eq:summary_quadratic_noncentered_eigenvalue},
\eqref{eq:summary_quadratic_noncentered_eigenfunction},
\eqref{eq:summary_quadratic_noncentered_overlap}
and~\eqref{eq:summary_quadratic_ratios} of the main text
and confirm the absence of a transition predicted in
Sec.~\ref{sec:spiked model_non_centered}.
Indeed, as established above, the unperturbed quadratic
distance operator $T_f$ has a strictly positive, simple
principal eigenvalue, and its principal eigenfunction
has nonzero $P$-average.
The resulting condition $\Pi_{E_1}\mathbf1\neq0$ places the
principal eigenpair of $T_\theta$ on the branch
$q[\Psi]\neq0$ for every $\theta>0$ when
$\langle y\rangle_G\neq0$.

\section{Rank-two kernel: detailed computations}\label{appendix:rank_two_kernel}
In this Appendix, we derive the results summarized in Sec.~\ref{sec:results_rank2} for the kernel
\begin{equation}\label{eq:appendix_rank_two_kernel}
    f(\x,\x')=a+b\,h(\x)h(\x')
    \ ,
\end{equation}
where $h: \R^d \to \R$ is real-valued and 
\begin{equation}\label{eq:appendix_rank_two_assumptions}
    \langle h\rangle_P=0
    \ ,\qquad
    \langle h^2\rangle_P=1
    \ ,\qquad
    0<a<b
    \ .
\end{equation}
These assumptions ensure that
\begin{align}
    \int \dd\x\,\dd\x'\,
    P(\x)P(\x')|f(\x,\x')|^2
    &=
    a^2+2ab\langle h\rangle_P^2
    +b^2\langle h^2\rangle_P^2
    =a^2+b^2<\infty
    \ ,
\end{align}
so that Eq.~\eqref{eq:kernel_L2_condition} holds.
The corresponding operator $T_f$ is therefore compact and
self-adjoint on $L^2(P)$, and its spectral problem is well posed.
We first solve the associated unperturbed spectral problem,
the eigenvalue equation reads
\begin{align}\label{eq:results_rank_two_eigenvalue_equation}
    \mu\psi(\x)
    &=
    \int \dd\x'\,P(\x')
    \left[a+b\,h(\x)h(\x')\right]\psi(\x')
    \nonumber\\
    &=
    a\langle\psi\rangle_P
    +b\,h(\x)\langle h\psi\rangle_P
    \ .
\end{align}
Every eigenfunction associated with a nonzero eigenvalue
$\mu$ therefore has the form
\begin{equation}\label{eq:results_rank_two_psi}
    \psi(\x)=A+B\,h(\x)
    \ ,
\end{equation}
where the coefficients are defined by
\begin{equation}\label{eq:results_rank_two_coefficients}
    A\coloneqq\frac{a}{\mu}\langle\psi\rangle_P
    \ ,\qquad
    B\coloneqq\frac{b}{\mu}\langle h\psi\rangle_P
    \ .
\end{equation}
Inserting Eq.~\eqref{eq:results_rank_two_psi} into these
definitions and using
Eq.~\eqref{eq:appendix_rank_two_assumptions} yields
\begin{equation}\label{eq:results_rank_two_system}
    \left\{
    \begin{aligned}
        \mu A&=aA
        \ ,\\
        \mu B&=bB
        \ .
    \end{aligned}
    \right.
\end{equation}
A non-trivial solution exists if and only if
\begin{equation}
    (\mu-a)(\mu-b)=0
    \ .
\end{equation}
Using the scalar product defined in 
Eq.~\eqref{eq:appendix_scalar_product_P}, the squared norm of $\psi$ is 
\begin{equation} 
        \|\psi\|_{L^2(P)}^2
    =(\psi,\psi)_P
    =\langle\psi^2\rangle_P
    =A^2+B^2
    \ . 
\end{equation} 
The two nonzero eigenvalues and their corresponding 
normalized eigenfunctions are therefore 
\begin{equation}\label{eq:results_rank_two_eigenpairs} 
    \mu_1=b 
    \ ,\qquad e_1(\x)=h(\x) 
    \ ,\qquad 
    \mu_2=a 
    \ ,\qquad e_2(\x)=\mathbf 1 
    \ . 
\end{equation}
For $\mu=0$, Eq.~\eqref{eq:results_rank_two_eigenvalue_equation}
instead gives
\begin{equation}\label{eq:results_rank2_ON_zero_eigenfunctions}
    \langle\psi\rangle_P=0
    \ ,\qquad
    \langle h\psi\rangle_P=0
    \ .
\end{equation}
These two conditions characterize $\ker T_f$, whose nonzero
elements are eigenfunctions associated with the zero eigenvalue.
Let $\{e_k\}_{k\in\mathcal I_0}$ be an orthonormal basis of
this subspace, where
$\mathcal I_0\subseteq\{3,4,\ldots\}$ may be finite,
countably infinite or empty.
Together with $e_1$ and $e_2$, these functions form a complete
orthonormal eigenbasis
$\{e_k\}_{k\in\{1,2\}\cup\mathcal I_0}$ of $L^2(P)$,
with $\mu_k=0$ for $k\in\mathcal I_0$ and
$(e_k,e_\ell)_P=\delta_{k\ell}$, as guaranteed by the spectral
theorem for compact self-adjoint operators. Thus, $T_f$ has rank two and, since $0<a<b$, its principal
eigenvalue is
\begin{equation}\label{eq:results_rank_two_Lambda_0}
    \Lambda_0=\lambda_{\max}(T_f)=b>0
    \ .
\end{equation}
For the associated principal eigenfunction, the replica normalization condition in
Eq.~\eqref{eq:psi_normalisation} requires
\begin{equation}
    \|\psi_{\max}\|_{L^2(P)}^2
    =\lambda_{\max}(T_f)^2
    =b^2
    \ .
\end{equation}
Since $e_1(\x)=h(\x)$ has norm one, the corresponding
principal eigenfunction with this normalization is
\begin{equation}\label{eq:results_rank_two_psi_max_unperturbed}
    \psi_{\max}(\x)=b\,h(\x)
    \ ,
\end{equation}
up to an overall sign.
More generally, every eigenfunction in the principal
eigenspace $E_1$ is proportional to $h$, so that the projection of the constant function $\mathbf1$ onto $E_1$ satisfies
\begin{equation}\label{eq:results_rank_two_projection}
    \Pi_{E_1}\mathbf1
    =(\mathbf1,h)_P\,h
    =\langle h\rangle_P\,h
    =0
    \ .
\end{equation}
Thus, every eigenfunction in the principal eigenspace has
zero $P$-average. Since $\Lambda_0=b>0$ and
$\Pi_{E_1}\mathbf1=0$, the results of
Appendix~\ref{appendix:transition_thresholds} predict a transition for the
spiked kernel
\begin{equation}
    F_\theta\bigl((\x,y),(\x',y')\bigr)
    =
    a+b\,h(\x)h(\x')+\theta yy'
    \ ,
\end{equation}
both for centered and non-centered spikes. In both cases, below a positive threshold
$\theta_c$, the principal eigenpair belongs to the branch
$q[\Psi]=0$, and its eigenvalue remains equal to $b$.
Above $\theta_c$, it belongs to the branch $q[\Psi]\neq0$,
with an eigenvalue strictly greater than $b$ and a nonzero
overlap with the spike. We assume $0<\sigma^2_y<\infty$, so that
Eq.~\eqref{eq:spiked_kernel_L2_condition} holds and the
spectral problem for $T_\theta$ is well posed.
The eigenvalue equation reads
\begin{align}\label{eq:results_separable_spiked_eigenvalue_equation}
    \lambda\Psi(\x,y)=(T_{\theta}\Psi)(\x,y)
    &=
    \int \dd\x'\,\dd y'\,P(\x')G(y')
    \left[a+b\,h(\x)h(\x')+\theta yy'\right]
    \Psi(\x',y')
    \nonumber\\
    &=
    a\left\langle\bar{\Psi}\right\rangle_P
    +b\,h(\x)\left\langle h\bar{\Psi}\right\rangle_P
    +\theta yq[\Psi]
    \ ,
\end{align}
where $\bar{\Psi}$ and $q[\Psi]$ are defined in
Eqs.~\eqref{eq:barpsi} and~\eqref{eq:q_psi}.
Every eigenfunction associated with a nonzero eigenvalue
$\lambda$ therefore has the form
\begin{equation}\label{eq:results_separable_spiked_Psi}
    \Psi(\x,y)=A+B\,h(\x)+\theta Dy
    \ ,
\end{equation}
with coefficients
\begin{equation}\label{eq:results_separable_spiked_coefficients}
    A\coloneqq
    \frac{a}{\lambda}\left\langle\bar{\Psi}\right\rangle_P
    \ ,\qquad
    B\coloneqq
    \frac{b}{\lambda}\left\langle h\bar{\Psi}\right\rangle_P
    \ ,\qquad
    D\coloneqq\frac{q[\Psi]}{\lambda}
    \ .
\end{equation}
Averaging Eq.~\eqref{eq:results_separable_spiked_Psi}
over $y$ gives
\begin{equation}
    \bar{\Psi}(\x)
    =A+B\,h(\x)+\theta D\langle y\rangle_G
    \ .
\end{equation}
Inserting this expression into the definitions of $A$ and
$B$ in Eq.~\eqref{eq:results_separable_spiked_coefficients},
and inserting Eqs.~\eqref{eq:q_psi}
and~\eqref{eq:results_separable_spiked_Psi} into the
definition of $D$, yields, using
Eq.~\eqref{eq:appendix_rank_two_assumptions}
\begin{equation}\label{eq:results_separable_spiked_system}
    \left\{
    \begin{aligned}
        \lambda A
        &=a\left[A+\theta D\langle y\rangle_G\right]
        \ ,\\
        \lambda B
        &=bB
        \ ,\\
        \lambda D
        &=A\langle y\rangle_G
        +\theta D\langle y^2\rangle_G
        \ .
    \end{aligned}
    \right.
\end{equation}
We determine the principal eigenpair
$\bigl(\lambda_{\max}(T_\theta),\Psi_{\max}\bigr)$ by finding
the largest value of $\lambda$ for which
Eq.~\eqref{eq:results_separable_spiked_system} admits a
non-trivial solution. For the principal eigenfunction
$\Psi_{\max}$, we impose the normalization condition in
Eq.~\eqref{eq:spiked_psi_normalisation}. Combining this
normalization condition with the representation in
Eq.~\eqref{eq:results_separable_spiked_Psi} and the coefficient
relations in Eq.~\eqref{eq:results_separable_spiked_coefficients},
we obtain
\begin{equation}\label{eq:results_separable_norm_identity}
\begin{aligned}
    \lambda_{\max}(T_\theta)^2
    &=
    A\left\langle\Psi_{\max}\right\rangle_{P\otimes G}
    +B\left\langle h\Psi_{\max}\right\rangle_{P\otimes G}
    +\theta D\left\langle y\Psi_{\max}\right\rangle_{P\otimes G}
    \\
    &=
    \lambda_{\max}(T_\theta)
    \left(
        \frac{A^2}{a}+\frac{B^2}{b}+\theta D^2
    \right)
    \ .
\end{aligned}
\end{equation}
The normalization
constraint for the principal eigenfunction therefore
reduces to
\begin{equation}\label{eq:results_separable_normalization}
    \lambda_{\max}(T_\theta)=\frac{A^2}{a}+\frac{B^2}{b}+\theta D^2
    \ .
\end{equation}
With this normalization, the coefficient $D$ defined in
Eq.~\eqref{eq:results_separable_spiked_coefficients} gives
the limiting average overlap between the top eigenvector
and the spike, as shown in Appendix~\ref{appendix_replica_der}
\begin{equation}\label{eq:results_separable_overlap}
    D
    =
    \frac{q[\Psi_{\max}]}{\lambda_{\max}(T_\theta)}
    =
    \lim_{N\to\infty}
    \left\langle
        \frac{(\vec y,\vec v^{\, (1)})}{N}
    \right\rangle
    \ .
\end{equation}
The system in Eq.~\eqref{eq:results_separable_spiked_system}
and the normalization constraint leave an overall sign
ambiguity: replacing $(A,B,D)$ by $(-A,-B,-D)$ preserves
both. To fix this ambiguity, whenever possible, we choose the overall sign so that all
nonzero coefficients are positive. Otherwise, we require
$D>0$ when $D\neq0$; if $D=0$, we take $B>0$ when
$B\neq0$, and $A>0$ otherwise. We now solve the system in
Eq.~\eqref{eq:results_separable_spiked_system} explicitly,
treating centered and non-centered spikes separately to
recover the results of Sec.~\ref{sec:results_rank2}.

\subsection{Centered spike}\label{sec:res_rank2_centered}
When $\langle y\rangle_G=0$, the system in
Eq.~\eqref{eq:results_separable_spiked_system} becomes
\begin{equation}\label{eq:results_separable_centered_system}
    \left\{
    \begin{aligned}
        \lambda A&=aA
        \ ,\\
        \lambda B&=bB
        \ ,\\
        \lambda D&=\theta\sigma_y^2D
        \ .
    \end{aligned}
    \right.
\end{equation}
This diagonal system gives the eigenvalues
\begin{equation}\label{eq:results_rk2_explicit_roots_centered}
\begin{aligned}
    &\lambda_1
    =
    a
    \ ,\\
    &\lambda_2
    =
    \theta\sigma_y^2
    \ ,\\
    &\lambda_3
    =
    b
    \ .
\end{aligned}
\end{equation}
Since $0<a<b$, the principal eigenvalue is
\begin{equation}\label{eq:results_separable_centered_eigenvalue}
    \lambda_{\max}(T_\theta)
    =\max\{b,\theta\sigma_y^2\}
    \ .
\end{equation}
The two competing eigenvalues coincide at
\begin{equation}\label{eq:results_separable_centered_threshold}
    \theta_c=\frac{b}{\sigma_y^2}
    \ .
\end{equation}

For $0<\theta<\theta_c$, the principal eigenvalue is $b$,
and Eq.~\eqref{eq:results_separable_centered_system}
gives $A=D=0$. The normalization in
Eq.~\eqref{eq:results_separable_normalization} and our sign
convention then give $B=b$. Thus
\begin{equation}
    \lambda_{\max}(T_\theta)=b
    \ ,\qquad
    \Psi_{\max}(\x,y)=b\,h(\x)
    \ ,\qquad
    D=0
    \ .
\end{equation}
For $\theta>\theta_c$, the principal eigenvalue is
$\theta\sigma_y^2$, and the system gives $A=B=0$.
The normalization reduces to $D^2=\sigma_y^2$, so that
$D=\sigma_y$ and
\begin{equation}
    \lambda_{\max}(T_\theta)=\theta\sigma_y^2
    \ ,\qquad
    \Psi_{\max}(\x,y)=\theta\sigma_y y
    \ ,\qquad
    D=\sigma_y
    \ .
\end{equation}

These expressions recover Eqs.~\eqref{eq:summary_rank2_centered_eigenvalue}, \eqref{eq:summary_rank2_centered_eigenfunction} and  \eqref{eq:summary_rank2_centered_overlap} of the main text, together with the centered-spike results of
Appendix~\ref{appendix:transition_thresholds}. Indeed, the principal eigenpair
of $T_f$ is
\begin{equation}
    \bigl(\lambda_{\max}(T_f),\psi_{\max}(\x)\bigr)
    =\bigl(b,b\,h(\x)\bigr)
    \ ,
\end{equation}
as given in Eqs.~\eqref{eq:results_rank_two_Lambda_0}
and~\eqref{eq:results_rank_two_psi_max_unperturbed}, so that
$\theta_c=\lambda_{\max}(T_f)/\sigma_y^2$.
Below this threshold, the principal eigenpair is unchanged,
with $\Psi_{\max}(\x,y)=\psi_{\max}(\x)$ and zero overlap
$D=0$. Above it, the principal eigenfunction is proportional
to $y$, with overlap $D=\sigma_y$.

At $\theta=\theta_c$, the eigenvalues $b$ and
$\theta_c\sigma_y^2$ coincide, and the principal eigenvalue
is doubly degenerate. Since $a<b$, the first equation of
Eq.~\eqref{eq:results_separable_centered_system} gives
$A=0$, while the remaining equations place no constraint
on $B$ and $D$. The normalization condition in
Eq.~\eqref{eq:results_separable_normalization} becomes
\begin{equation}\label{eq:results_separable_centered_critical_normalization}
    \frac{B^2}{b}+\theta_cD^2=b
    \ .
\end{equation}
Using $\theta_c=b/\sigma_y^2$, the normalized principal
eigenfunctions therefore form a family with coefficients
\begin{equation}\label{eq:results_separable_centered_critical_coefficients}
    A=0
    \ ,\qquad
    B=b\sqrt{1-\frac{D^2}{\sigma_y^2}}
    \ ,\qquad
    0\leq D\leq\sigma_y
    \ .
\end{equation}
The range of $D$ follows from the normalization and our
sign convention, which also selects the solution with
$B\geq0$. The corresponding eigenfunctions are
\begin{equation}\label{eq:results_separable_centered_critical_Psi}
    \Psi_{\max}(\x,y)
    =
     b\sqrt{1-\frac{D^2}{\sigma_y^2}}\,h(\x)
    +\frac{bD}{\sigma_y^2}\,y
    \ .
\end{equation}
The endpoints $D=0$ and $D=\sigma_y$ give, respectively,
the unperturbed principal eigenfunction and the
eigenfunction proportional to $y$. Intermediate values
give normalized linear combinations of these two
eigenfunctions.

Thus, at the threshold, the limiting operator has no
unique normalized principal eigenfunction, and the system
does not select a unique overlap $D$. As in the quadratic
kernel case, determining the limiting average overlap
of the finite-size top eigenvector at this point would require
an analysis of how finite-size fluctuations resolve
the degeneracy.

\subsection{Non-centered spike}\label{sec:res_rank2_noncentered}
When $\langle y\rangle_G\neq0$, the system in
Eq.~\eqref{eq:results_separable_spiked_system} becomes
\begin{equation}\label{eq:results_separable_noncentered_system}
    \left\{
    \begin{aligned}
        (\lambda-a)A-a\theta\langle y\rangle_GD&=0
        \ ,\\
        (\lambda-b)B&=0
        \ ,\\
        -\langle y\rangle_GA
        +(\lambda-\theta\langle y^2\rangle_G)D&=0
        \ .
    \end{aligned}
    \right.
\end{equation}
The coefficient $B$ remains decoupled, while $A$ and $D$
are coupled through the spike mean.  A non-trivial solution
requires
\begin{equation}\label{eq:results_separable_characteristic}
    (\lambda-b)
    \left[
        \lambda^2
        -(a+\theta\langle y^2\rangle_G)\lambda
        +a\theta\sigma_y^2
    \right]
    =0
    \ .
\end{equation}
Thus, the eigenvalue $b$ persists for every $\theta>0$, while the quadratic factor
gives the other two roots
\begin{equation}\label{eq:results_rk2_explicit_roots_noncentered}
\begin{aligned}
    &\lambda_-(\theta)
    =
    \frac{
        a+\theta\langle y^2\rangle_G
        -\sqrt{
            (a+\theta\langle y^2\rangle_G)^2
            -4a\theta \sigma_y^2
        }
    }{2}
    =
    \frac{
        a+\theta\langle y^2\rangle_G
        -\sqrt{
            (a-\theta\langle y^2\rangle_G)^2
            +4a\theta\langle y\rangle_G^2
        }
    }{2}
    \ ,\\
    &\lambda_+(\theta)
    =
    \frac{
        a+\theta\langle y^2\rangle_G
        +\sqrt{
            (a+\theta\langle y^2\rangle_G)^2
            -4a\theta\sigma_y^2
        }
    }{2}
    =
    \frac{
        a+\theta\langle y^2\rangle_G
        +\sqrt{
            (a-\theta\langle y^2\rangle_G)^2
            +4a\theta\langle y\rangle_G^2
        }
    }{2}
    \ ,\\
    &\lambda_3=b
    \ .
\end{aligned}
\end{equation}
All three eigenvalues are real and strictly positive.
To compare them, we denote the quadratic factor in
Eq.~\eqref{eq:results_separable_characteristic} by
\begin{equation}\label{eq:results_separable_p_theta}
    p_\theta(\lambda)
    \coloneqq
    \lambda^2
    -(a+\theta\langle y^2\rangle_G)\lambda
    +a\theta\sigma_y^2=(\lambda-\lambda_+(\theta))(\lambda-\lambda_-(\theta))
    \ .
\end{equation}
Since
\begin{equation}
    p_\theta(a)=(a-\lambda_+(\theta))(a-\lambda_-(\theta))=-a\theta\langle y\rangle_G^2<0
    \ ,
\end{equation}
the roots satisfy
$0<\lambda_-(\theta)<a<\lambda_+(\theta)$.
Together with $a<b$, this implies
\begin{equation}\label{eq:results_separable_noncentered_eigenvalue}
    \lambda_{\max}(T_\theta)
    =\max\{b,\lambda_+(\theta)\}
    \ .
\end{equation}
There is a transition when $\lambda_+(\theta)=b$, or
equivalently when $p_\theta(b)=0$. Evaluating
Eq.~\eqref{eq:results_separable_p_theta} at $\lambda=b$ yields
\begin{equation}
    p_\theta(b)
    =
    b(b-a)
    -\theta\left[
        b\langle y\rangle_G^2+(b-a)\sigma_y^2
    \right]
    \ ,
\end{equation}
which vanishes at the unique positive threshold
\begin{equation}\label{eq:results_separable_noncentered_threshold}
    \theta_c
    =
    \frac{b(b-a)}
    {b\langle y\rangle_G^2+(b-a)\sigma_y^2}
    =
    \left[
        \frac{\langle y\rangle_G^2}{b-a}
        +\frac{\sigma_y^2}{b}
    \right]^{-1}
    \ .
\end{equation}
Because $b>\lambda_-(\theta)$, the sign of $p_\theta(b)$
is the sign of $b-\lambda_+(\theta)$. Consequently
\begin{equation}
    \lambda_{\max}(T_\theta)
    =
    \begin{cases}
        b, & 0<\theta\leq\theta_c,\\
        \lambda_+(\theta), & \theta>\theta_c.
    \end{cases}
\end{equation}

For $0<\theta<\theta_c$, the principal eigenvalue is
$\lambda_{\max}(T_\theta)=b$. The determinant of the
$(A,D)$ system in
Eq.~\eqref{eq:results_separable_noncentered_system}
is $p_\theta(b)>0$, so that $A=D=0$.
The normalization in
Eq.~\eqref{eq:results_separable_normalization} then gives
$B^2=b^2$, and our sign convention selects $B=b$.
The principal eigenpair and the overlap are therefore
\begin{equation}\label{eq:results_separable_noncentered_below}
    \lambda_{\max}(T_\theta)=b
    \ ,\qquad
    \Psi_{\max}(\x,y)=b\,h(\x)
    \ ,\qquad
    D=0
    \ .
\end{equation}
Thus, below the threshold, the principal eigenpair
coincides with that of the unperturbed operator $T_f$ and
the overlap with the spike vanishes.

For $\theta>\theta_c$, the principal eigenvalue is
$\lambda_{\max}(T_\theta)=\lambda_+(\theta)>b$.
The second equation of
Eq.~\eqref{eq:results_separable_noncentered_system}
gives $B=0$, while the first gives
\begin{equation}\label{eq:results_separable_noncentered_rA}
    A=r_A D
    \ ,\qquad
    r_A\coloneqq
    \frac{a\theta\langle y\rangle_G}
         {\lambda_+(\theta)-a}
    \ .
\end{equation}
Inserting this relation into the third equation yields
\begin{equation}
    \left[
        \lambda_+(\theta)-\theta\langle y^2\rangle_G
        -\langle y\rangle_G r_A
    \right]D=0
    \ .
\end{equation}
The coefficient multiplying $D$ vanishes, since by Eq.~\eqref{eq:results_separable_p_theta}
\begin{equation}
\begin{aligned}
    \lambda_+(\theta)-\theta\langle y^2\rangle_G
    -\langle y\rangle_G r_A
    &=
    \frac{
        \lambda_+(\theta)^2
        -(a+\theta\langle y^2\rangle_G)\lambda_+(\theta)
        +a\theta\sigma_y^2
    }{\lambda_+(\theta)-a}=
    \frac{p_\theta(\lambda_+(\theta))}
         {\lambda_+(\theta)-a}
    =0
    \ .
\end{aligned}
\end{equation}
The system therefore fixes the ratio $A/D=r_A$ but
leaves the magnitude of $D$ undetermined. For the principal
eigenfunction, the normalization condition in Eq.~\eqref{eq:results_separable_normalization}
reduces to
\begin{equation}
    \left(\frac{r_A^2}{a}+\theta\right)D^2
    =\lambda_+(\theta)
    \ ,
\end{equation}
and our sign convention selects the positive solution.
The principal eigenpair and the overlap are consequently
\begin{equation}\label{eq:results_separable_noncentered_above}
    \lambda_{\max}(T_\theta)=\lambda_+(\theta)
    \ ,\qquad
    \Psi_{\max}(\x,y)=D\left(r_A+\theta y\right)
    \ ,\qquad
    D=\sqrt{\frac{a\lambda_+(\theta)}{a\theta+r_A^2}}>0
    \ ,
\end{equation}
where $\lambda_+(\theta)$ is given in Eq.~\eqref{eq:results_rk2_explicit_roots_noncentered}. Above the threshold, the principal eigenfunction is
independent of $\x$ and the overlap $D$ in
Eq.~\eqref{eq:results_separable_overlap} is strictly positive.

At $\theta=\theta_c$, the eigenvalues $b$ and
$\lambda_+(\theta_c)$ coincide, and the principal eigenvalue
is doubly degenerate. Setting $\lambda=b$ in
Eq.~\eqref{eq:results_separable_noncentered_system}, the
second equation places no constraint on $B$, while the
first gives
\begin{equation}\label{eq:results_separable_noncentered_critical_rA}
    A=r_{A,c}D
    \ ,\qquad
    r_{A,c}\coloneqq
    \frac{a\theta_c\langle y\rangle_G}{b-a}
    \ .
\end{equation}
The third equation is then automatically satisfied,
since its coefficient multiplying $D$ is
$p_{\theta_c}(b)/(b-a)=0$. The normalization condition
therefore reduces to
\begin{equation}\label{eq:results_separable_noncentered_critical_normalization}
    \frac{B^2}{b}
    +\left(\frac{r_{A,c}^2}{a}+\theta_c\right)D^2
    =b
    \ .
\end{equation}
Defining
\begin{equation}\label{eq:results_separable_noncentered_Dc}
    D_c\coloneqq
    \sqrt{\frac{ab}{a\theta_c+r_{A,c}^2}}
    \ ,
\end{equation}
we obtain, with our convention selecting $D\geq0$ and
$B\geq0$, the family
\begin{equation}\label{eq:results_separable_noncentered_critical_coefficients}
    A=r_{A,c}D
    \ ,\qquad
    B=b\sqrt{1-\frac{D^2}{D_c^2}}
    \ ,\qquad
    0\leq D\leq D_c
    \ .
\end{equation}
The sign of $A$ is fixed by $\langle y\rangle_G$ whenever
$D>0$. The corresponding principal eigenfunctions are
\begin{equation}\label{eq:results_separable_noncentered_critical_Psi}
    \Psi_{\max}(\x,y)
    =
    b\sqrt{1-\frac{D^2}{D_c^2}}\,h(\x)
    +D\left(r_{A,c}+\theta_c y\right)
    \ .
\end{equation}
At $D=0$, this expression gives the unperturbed principal
eigenfunction $b\,h(\x)$, while at $D=D_c$ it gives the
limit of the principal eigenfunction in
Eq.~\eqref{eq:results_separable_noncentered_above} as
$\theta$ approaches $\theta_c$ from above. Intermediate
values give normalized linear combinations of these two
eigenfunctions.

Thus, the overlap vanishes below the threshold but
approaches the strictly positive value $D_c$ from above.
At the threshold itself, the limiting operator admits
no unique normalized principal eigenfunction, and the
system does not select a unique overlap. As in the
centered case, determining the limiting average overlap
of the finite-size top eigenvector at this point would
require an analysis of how finite-size fluctuations
resolve the degeneracy.

With the identification $\lambda_c(\theta)=\lambda_+(\theta)$,
these results recover
Eqs.~\eqref{eq:summary_rank2_noncentered_eigenvalue},
\eqref{eq:summary_rank2_noncentered_eigenfunction},
\eqref{eq:summary_rank2_noncentered_overlap}
and~\eqref{eq:summary_rank2_rA} of the main text, together
with the results of Sec.~\ref{sec:spiked model_non_centered}
for the case $\Lambda_0>0$ and $\Pi_{E_1}\mathbf1=0$.
We justify this identification below by showing that
$\lambda_+(\theta)$ satisfies the secular
equation~\eqref{eq:F_secular_equation}. In fact, here
$\Lambda_0=b>0$ by Eq.~\eqref{eq:results_rank_two_Lambda_0},
and the vanishing projection follows from the zero
$P$-average of every principal eigenfunction, as expressed
in Eq.~\eqref{eq:results_rank_two_projection}. The orthonormal eigenbasis constructed above consists of
$e_1=h$ and $e_2=\mathbf1$, given in
Eq.~\eqref{eq:results_rank_two_eigenpairs}, together with
the orthonormal basis $\{e_k\}_{k\in\mathcal I_0}$ of
$\ker T_f$, whose elements have zero $P$-average by
Eq.~\eqref{eq:results_rank2_ON_zero_eigenfunctions}.
Therefore, the coefficients $c_k$ of the constant function
in this basis, defined in Eq.~\eqref{eq:coeffs_ck}, are
\begin{equation}
    c_k=(\mathbf1,e_k)_P=\langle e_k\rangle_P
    =\delta_{k2}
    \ .
\end{equation}
Using $\mu_2=a$ from
Eq.~\eqref{eq:results_rank_two_eigenpairs}, the function
in Eq.~\eqref{eq:F_definition} therefore reduces to
\begin{equation}\label{eq:results_separable_F}
    F(\lambda)
    =\langle y\rangle_G^2
        \sum_k
        \frac{c_k^2}{\lambda-\mu_k}
        +
        \frac{\sigma_y^2}{\lambda}
    =\frac{\langle y\rangle_G^2}{\lambda-a}
    +\frac{\sigma_y^2}{\lambda}
    \ .
\end{equation}
Since $b>a$, its limit at $\lambda=b$ is finite, and
Eq.~\eqref{eq:theta_critical} gives
\begin{equation}
    \theta_c
    =\frac{1}{F(b^+)}
    =
    \left[
        \frac{\langle y\rangle_G^2}{b-a}
        +\frac{\sigma_y^2}{b}
    \right]^{-1}
    \ ,
\end{equation}
which coincides with Eq.~\eqref{eq:summary_rank2_noncentered_threshold} of the main text and
Eq.~\eqref{eq:results_separable_noncentered_threshold}. Inserting Eq.~\eqref{eq:results_separable_F} into
the secular equation \eqref{eq:F_secular_equation} gives
\begin{equation}
    \frac{\langle y\rangle_G^2}{\lambda-a}
    +\frac{\sigma_y^2}{\lambda}
    =\frac{1}{\theta}
    \ .
\end{equation}
Multiplying by $\theta\lambda(\lambda-a)$ recovers
\begin{equation}
    \lambda^2
    -(a+\theta\langle y^2\rangle_G)\lambda
    +a\theta\sigma_y^2=0
    \ ,
\end{equation}
the quadratic factor in
Eq.~\eqref{eq:results_separable_characteristic}, whose
roots are $\lambda_\pm(\theta)$ in
Eq.~\eqref{eq:results_rk2_explicit_roots_noncentered}.
We proved above that only $\lambda_+(\theta)$ can exceed $\Lambda_0=b$,
and it does so exactly when $\theta>\theta_c$.
Thus, the general analysis reproduces the transition
between the solutions in
Eqs.~\eqref{eq:results_separable_noncentered_below}
and~\eqref{eq:results_separable_noncentered_above}:
below the threshold, the principal eigenpair belongs
to the branch $q[\Psi]=0$ with eigenvalue $b$; above it,
the principal eigenpair belongs to the branch
$q[\Psi]\neq0$ with eigenvalue $\lambda_+(\theta)>b$.

\clearpage
\bibliography{biblio}
\end{document}